\documentclass[3p]{elsarticle} %

\usepackage{hyperref}
\usepackage{changepage}
\journal{Computer Methods in Applied Mechanics and Engineering}

\usepackage{lmodern}

\usepackage[T1]{fontenc} 

\usepackage[mathlines]{lineno}

\usepackage{amsmath} %
\usepackage{amssymb}
\usepackage{upgreek}
\usepackage{mathtools, nccmath} %
\biboptions{sort&compress}
\usepackage{makecell}  
\usepackage{float} %
\usepackage{algorithm}
\usepackage{algpseudocode}
\usepackage{tablefootnote}
\usepackage{subcaption} %
\usepackage[capitalise, nameinlink]{cleveref}

\usepackage{newtxmath}
\usepackage{bm} %

\usepackage{placeins}

\renewcommand{\v}[1]{\bm{\mathrm{#1}}}
\newcommand{\m}[1]{\bm{#1}} %

\usepackage{todonotes}
\let\oldtodo\todo
\renewcommand{\todo}[1]{\oldtodo[color=orange!40, inline]{\color{black}#1}}
\usepackage{etoolbox}
\makeatletter
\@for\reviewenv:=equation,equation*,align,align*,gather,gather*,multline,multline*,subequations\do{%
  \expandafter\AtBeginEnvironment\expandafter{\reviewenv}{\color{black}}}
\makeatother

\newcommand*\linenomathpatch[1]{%
  \cspreto{#1}{\linenomath}%
  \cspreto{#1*}{\linenomath}%
  \csappto{end#1}{\endlinenomath}%
  \csappto{end#1*}{\endlinenomath}%
}
\newcommand*\linenomathpatchAMS[1]{%
  \cspreto{#1}{\linenomathAMS}%
  \cspreto{#1*}{\linenomathAMS}%
  \csappto{end#1}{\endlinenomath}%
  \csappto{end#1*}{\endlinenomath}%
}
\expandafter\ifx\linenomath\linenomathWithnumbers
  \let\linenomathAMS\linenomathWithnumbers
  \patchcmd\linenomathAMS{\advance\postdisplaypenalty\linenopenalty}{}{}{}
\else
  \let\linenomathAMS\linenomathNonumbers
\fi
\linenomathpatch{equation}
\linenomathpatchAMS{gather}
\linenomathpatchAMS{multline}
\linenomathpatchAMS{align}
\linenomathpatchAMS{alignat}
\linenomathpatchAMS{flalign}

\begin{document}

\begin{frontmatter}
\title{Fracture and creep as a single constitutive problem: strength-based phase-field modelling of nucleation, propagation and creep relaxation}

\author[1]{Tim Hageman\corref{A1}}
\cortext[A1]{Corresponding author}
\ead{tim.hageman@eng.ox.ac.uk}

\address[1]{Department of Engineering Science, University of Oxford, Oxford OX1 3PJ, UK}

\begin{abstract} 
    Materials such as ice and high-temperature alloys creep under sustained loading while also developing fractures, and whether this leads to gradual deformation or sudden failure depends on how the two mechanisms interact. While traditional phase-field fracture models can describe such materials, the coupling between creep and fracture depends on constitutive choices and the material strength cannot be prescribed independently of the fracture energy and regularisation length. This paper presents a cohesive phase-field formulation in which power-law creep strains and fracture eigenstrains are internal variables of a single material model, capturing both strength-based crack nucleation, dissipation-based propagation, and bulk creep deformation. Fracture nucleates once the stress reaches an explicit strength surface, with the phase-field governing the subsequent loss of cohesion, while the creep responds to the same stress state without additional degradation functions or creep-dependent fracture driving forces. Both are updated together by a local return-mapping scheme with a consistent tangent, so no global degrees of freedom are added to a standard phase-field implementation while also not placing any priority on either dissipation mechanism. Benchmark cases are used to demonstrate the mesh independence, as well as provide guidelines for the element size. These same cases also demonstrate the effects of the creep-fracture interactions, across a range of external loading rates. Using adaptive mesh refinement, application to two large-scale settings is presented, considering fragmentation of sea-ice and crevasse growth in glacial ice-cliffs, demonstrating the ability of the model to capture large-scale phenomena. The finite element code is openly available.
\end{abstract}

\begin{keyword}
Cohesive fracture, phase-field, creep, crack nucleation, finite element method
\end{keyword}

\end{frontmatter}

\section{Introduction}
\label{sec:intro}

Materials subjected to sustained loading can deform through creep while also developing fractures, with the interaction between these mechanisms determining when and how failure occurs \citep{ElmukashfiCocks2017}. This interaction is relevant to high-temperature engineering applications, where the temperature being close to the melting point allows creep to gradually redistribute stresses and accumulate deformation during the material's lifetime \citep{Riedel1981}. For instance, components used in fusion devices must resist both high-temperature deformation and cracking induced by thermal gradients \citep{RiethDafferner2005,Rieth2013,CrosbyGhoniem2012}. Predicting their behaviour requires differentiating slow deformation and time-dependent crack growth from sudden failure due to fracture \citep{Xie2023,Si2025}. At these temperatures, creep can relax the stresses surrounding a crack or defect, delaying further fracture, but also redistributes the stresses to potentially be more favourable for fracture \citep{RiedelRice1980,Riedel1981}. A model for these materials therefore needs to capture both mechanisms and their interaction over the relevant loading timescales.

A different application showing similar interactions is ice sheets and glaciers. Although their absolute temperatures are low compared to engineering metals, ice is close to its melting point and therefore undergoes substantial viscous creep \citep{Glen1955,SchulsonDuval2009,CuffeyPaterson2010}. This slow deformation occurs alongside crevasse formation and calving, potentially causing failure and ice mass-loss. Continuum damage models have been developed to describe the creep and progressive failure of ice \citep{Duddu2012a,Jimenez2017}, including hydrofracture in grounded glaciers and rift propagation through ice shelves \citep{Duddu2020,Huth2023}. However, the importance of creep is not restricted to long-term processes: viscous deformation can also strongly influence hydraulic fracture during supraglacial lake drainage, over timescales of minutes to hours \citep{Hageman2024}. These fracture processes have also been studied using phase-field models, with applications to glacier hydrofracture \citep{Sun2021,Clayton2022}, viscoelastic fracture of floating and freshwater ice \citep{Sondershaus2023,Montmayeur2023}, and mixed-mode ice-cliff failure \citep{Clayton2025,Hageman2026}. As these applications cover large domains, adaptive mesh refinement schemes have also been used to accelerate computation in a three-dimensional setting \citep{Gupta2026}.

Phase-field fracture methods approximate cracks as a continuous field, allowing their paths to evolve without explicitly inserting or tracking discrete crack faces. This method builds on the variational formulation of brittle fracture \citep{FrancfortMarigo1998}, using a regularized surface-energy functional to distribute a discrete crack over a finite region \citep{Ambrosio1990,Bourdin2000}. As a result, crack propagation and branching can be captured without special considerations, including under dynamic loading \citep{Borden2012,Hofacker2013}. The resulting phase-field equations can be implemented in standard finite element software \citep{Molnar2017,Navidtehrani2021}, while energy decompositions and staggered solution schemes allow for tuning the apparent strength of materials and efficient numerical treatment \citep{Amor2009,Miehe2010,Miehe2010a,Ambati2015b}. 

Extending phase-field methods to include viscous creep or plastic deformations requires additional constitutive choices for how these deformation mechanisms interact with existing cracks and contribute to new cracks forming. For instance, the phase-field method has been combined with plasticity to describe fracture in elastic-plastic solids \citep{Duda2015}, with accumulated plastic strain used to couple plastic deformation and damage \citep{Ambati2015,Ambati2016}. Other formulations include degradation of the plastic resistance and stress-triaxiality effects \citep{Borden2016}, or combine gradient plasticity with phase-field fracture \citep{Miehe2016Plasticity}. These models can capture transitions between brittle fracture, plasticity-driven failure, and ductile flow \citep{Alessi2018a,Choo2018}, but their predictions depend on the constitutive choices used to couple these mechanisms \citep{Alessi2018}. Their interaction can also introduce difficulties in the numerical convergence \citep{Hageman2021d}. Similar extensions have been made for viscous materials, capturing viscoelastic fracture and relaxation \citep{Shen2019,Yin2020a,Dammas2021,Kamarei2025}, as well as creep fracture and creep-fatigue interactions \citep{Xie2023,Xue2024,Si2025}.

A first modelling choice is which stress drives the creep or plastic deformation \citep{Alessi2018,Shen2019}. In standard phase-field formulations, damage reduces the elastic stiffness, such that the stress transferred through the damaged material differs from the effective stress obtained using the undamaged elastic law \citep{Miehe2010,Amor2009}. At a given elastic strain, using the damaged stress to drive creep reduces the creep rate as damage lowers the stress, whereas using the undamaged stress can retain substantial creep within the damaged region. The choice of degradation functions for the elastic energy and viscous deformation similarly determines how damage influences stress relaxation \citep{Dammas2021}. Furthermore, these choices impact the direction of the plastic strains, determining how plastic deformation continues within the area around a developing fracture \citep{Borden2016,Choo2018,Miehe2016Plasticity}. Inversely, a choice needs to be made for how the plastic deformation contributes to the driving force for fractures. Existing formulations use elastic energy, hardening, accumulated plastic work, or stress-based criteria \citep{Miehe2015,Alessi2018,Clayton2022}. For viscous materials, part of the accumulated viscous dissipation can be included in this driving force \citep{Shen2019,Dammas2021}, while other models only use stored elastic contributions \citep{Yin2020a,Kamarei2025}. Alternatively, the fracture resistance can be made dependent on the accumulated creep strain \citep{Xie2023,Si2025}. These choices describe different assumptions about the material behaviour \citep{Alessi2018}. Although thermodynamic arguments constrain the possible formulations, they do not uniquely determine which coupling should be used. Additional driving forces and degradation rules therefore need physical justification and calibration \citep{Dammas2021,MengWang2019}.

A further limitation of standard brittle phase-field models is the difficulty of prescribing a material strength independently of the fracture energy. As phase-field models are based on Griffith's brittle fracture mechanics, they do not contain separate criteria for nucleation and propagation. The stress for tensile fracture within phase-field frameworks scales as $f_\text{t} \propto \sqrt{E G_\text{c}/\ell}$, with $E$ the elastic modulus, $G_\text{c}$ the fracture energy, and $\ell$ the regularization length \citep{Pham2011,Tanne2018}. Treating $\ell$ as a material parameter allows a tensile strength to be defined, but also fixes the width over which the fracture is distributed \citep{Kristensen2021,Mandal2019}. This departs from the mathematical concept of having $\ell$ be a purely numerical indicator of the area the phase-field is spread over, instead fixing the element-size required, and thus the scale of the problems that can be simulated. Furthermore, under multiaxial loading, the energy decomposition determines both the shape of the implied strength surface and the residual stresses in fractured material \citep{DeLorenzis2021,Vicentini2024}. Parametrised decompositions give some control over the ratio of compressive to tensile strength, but the strength under other stress states remains a consequence of the chosen decomposition, which simultaneously governs the behaviour of the fractured material. The strength surface can therefore not be prescribed independently of the crack behaviour as long as the fracture driving force is derived from the decomposed elastic energy \citep{Lopez-Pamies2025}.

Cohesive phase-field formulations have been developed to separate the material strength from the energy dissipated as cracks open. Gradient-damage and cohesive approximations establish connections between the diffuse phase-field description and traction-separation laws \citep{Lorentz2011,Conti2016,Wu2017}. Analytical constructions can also determine the energy density from a prescribed cohesive law \citep{Feng2021}, with directional energy decomposition extending this treatment to mixed-mode fracture \citep{Feng2023}. Through suitable choices of constitutive functions, the sensitivity to the regularization length can be substantially reduced \citep{WuNguyen2018,Geelen2019,Hai2025}, although both the softening response and the associated energy dissipation need to be assessed \citep{Chen2021}. Alternatively, auxiliary fields can be used to describe the crack opening and introduce cohesive interactions directly \citep{Verhoosel2013,GhaffariMotlagh2020}, with recent formulations also allowing strong displacement discontinuities within the diffuse description \citep{Feng2026}. Cohesive phase-field models have also been extended to ductile failure \citep{Talamini2021} and thermo-viscoelastic materials, combining a generalized Maxwell bulk response with a fracture driving force based on elastic energy and viscous dissipation \citep{MunetonLopez2026}.

Recent work by Bourdin et al. \citep{Bourdin2025} and Vicentini et al. \citep{Vicentini2026} introduces an explicit strength surface within a variationally consistent cohesive phase-field formulation. Fracture eigenstrains represent crack opening and a strength potential describes the associated cohesive resistance. The strength surface determines when these eigenstrains start to develop, allowing cracks to nucleate in initially intact material without a pre-inserted crack. The phase-field then describes the loss of cohesion, with the fracture energy governing the energy dissipated during this process. This separates the prescribed nucleation strength from both the fracture energy and the regularization length. Since the eigenstrains require no spatial gradients, they can be resolved locally using a return-mapping scheme, without adding global degrees of freedom \citep{Hageman2026b}. Recent extensions to dynamic fracture also show the ability of these methods to properly capture dynamic cracks and reflect stress waves at pre-existing cracks \citep{Heinzmann2026a}.

The aim of this work is to extend this cohesive strength-surface formulation to creep, retaining the explicit description of crack nucleation while allowing both mechanisms to act within a single material model. Creep strains are introduced alongside the fracture eigenstrains, with constitutive relations derived from the elastic and cohesive energies together with a power-law creep dissipation potential. Both constitutive relations are solved simultaneously, so that creep and fracture respond to the same updated stress state without imposing priority on either mechanism: creep can relax the stresses that would otherwise cause fracture, while the loss of cohesion reduces the stresses available to drive further creep. Material strengths and fracture energy are kept constant, with no additional fracture driving force or degradation based on accumulated creep, and the rate dependence follows from the bulk creep law alone. Inertia is included to allow rapid fracture under slowly applied loading, with damping limiting the effects of stress waves. The governing equations and numerical implementation are described in \cref{sec:methods}, followed by the benchmark studies in \cref{sec:results_2D_plate} to show that the results of the presented scheme converge to a state that is independent of the element size and the phase-field length scale, once this length scale is sufficiently small relative to the geometric features. Finally, applicability of the model to large-scale application cases is shown in \cref{sec:applications}.

\section{Governing equations and numerical implementation}

\label{sec:methods}

We consider a material described by its displacement field $\v{u}$, and the phase-field $\phi$. The material is assumed to be isotropic, and allows for viscous creep described by the creep strain $\m{\varepsilon}_\text{v}$, which is treated as a local variable. To describe the jump in displacements across cracks, we furthermore consider the fracture eigenstrains $\m{\eta}$ \citep{Vicentini2026}, also treated as a local variable \citep{Hageman2026b}. As a result of these viscous and fracture-related strains, the elastic strain $\m{\varepsilon}_\text{el}$ is given by the decomposition:
\begin{equation}
    \label{eq:strain_decomposition}
    \m{\varepsilon}_\text{el} = \m{\varepsilon}-\m{\varepsilon}_\text{v}-\m{\eta}= \nabla^s\v{u}-\m{\varepsilon}_\text{v}-\m{\eta}
\end{equation}
where $\nabla^\text{s}$ indicates the symmetric gradient operator, $\nabla^\text{s}\v{u}=\frac{1}{2}(\nabla\v{u}+\nabla\v{u}^\top)$. Small strains are adopted throughout, limiting ourselves to timescales where the creep has a noticeable impact on the stress state but does not yet result in significant changes in geometry or large strains. This allows us to focus on the coupling between bulk creep and cohesive fracture, and the interactions between these two dissipative phenomena. Extension to large strains is possible, but left for future work.  

To obtain a thermodynamically consistent set of equations, the governing equations are derived from a variational principle, based on the energy and dissipation potentials of the system \citep{Houlsby2007}. The elastic energy is given based on the elastic strains from \cref{eq:strain_decomposition} as:
\begin{equation}
    \label{eq:elastic_energy}
    \Psi_\text{el} = \int_\Omega\frac{1}{2}K\text{tr}^2(\m{\varepsilon}-\m{\varepsilon}_\text{v}-\m{\eta})+\mu\lVert\text{dev}(\m{\varepsilon}-\m{\varepsilon}_\text{v}-\m{\eta})\rVert^2 \;\text{d}\Omega
\end{equation}
where $K$ and $\mu$ are the bulk and shear moduli, respectively, $\text{dev}(\m{\varepsilon})=\m{\varepsilon}-\frac{1}{3}\text{tr}(\m{\varepsilon})\m{I}$ is the deviatoric part of the elastic strain tensor, and $\lVert \m{\varepsilon}\rVert$ indicates the Frobenius norm. The kinetic energy, damping dissipation potential, and external load potential are given by:
\begin{align}
    \label{eq:kinetic_external_energy}
    \Psi_\text{i} &= \int_\Omega\frac{1}{2}\rho \dot{\v{u}}^\top\dot{\v{u}}\;\text{d}\Omega \\
    \label{eq:damping_potential}
    \Phi_\text{d} &= \int_\Omega \frac{1}{2}\rho c\,\dot{\v{u}}^\top\dot{\v{u}}\;\text{d}\Omega \\
    \label{eq:external_work}
    \Psi_\text{b} &= \int_\Omega \v{b}^\top\v{u}\;\text{d}\Omega+\int_\Gamma\v{\tau}^\top\v{u}\;\text{d}\Gamma
\end{align}
where $\rho$ is the density, $c$ is the damping coefficient, $\v{b}$ is the body forces, and $\v{\tau}$ is the external traction on boundary $\Gamma$.  Inertia is included to capture the difference in time-scales between the slow effects of creep and the rapid stress changes due to fracture, allowing for a clear distinction between dissipation-driven propagation and stress-driven propagation. Damping is added to mitigate the effect of travelling stress waves.

Viscous creep is described by the dissipation potential $\Phi_\text{v}$, using a power-law creep model with parameters $A$ and $n$ \citep{Hutchinson1976, Kamarei2025}:
\begin{equation}
    \label{eq:creep_potential}
    \Phi_\text{v} = \int_\Omega \frac{n}{n+1}A^{-1/n}\left(\frac{2}{3}\m{\dot{\varepsilon}}_\text{v}:\m{\dot{\varepsilon}}_\text{v}\right)^{\frac{n+1}{2n}}\;\text{d}\Omega \qquad \text{subject to} \qquad \text{tr}(\m{\dot{\varepsilon}}_\text{v})=0
\end{equation}
with the incompressibility constraint added to the creep strains to ensure that the creep becomes purely driven by deviatoric stresses. We note that, similarly to the damping dissipation potential from \cref{eq:damping_potential}, the creep dissipation potential is expressed in terms of a rate and has units of power, whereas the $\Psi$ contributions have units of energy. The derivatives of these dissipation potentials with respect to $\dot{\v{u}}$ and $\m{\dot{\varepsilon}}_\text{v}$ give the damping force and deviatoric creep stress, respectively. Consequently, these derivatives act on variations of the associated state variables in the variational principle, while the energy contributions enter through their direct variations.

The total energy dissipated by the loss of cohesion due to cracks is given in terms of the fracture energy $G_\text{c}$ as \citep{Bourdin2000,Amor2009,Miehe2010}:
\begin{equation}
    \label{eq:fracture_energy}
    \Psi_\text{c} = \int_{\Gamma_\text{c}} G_\text{c} \;\text{d}\Gamma \approx \int_\Omega G_\text{c} \gamma(\phi) \;\text{d}\Omega \qquad \gamma(\phi) = \frac{1}{2\ell}(\phi^2 + \ell^2 |\nabla \phi|^2)
\end{equation}
where a quadratic (AT-2) crack density function $\gamma(\phi)$ is used to approximate the fracture energy over the diffuse crack region \citep{Ambrosio1990}, with $\ell$ the length scale parameter controlling the width of the diffuse crack. Unlike standard phase-field schemes, where $\phi$ indicates the presence of cracks, the eigenstrain-based phase-field model uses $\phi$ to indicate the loss of cohesion, with $\phi=0$ indicating a fully cohesive material and $\phi=1$ indicating a fully fractured material. The fracture energy is therefore dissipated progressively as cohesion is lost during crack opening or sliding.

Finally, following strength-based variational formulations of cohesive fracture \citep{Bourdin2025,Vicentini2026}, the material strength is included through the strength potential:
\begin{equation}
    \label{eq:strength_energy}
    \Psi_\text{f} = \int_\Omega d(\phi) F_\text{d}(\m{\eta}) + F_\text{i}(\m{\eta}) \;\text{d}\Omega
\end{equation}
The strength potential represents the energy associated with cohesive forces resisting the fracture eigenstrain $\m{\eta}$. Developing this eigenstrain reduces the elastic strain in \cref{eq:strain_decomposition}, and therefore the elastic energy, but introduces an energetic cost through $\Psi_\text{f}$. For initially intact material, the fracture eigenstrain remains zero until the stress reaches the strength surface prescribed by the potential. The damageable component $F_\text{d}(\m{\eta})$ describes the material strength related to crack opening and sliding, and also provides the driving force for loss of cohesion. The undamageable component $F_\text{i}(\m{\eta})$ penalizes interpenetration of crack faces. As the phase-field increases, the degradation function $d(\phi)$ reduces the material strength while providing a contribution to the phase-field driving energy. A standard quadratic degradation function is used:
\begin{equation}
    \label{eq:strength_degradation}
    d(\phi) = (1-\kappa)(1-\phi)^2+\kappa
\end{equation}
using a small residual strength factor $\kappa$ to avoid numerical ill-conditioning when the phase-field approaches one. At a fixed phase-field, the strength potential describes the limits of elastic deformation \citep{Vicentini2026}, while irreversible loss of cohesion is represented through the evolution of the phase-field and the associated fracture energy $\Psi_\text{c}$. The strength potential therefore prescribes the initial strength surface associated with crack nucleation, while $G_\text{c}$ governs the energy dissipated during propagation through loss of cohesion.

\subsection{Variational formulation and weak forms}
Collecting the energy contributions as $\Psi=\Psi_\text{el}+\Psi_\text{c}+\Psi_\text{f}-\Psi_\text{b}-\Psi_\text{i}$, the governing equations and constitutive models follow from the variational principle:
\begin{equation}
    \label{eq:variational_principle}
    0=\delta\int_{t_0}^{t_1}\Psi\;\text{d}t +\int_{t_0}^{t_1}\left( \frac{\delta\Phi_\text{d}}{\delta\dot{\v{u}}}\cdot\delta\v{u} +\frac{\delta\Phi_\text{v}}{\delta\m{\dot{\varepsilon}}_\text{v}} :\delta\m{\varepsilon}_\text{v} \right)\;\text{d}t
\end{equation}
Here, the dissipative terms are differentiated with respect to the rates, while the resulting derivatives act on the state variations $\delta\v{u}$ and $\delta\m{\varepsilon}_\text{v}$. We note that the time-integrals are included to allow for the inclusion of inertia, re-writing the variations in this term as:
\begin{equation}
    \label{eq:kinetic_variation}
    \delta\int_{t_0}^{t_1}\Psi_\text{i}\;\text{d}t =\int_{t_0}^{t_1}\delta\dot{\v{u}}^\top\frac{\delta\Psi_\text{i}}{\delta\dot{\v{u}}}\;\text{d}t =\int_{t_0}^{t_1}\int_\Omega\rho\,\delta\dot{\v{u}}^\top\dot{\v{u}}\;\text{d}\Omega\;\text{d}t = -\int_{t_0}^{t_1}\int_\Omega\rho\,\delta\v{u}^\top\ddot{\v{u}}\;\text{d}\Omega\;\text{d}t + \left[ \int_\Omega \rho \delta \v{u}^\top \dot{\v{u}}\;\text{d}\Omega \right]_{t_0}^{t_1}
\end{equation}
noting that this used integration-by-parts to reduce the variations in velocity to variations in displacement, with the temporal boundary terms vanishing. For brevity, we will not explicitly write the time-integral terms in the following, but they are implied in all variations of the kinetic and dissipative contributions.

Taking the variation of the displacement $\v{u}$ in \cref{eq:variational_principle} gives the weak form of the momentum balance:
\begin{equation}
    \label{eq:momentum_weak}
    0=\left(\frac{\delta\Psi}{\delta\v{u}}+\frac{\delta\Phi_\text{d}}{\delta\dot{\v{u}}}\right)\cdot\delta\v{u}\\=\int_\Omega\nabla^\text{s}\delta\v{u}^\top:\m{\sigma}+\delta\v{u}^\top\left(\rho\ddot{\v{u}}+\rho c\dot{\v{u}}-\v{b}\right)\;\text{d}\Omega-\int_\Gamma\delta\v{u}^\top\v{\tau}\;\text{d}\Gamma
\end{equation}
and variations of the phase-field $\phi$ result in the evolution equation for the phase-field:
\begin{equation}
    \label{eq:phase_weak}
    0=\frac{\delta\Psi}{\delta\phi}\,\delta\phi=\int_\Omega\frac{G_\text{c}}{\ell}\phi\,\delta\phi+G_\text{c}\ell\,\nabla\delta\phi\cdot\nabla\phi-2(1-\kappa)(1-\phi)F_\text{d}(\m{\eta})\,\delta\phi\;\text{d}\Omega
\end{equation}
To enforce irreversibility of the phase-field, a history variable is used to replace $F_\text{d}(\m{\eta})$ in \cref{eq:phase_weak}, with this history variable defined as \citep{Miehe2010a}:
\begin{equation}
    \mathcal{F}=\max_{\tau\in[0,t]}F_\text{d}(\m{\eta}(\tau))
\end{equation}  
The use of this history variable ensures that the phase-field can only increase, and therefore that the fracture energy is a purely dissipative contribution. 

While the finite element scheme used in this work directly uses the weak forms from \cref{eq:momentum_weak,eq:phase_weak}, the strong forms of these equations, provided for completeness and readability, are given by:
\begin{align}
    &\rho \ddot{\v{u}} + \rho c \dot{\v{u}}- \nabla \cdot \m{\sigma}(\nabla^\text{s}\v{u}, \m{\varepsilon}_\text{v}, \m{\eta}) = \v{b} \\
    &\frac{G_\text{c}}{\ell}\left(\phi - \ell^2 \nabla^2 \phi\right) = 2(1-\kappa)(1-\phi)\mathcal{F}(\m{\eta}) 
\end{align}
with these equations solving for the displacement $\v{u}$ and phase-field $\phi$. In addition to these equations, constitutive models for the stress $\m{\sigma}$, creep strain $\m{\varepsilon}_\text{v}$, and fracture eigenstrain $\m{\eta}$ are required to close the system of equations. These constitutive models are obtained from the variational principle in \cref{eq:variational_principle}, as described in the remainder of this subsection.

The stresses in \cref{eq:momentum_weak} are obtained from the elastic energy in \cref{eq:elastic_energy} via:
\begin{equation}
\frac{\delta\Psi_\text{el}}{\delta\m{\varepsilon}}:\delta\m{\varepsilon}=\int_\Omega\frac{\partial\psi_\text{el}}{\partial\m{\varepsilon}}:\delta\m{\varepsilon}\;\text{d}\Omega=\int_\Omega\m{\sigma}:\delta\m{\varepsilon}\;\text{d}\Omega
\end{equation}
where the symbol $\psi_\text{el}$ indicates the local elastic energy density. This then leads to the definition of the stress tensor $\m{\sigma}$ as:
\begin{equation}
    \label{eq:stress_constitutive}
    \m{\sigma}=\frac{\partial\psi_\text{el}}{\partial\m{\varepsilon}} =K\text{tr}(\m{\varepsilon}-\m{\varepsilon}_\text{v}-\m{\eta})\m{I} +2\mu\text{dev}(\m{\varepsilon}-\m{\varepsilon}_\text{v}-\m{\eta})
\end{equation}
with $\m{I}$ the identity matrix. 

\subsubsection{Creep constitutive model}
The constitutive model for the viscous creep is obtained from \cref{eq:variational_principle} by taking the variation of the creep strain $\m{\varepsilon}_\text{v}$:
\begin{equation}
    \label{eq:creep_variation}
    0=\left( \frac{\delta\Psi}{\delta\m{\varepsilon}_\text{v}} +\frac{\delta\Phi_\text{v}}{\delta\m{\dot{\varepsilon}}_\text{v}} \right):\delta\m{\varepsilon}_\text{v}
\end{equation}
For the first term, we note that $\delta \psi_\text{el}/\delta \m{\varepsilon}_\text{v} = -\delta \psi_\text{el}/\delta \m{\varepsilon}_\text{el} = -\m{\sigma}$. The second term is obtained by differentiating the creep potential in \cref{eq:creep_potential} with respect to the creep-rate tensor, giving a power-law relation between the deviatoric stress and the creep rate. The resulting constitutive model for the creep is therefore:
\begin{equation}
    \label{eq:creep_weak}
    0=\int_\Omega\left( -\m{\sigma}+\frac{2}{3}A^{-1/n} \left(\frac{2}{3}\m{\dot{\varepsilon}}_\text{v}:\m{\dot{\varepsilon}}_\text{v}\right)^{\frac{1-n}{2n}} \m{\dot{\varepsilon}}_\text{v} \right):\delta\m{\varepsilon}_\text{v}\;\text{d}\Omega
\end{equation}
Following from the incompressibility of the creep strain, $\m{\sigma}:\delta \m{\varepsilon}_\text{v} = \m{\sigma}_\text{dev}:\delta \m{\varepsilon}_\text{v}$, the first term in \cref{eq:creep_weak} can be expressed in terms of the deviatoric stress. This simplification also directly enforces the incompressibility condition, removing the need for additional constraints. Writing the resulting equation in its strong form, we obtain the local constitutive relation for the creep:
\begin{equation}
    \label{eq:creep_stress_rate}
    0=-\m{\sigma}_\text{dev}+\frac{2}{3}A^{-1/n} \left(\frac{2}{3}\m{\dot{\varepsilon}}_\text{v}:\m{\dot{\varepsilon}}_\text{v}\right)^{\frac{1-n}{2n}} \m{\dot{\varepsilon}}_\text{v}
\end{equation}
Taking the self-contraction of \cref{eq:creep_stress_rate} gives
\begin{equation}
    \frac{3}{2}\m{\sigma}_\text{dev}:\m{\sigma}_\text{dev} =A^{-2/n} \left(\frac{2}{3}\m{\dot{\varepsilon}}_\text{v}:\m{\dot{\varepsilon}}_\text{v}\right)^{1/n}
\end{equation}
which upon substitution back into the tensor relation gives the explicit power-law creep flow rule:
\begin{equation}
    \label{eq:creep_constitutive}
    \m{\dot{\varepsilon}}_\text{v} =\frac{3}{2}A \left(\frac{3}{2}\m{\sigma}_\text{dev}:\m{\sigma}_\text{dev}\right)^{\frac{n-1}{2}} \m{\sigma}_\text{dev}
\end{equation}
The creep rate is therefore aligned with the deviatoric stress. To express this flow rule in terms of a scalar viscoplastic strain multiplier $\lambda_\text{v}$, we define the stress function $f_\text{v}$ and its derivative as:
\begin{equation}
    \label{eq:creep_flow_direction}
    f_\text{v}(\m{\sigma}) =\sqrt{\frac{3}{2}\m{\sigma}_\text{dev}:\m{\sigma}_\text{dev}} \qquad \frac{\partial f_\text{v}}{\partial\m{\sigma}} =\frac{3}{2f_\text{v}}\m{\sigma}_\text{dev}
\end{equation}
allowing \cref{eq:creep_constitutive} to be written equivalently as
\begin{equation}
    \label{eq:creep_multiplier}
    \m{\dot{\varepsilon}}_\text{v} =\dot{\lambda}_\text{v}\frac{\partial f_\text{v}}{\partial\m{\sigma}} \qquad 0=\dot{\lambda}_\text{v}-A\left(f_\text{v}(\m{\sigma})\right)^n
\end{equation}
where $\dot{\lambda}_\text{v}$ is the multiplier rate, required to be non-negative to enforce creep to act as a dissipative mechanism. While the viscous creep model here is aimed to capture power-law creep, the above formulation can be readily extended to visco-plasticity models through the introduction of a yield strength. One interesting note is that the creep rate has no direct interactions with the damage functions or fracture eigenstrains. Instead, these interactions are indirectly imposed through the stress state, being restricted by the strength surface and the fracture eigenstrains. This removes the need to consider a constitutive model for the driving stresses, which in standard phase-field models requires choosing the damaged or undamaged stress state, as well as considering whether creep is driven by a consistent or inconsistent stress split. The cohesive fracture model presented here therefore allows for a more straightforward and physically consistent treatment of the creep-fracture interaction, retaining a clear distinction between creep being the dissipative mechanism for the bulk material, and cohesive fracture being the dissipative mechanism for the crack surfaces. 

\subsubsection{Fracture constitutive model}
\label{sec:fracture_constitutive}

The constitutive model for the fracture eigenstrains follows from taking variations in the eigenstrain $\m{\eta}$ in \cref{eq:variational_principle}. The fracture eigenstrains require that admissible variations cannot reduce the total energy, obtaining the inequality:
\begin{equation}
    \label{eq:fracture_variation}
    0\leq\frac{\delta\Psi}{\delta\m{\eta}}:\delta\m{\eta} = \left(\frac{\delta \Psi_\text{el}}{\delta \m{\eta}} + \frac{\delta \Psi_\text{f}}{\delta \m{\eta}}\right):\delta\m{\eta}
\end{equation}
Similar as for the viscous creep, the first term can be expressed in terms of the stress tensor as $\delta \psi_\text{el}/\delta \m{\eta} = -\delta \psi_\text{el}/\delta \m{\varepsilon}_\text{el} = -\m{\sigma}$. The second term is obtained by differentiating the strength potential in \cref{eq:strength_energy} with respect to the fracture eigenstrain, giving:
\begin{equation}
    \label{eq:fracture_weak}
    0\leq\int_\Omega\left(-\m{\sigma}+d(\phi)\frac{\partial F_\text{d}}{\partial\m{\eta}}+\frac{\partial F_\text{i}}{\partial\m{\eta}}\right):\delta\m{\eta}\;\text{d}\Omega
\end{equation}

The inequality of \cref{eq:fracture_weak} needs to hold for all admissible variations of the eigenstrain tensor. Since the stress and the strength potentials are isotropic, the eigenstrain resulting from this minimization shares its principal directions with the strain remaining after viscous creep, $\m{\varepsilon}-\m{\varepsilon}_\text{v}$, as shown in \ref{app:directions}. The directions of the fracture eigenstrain are therefore known from $\m{\varepsilon}-\m{\varepsilon}_\text{v}$, and only its magnitude in these directions needs to be determined. This allows the fracture eigenstrain $\m{\eta}$ to be defined in terms of a set of multipliers $\lambda_i$ and their associated direction tensors $\m{G}_i$ \citep{Hageman2026b}. The eigenstrains are then given as:
\begin{equation}
    \label{eq:eigenstrain_directions}
    \m{\eta}=\sum_i\lambda_i\m{G}_i\qquad\lambda_i\geq0\qquad\frac{\partial\m{\eta}}{\partial\lambda_i}=\m{G}_i
\end{equation}
and enables the fracture constitutive model to be expressed in terms of the variations in the multipliers $\lambda_i$:
\begin{equation}
    \label{eq:fracture_multiplier_variation}
    0\leq\sum_i \int_\Omega\left(
    -\m{\sigma}
    +d(\phi)\frac{\partial F_\text{d}}{\partial\m{\eta}}
    +\frac{\partial F_\text{i}}{\partial\m{\eta}}
    \right):\frac{\partial \m{\eta}}{\partial \lambda_i}\delta\lambda_i\;\text{d}\Omega
\end{equation}
where the direction tensors that correspond to the strength potentials used in this work are given below.

Finally, \cref{eq:fracture_multiplier_variation} is written in its strong form, giving a set of $i$ constitutive relations for the fracture multipliers:
\begin{equation}
    \label{eq:fracture_multiplier_derivative}
    0\leq\frac{\partial\m{\eta}}{\partial\lambda_i}:
    \left(
    -\m{\sigma}
    +d(\phi)\frac{\partial F_\text{d}}{\partial\m{\eta}}
    +\frac{\partial F_\text{i}}{\partial\m{\eta}}
    \right)
\end{equation}
For this inequality to hold, the stress state must remain within the strength surface defined through the strength potential, in which case $\lambda_i=0$, or be on the strength surface, resulting in $\lambda_i>0$ and non-zero fracture eigenstrains. Similarly as for the creep constitutive model, interactions between the fracture and creep are solely through the stress state, which depends on both the fracture eigenstrain and viscous creep strain. The fracture constitutive model therefore allows for a clear distinction between the dissipative mechanisms of bulk creep and cohesive fracture, while still allowing for a physically consistent interaction between these two phenomena.

Two strength criteria are considered in this work, which share their fracture potential but differ in the fracture eigenstrains that are admissible. The first is a two-multiplier criterion that only prevents the trace of the eigenstrain from becoming negative. The second restricts the eigenstrain to opening, and is given in \cref{sec:opening_criterion}. For the two-multiplier criterion, the fracture potential and interpenetration potential are given by:
\begin{equation}
    \label{eq:r2_2l_potentials}
    F_\text{d}(\m{\eta})=\sqrt{{f_\text{t}^*}^2{\left\langle\text{tr}(\m{\eta})\right\rangle^+}^2+{f_\text{s}^*}^2\left\lVert\text{dev}(\m{\eta})\right\rVert^2} \qquad F_\text{i}(\m{\eta})=-10^6f_\text{t}^* \left\langle\text{tr}(\m{\eta})\right\rangle^-
\end{equation}
where the Macaulay brackets are used to select the positive and negative parts of the trace, capturing tensile fractures while preventing crack interpenetration under compression. The potential coefficients $f_\text{t}^*$ and $f_\text{s}^*$ are calibrated from the prescribed uniaxial tensile and compressive strengths $f_\text{t}$ and $f_\text{c}$, with $f_\text{c}>f_\text{t}$. For a three-dimensional setting, the coefficients are:
\begin{equation}
    \label{eq:r2_strength_scales_3d}
    f_\text{t}^* = \frac{f_\text{t}}{3\sqrt{1-(f_\text{t}/f_\text{c})^2}} \qquad f_\text{s}^* = \sqrt{\frac{2}{3}}f_\text{c}
\end{equation}

However, to more clearly show evolutions of the stress state, most of the results in this work are presented in a ``true'' two-dimensional setting, explicitly assuming a two-dimensional material (thus no out-of-plane strains) with a two-dimensional stress state, $\m{\sigma}=[\sigma_{xx} \;\;\sigma_{xy}; \sigma_{yx} \;\; \sigma_{yy}]$. In this case, the coefficients are instead given by:
\begin{equation}
    \label{eq:r2_strength_scales_2d}
    f_\text{t}^* = \frac{f_\text{t}}{2\sqrt{1-(f_\text{t}/f_\text{c})^2}} \qquad f_\text{s}^* = \frac{f_\text{c}}{\sqrt{2}}
\end{equation}
to calibrate the strength surface against uniaxial tension and compression. This two-dimensionality carries through in the definitions of the trace operators and deviatoric operators as well, e.g. $\text{dev}(\m{\eta})=\m{\eta}-\frac{1}{2}\text{tr}(\m{\eta})\m{I}$. It is stressed that this is a purely mathematical simplification to allow the results to be studied more clearly, without the need to consider out-of-plane effects such as thinning or stress accumulation in the out-of-plane direction under the influence of creep. It should not, however, be interpreted as using either a plane-stress or plane-strain reduction of a three-dimensional material, as the resulting stress states are distinct from these reductions.

The two direction tensors are constructed from the strain remaining after viscous creep, $\m{\varepsilon}-\m{\varepsilon}_\text{v}$, before subtracting the fracture eigenstrain:
\begin{equation}
    \label{eq:r2_2l_directions}
    \m{G}_1=\begin{cases}\frac{\m{I}}{\sqrt{n_\text{d}}} & \text{tr}(\m{\varepsilon}-\m{\varepsilon}_\text{v})\geq0\\ -\frac{\m{I}}{\sqrt{n_\text{d}}} & \text{tr}(\m{\varepsilon}-\m{\varepsilon}_\text{v})<0 \end{cases} \qquad 
    \m{G}_2=\frac{\text{dev}(\m{\varepsilon}-\m{\varepsilon}_\text{v})}{\left\lVert\text{dev}(\m{\varepsilon}-\m{\varepsilon}_\text{v})\right\rVert}
\end{equation}
where $n_\text{d}$ is the number of dimensions considered, either $n_\text{d}=2$ for the intrinsic two-dimensional model or $n_\text{d}=3$ for all other approximations. Both tensors have unit norm and are orthogonal, following the volumetric and deviatoric components of the strain remaining after viscous creep. These directions are a result of the potentials of \cref{eq:r2_2l_potentials} depending on the fracture eigenstrain solely through its trace and the norm of its deviatoric part. The deviatoric part of the eigenstrain that minimises the energy is then aligned with the deviatoric part of $\m{\varepsilon}-\m{\varepsilon}_\text{v}$, while its trace takes the sign of $\text{tr}(\m{\varepsilon}-\m{\varepsilon}_\text{v})$, such that two multipliers are sufficient to describe the fracture eigenstrain (\ref{app:directions}).

\begin{figure}
    \centering
    \begin{subfigure}[t]{0.49\linewidth}
        \centering
        \includegraphics[width=\linewidth]{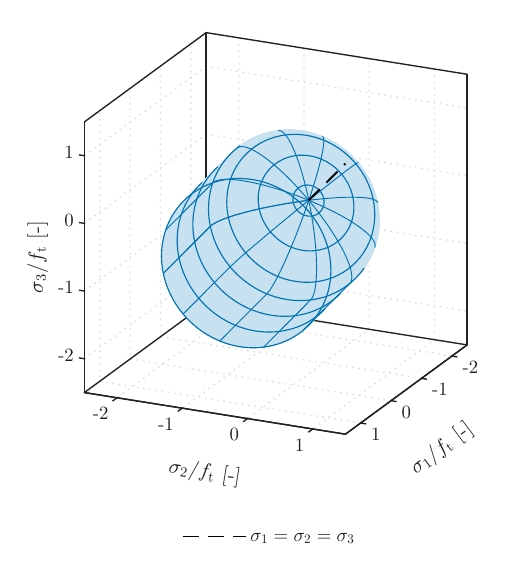}
        \caption{Three-dimensional fracture strength surface. The black dashed line marks the hydrostatic axis $\sigma_1=\sigma_2=\sigma_3$.}
        \label{fig:fracture_surface_3d}
    \end{subfigure}\hfill
    \begin{subfigure}[t]{0.49\linewidth}
        \centering
        \includegraphics[width=\linewidth]{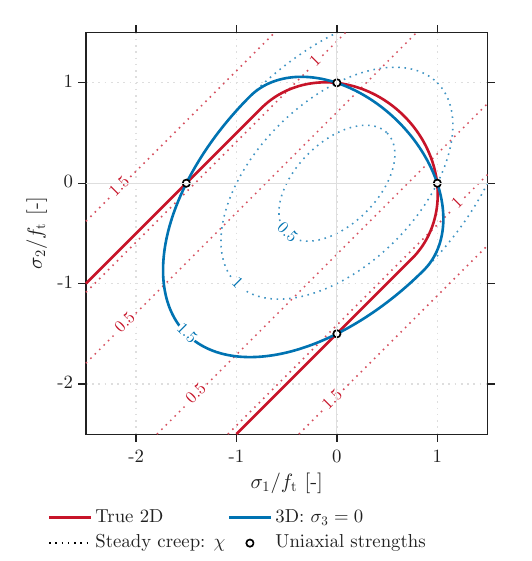}
        \caption{Fracture strength surfaces (solid): intrinsic two-dimensional model (red) and $\sigma_3=0$ slice of the three-dimensional model (blue). Dotted contours of matching colours show steady creep.}
        \label{fig:fracture_surface_2d}
    \end{subfigure}
    \caption{Undegraded failure surfaces ($\phi=0$) resulting from \cref{eq:r2_2l_potentials}, with $f_\text{c}/f_\text{t}=1.5$, and limits on the stress imposed via the combination of applied loading rate and viscous creep. All stress axes are normalized by the tensile strength $f_\text{t}$. The creep contours in (b) denote  loading rates defined through $\chi=(\dot{\varepsilon}_\text{eq}/A)^{1/n}/f_\text{t}$.}
    \label{fig:fracture_surface}
\end{figure}

\subsubsection{Resulting stress state limits}
The resulting strength surfaces are shown in \cref{fig:fracture_surface}, considering the behaviour of a single material point. These surfaces are obtained from the fracture potentials from \cref{eq:r2_2l_potentials}, assuming an undamaged material. Considering a trial eigenstrain given by $\m{\eta}^\text{trial} = \alpha \m{\eta}$, with $\alpha$ approaching zero, the strength criterion in \cref{eq:fracture_multiplier_derivative} can be expressed as:
\begin{equation}
    \label{eq:r2_strength_onset_energy}
    0\leq{}\frac{\partial}{\partial\alpha}\left(\psi_\text{el}(\m{\varepsilon}-\m{\varepsilon}_\text{v}-\alpha\m{\eta})+F_\text{d}(\alpha\m{\eta})+F_\text{i}(\alpha\m{\eta})\right)=-\m{\sigma}:\m{\eta}+F_\text{d}(\m{\eta})+F_\text{i}(\m{\eta})
\end{equation}
Consequently, the admissible stress states satisfy $\m{\sigma}:\m{\eta}\leq F_\text{d}(\m{\eta})+F_\text{i}(\m{\eta})$ for every admissible trial direction. The resistance is positive for every nonzero direction, so dividing by it and taking the maximum gives the equivalent stress-space criterion
\begin{equation}
    \label{eq:r2_strength_surface_polar}
    f_\text{f}(\m{\sigma})=
    \max_{\lambda_1,\lambda_2\geq0}
    \frac{\m{\sigma}:\m{\eta}}{F_\text{d}(\m{\eta})+F_\text{i}(\m{\eta})}-1
    \leq0
\end{equation}
with $f_\text{f}=0$ defining the strength surface and $f_\text{f}<0$ its interior. Using the definitions:
\begin{equation}
    \label{eq:r2_strength_trial_invariants}
    \begin{aligned}
        \text{tr}(\m{\eta})&=\begin{cases}
            \sqrt{n_\text{d}}\lambda_1 & \text{tr}(\m{\sigma})\geq0\\
            -\sqrt{n_\text{d}}\lambda_1 & \text{tr}(\m{\sigma})<0
        \end{cases}
        &\quad \lVert\text{dev}(\m{\eta})\rVert&=\lambda_2\\[1ex]
        \m{\sigma}:\m{\eta}&=\frac{|\text{tr}(\m{\sigma})|}{\sqrt{n_\text{d}}}\lambda_1
        +\lVert\m{\sigma}_\text{dev}\rVert\lambda_2
    \end{aligned}
\end{equation}
and considering the specific strength potentials from \cref{eq:r2_2l_potentials}, the resulting strength surface is given as:
\begin{equation}
    \label{eq:r2_strength_surface_explicit}
    f_\text{f}(\m{\sigma})=
    \begin{cases}
        \displaystyle\sqrt{
            \left(\frac{\text{tr}(\m{\sigma})}{n_\text{d}f_\text{t}^*}\right)^2+
            \left(\frac{\lVert\m{\sigma}_\text{dev}\rVert}{f_\text{s}^*}\right)^2}-1
        & \text{tr}(\m{\sigma})\geq0,\\[2ex]
            \frac{\lVert\m{\sigma}_\text{dev}\rVert}{f_\text{s}^*}-1
        & \text{tr}(\m{\sigma})<0
    \end{cases}
\end{equation}

The dotted contours in \cref{fig:fracture_surface_2d} show the stress limits under a prescribed loading rate, associated with steady creep. Before fracture, a steady deviatoric stress requires $\text{dev}(\m{\dot{\varepsilon}})=\m{\dot{\varepsilon}}_\text{v}$, requiring a balance between the applied strain rate, the stress state, and the creep rate. Following from \cref{eq:creep_constitutive}, this requires for an applied equivalent strain rate $\dot{\varepsilon}_\text{eq}$ that:
\begin{equation}
    \label{eq:steady_creep_contours}
        \dot{\varepsilon}_\text{eq}=\sqrt{\frac{2}{n_\text{d}}}\lVert\text{dev}(\m{\dot{\varepsilon}})\rVert
        =\dot{\lambda}_\text{v}=A\left(\sqrt{\frac{n_\text{d}}{2}}\lVert\m{\sigma}_\text{dev}\rVert\right)^n
\end{equation}
Or, expressing this in a non-dimensional form, we define the external loading parameter $\chi$, resulting in the stress contours being defined by:
\begin{equation}
    \label{eq:steady_creep_stress_limits}
    \frac{\lVert\m{\sigma}_\text{dev}\rVert}{f_\text{t}}=\sqrt{\frac{2}{n_\text{d}}}\chi\qquad \text{with}\qquad\chi=\frac{1}{f_\text{t}}\left(\frac{\dot{\varepsilon}_\text{eq}}{A}\right)^{1/n}
\end{equation}
The dotted curves in \cref{fig:fracture_surface_2d} are generated by tracing the deviatoric stress limit for $\chi=0.5,\,1,\,1.5$.

Comparing the stress limits resulting from the strength surface and the steady-state creep in \cref{fig:fracture_surface_2d} clearly shows the interactions between these two dissipative mechanisms. When loading is applied in a purely volumetric manner, the stress state is solely limited by the strength surface. However, once deviatoric loading is applied, there is a direct competition between the formation of fractures and the viscous creep, with the resulting mechanism being dependent on the loading rate. For loading paths that can be accommodated by deviatoric creep, rapid loading favours reaching the strength surface, whereas sufficiently slow loading can be accommodated without loss of cohesion. This interplay happens on a point-by-point basis, e.g. allowing cracks to nucleate near stress concentrators while the bulk material deforms under the influence of viscous creep. Subsequent fracture propagation can then slowly transfer material around the crack tip from the slow loading creep regime into the fracture propagation regime. It is also worth noting, that as the strength surface decreases in size due to loss of cohesion, damaged material becomes less susceptible to creep, thereby allowing decohesion to be solely driven by elastic crack opening. One important limitation here is that we have assumed the material strengths and fracture energy to be constant, and not impacted by any potential micromechanical degradation that sustained creep could induce \citep{Xie2023, Xue2024,Si2025}. 

\subsubsection{Opening-only criterion}
\label{sec:opening_criterion}

The strength surface of \cref{eq:r2_strength_surface_explicit} is closed in compression: for $\text{tr}(\m{\sigma})<0$, the deviatoric stress is limited to $d(\phi)f_\text{s}^*$, and this limit reduces with the loss of cohesion in the same manner as the tensile strength. Damaged material thereby loses its resistance to deviatoric stresses, irrespective of whether these stresses are tensile or compressive. This allows failure under compression and shear to be described, but it also implies that material which has lost its cohesion cannot carry a sustained compressive load, such as the weight of the material above it. For cases in which damaged material is loaded in compression (e.g. due to gravity), we therefore consider a second criterion, in which the fracture eigenstrain is restricted to opening: it is required to be positive semidefinite, so that crack faces can separate in any direction, while compressive stresses are transmitted as in the intact material. 

For the opening-only criteria, the directions of the fracture eigenstrain are the principal directions of the strain remaining after viscous creep (\ref{app:directions}), which are used as the direction tensors in \cref{eq:eigenstrain_directions}:
\begin{equation}
    \label{eq:opening_directions}
    \m{G}_i=\v{n}_i\otimes\v{n}_i \qquad i=1,2,3
\end{equation}
with $\v{n}_i$ the unit eigenvectors of $\m{\varepsilon}-\m{\varepsilon}_\text{v}$. The multipliers $\lambda_i\geq0$ are now the principal values of the fracture eigenstrain, none of which can become negative. As a result, $\text{tr}(\m{\eta})=\lambda_1+\lambda_2+\lambda_3\geq0$, and the potentials of \cref{eq:r2_2l_potentials} reduce to:
\begin{equation}
    \label{eq:opening_potentials}
    F_\text{d}(\m{\eta})=\sqrt{{f_\text{t}^*}^2\,\text{tr}(\m{\eta})^2+{f_\text{s}^*}^2\left\lVert\text{dev}(\m{\eta})\right\rVert^2} \qquad F_\text{i}(\m{\eta})=0
\end{equation}
with the coefficients $f_\text{t}^*$ and $f_\text{s}^*$ of \cref{eq:r2_strength_scales_3d}. The fracture potential is unchanged, whereas the interpenetration potential is no longer required, as interpenetration is excluded by the admissible eigenstrain directions themselves. The constitutive relations of \cref{eq:fracture_multiplier_derivative} apply as before, with three multipliers in place of two.

The strength surface of this criterion is obtained in the same manner as for the two-multiplier criterion. With three multipliers and without the interpenetration potential, the stress-space criterion of \cref{eq:r2_strength_surface_polar} becomes:
\begin{equation}
    \label{eq:opening_surface_polar}
    f_\text{f}(\m{\sigma})=
    \max_{\lambda_1,\lambda_2,\lambda_3\geq0}
    \frac{\m{\sigma}:\m{\eta}}{F_\text{d}(\m{\eta})}-1
    \leq0
\end{equation}
As the elastic response is isotropic, the principal directions of $\m{\varepsilon}-\m{\varepsilon}_\text{v}$ at the onset of the fracture eigenstrain are those of the stress. This results in the definitions:
\begin{equation}
    \label{eq:opening_trial_invariants}
    \text{tr}(\m{\eta})=\lambda_1+\lambda_2+\lambda_3 \qquad
    \lVert\text{dev}(\m{\eta})\rVert^2=\lambda_1^2+\lambda_2^2+\lambda_3^2-\frac{1}{3}\text{tr}(\m{\eta})^2 \qquad
    \m{\sigma}:\m{\eta}=\sigma_1\lambda_1+\sigma_2\lambda_2+\sigma_3\lambda_3
\end{equation}
As none of the multipliers can become negative, the maximum in \cref{eq:opening_surface_polar} is not necessarily attained with all three multipliers being non-zero. The number of directions with a non-zero multiplier is defined as $n_\text{o}$, following the directions of the $n_\text{o}$ largest principal stresses. Defining the mean stress over the opening directions, and a potential coefficient that depends on their number, as:
\begin{equation}
    \label{eq:opening_mean_stress}
    \sigma_\text{o}=\frac{1}{n_\text{o}}\sum_{i=1}^{n_\text{o}}\sigma_i \qquad
    f_\text{o}^*=\sqrt{{f_\text{t}^*}^2+\left(\frac{1}{n_\text{o}}-\frac{1}{3}\right){f_\text{s}^*}^2}
\end{equation}
with the resulting strength surface given as:
\begin{equation}
    \label{eq:opening_surface_explicit}
    f_\text{f}(\m{\sigma})=
    \sqrt{
        \left(\frac{\sigma_\text{o}}{f_\text{o}^*}\right)^2+
        \sum_{i=1}^{n_\text{o}}\left(\frac{\sigma_i-\sigma_\text{o}}{f_\text{s}^*}\right)^2}-1
\end{equation}

\begin{figure}
    \centering
    \begin{subfigure}[t]{0.49\linewidth}
        \centering
        \includegraphics[width=\linewidth]{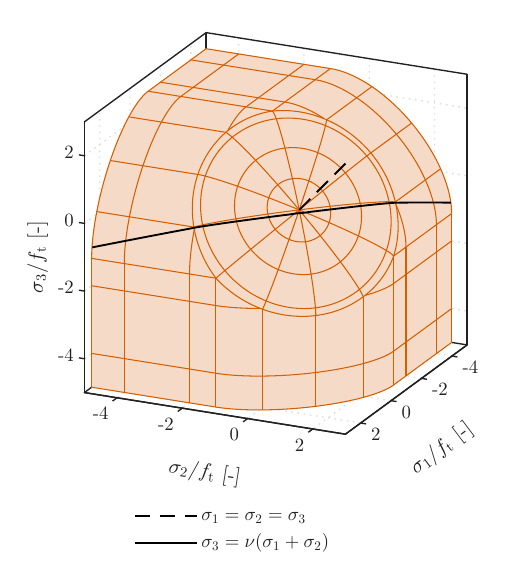}
        \caption{Three-dimensional strength surface of the opening-only criterion. The black dashed line marks the hydrostatic axis, the black solid line the elastic plane-strain section shown in (b).}
        \label{fig:opening_surface_3d}
    \end{subfigure}\hfill
    \begin{subfigure}[t]{0.49\linewidth}
        \centering
        \includegraphics[width=\linewidth]{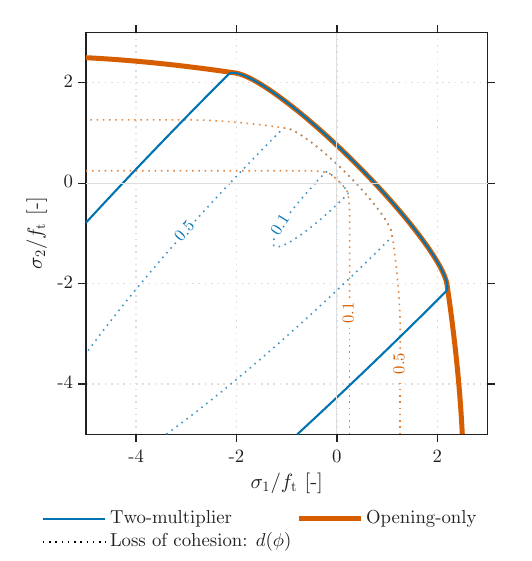}
        \caption{Plane-strain strength surfaces (solid): two-multiplier criterion (blue) and opening-only criterion (orange). Dotted contours of matching colours show the surfaces after loss of cohesion, labelled with $d(\phi)$.}
        \label{fig:opening_surface_2d}
    \end{subfigure}
    \caption{Strength surfaces of the opening-only criterion, \cref{eq:opening_surface_explicit}, and of the two-multiplier criterion, \cref{eq:r2_strength_surface_explicit}, for $f_\text{c}/f_\text{t}=3.75$, as used in \cref{sec:cliff}.}
    \label{fig:opening_surface}
\end{figure}

This failure surface is shown in \cref{fig:opening_surface}. While the criteria are the same under tension, the two criteria differ under compressive loading, and especially in what remains after the loss of cohesion. For the two-multiplier criterion, the admissible stress states reduce to hydrostatic compression, so fully damaged material responds to compression as a fluid. For the opening-only criterion, they reduce to all states without a tensile principal stress: fully damaged material is a no-tension material, which transmits compressive stresses as intact material does. A comparable response is obtained in phase-field models that degrade the elastic energy by restricting this degradation to the tensile principal components, using spectral decompositions of the strain energy \citep{Miehe2010a,Lo2019}, no-tension decompositions \citep{DelPiero1989,Freddi2010,Chambolle2018,DeLorenzis2021,Vicentini2024}, decompositions relative to the crack orientation \citep{Strobl2016,Steinke2019}, or driving forces based on the tensile principal stresses \citep{Miehe2015a}, with such approaches commonly used for the fracture of ice \citep{Sun2021,Clayton2022,Gupta2024,Hageman2026}. Here, it instead follows directly from the admissible directions of the fracture eigenstrain. In return, the opening-only criterion does not describe compressive failure, and $f_\text{c}$ is no longer a uniaxial compressive strength. It instead determines the increase of the tensile strength under lateral compression. As the two criteria coincide for tensile stress states, they are expected to give the same response for the tension-dominated cases of \cref{sec:results_2D_plate,sec:sea_ice}, for which the two-multiplier criterion is used. The choice between them becomes decisive once damaged material is loaded in compression, which is demonstrated in \cref{sec:cliff}.

\begin{figure}
    \centering
    \begin{subfigure}{8cm}
        \includegraphics[width=\linewidth]{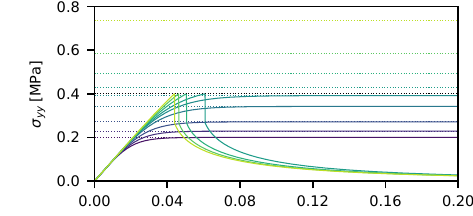}
        \caption{}
        \label{fig:material_point_stress}
    \end{subfigure}\hfill
    \begin{subfigure}{8cm}
        \includegraphics[width=\linewidth]{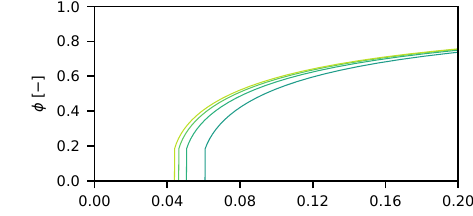}
        \caption{}
        \label{fig:material_point_phasefield}
    \end{subfigure}
    \begin{subfigure}{8cm}
        \includegraphics[width=\linewidth]{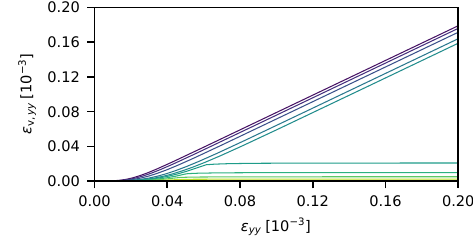}
        \caption{}
        \label{fig:material_point_creep}
    \end{subfigure}\hfill
    \begin{subfigure}{8cm}
        \includegraphics[width=\linewidth]{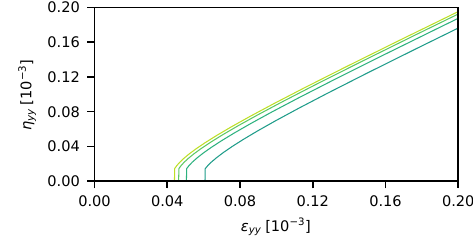}
        \caption{}
        \label{fig:material_point_fracture}
    \end{subfigure}
    \par\smallskip
    \begin{subfigure}{16cm}
        \includegraphics[width=\linewidth]{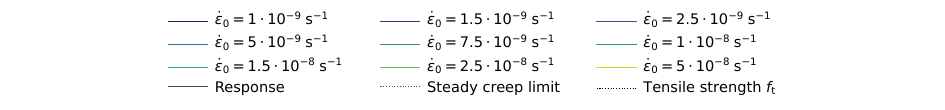}
    \end{subfigure}
    \caption{Response of a single material point to uniaxial tensile strain rates: (a) axial stress, (b) phase-field, (c) axial creep strain, and (d) axial fracture eigenstrain. Curves use an intrinsic two-dimensional material with $G_\text{c}=1.0\;\text{J/m}^2$ and $\ell=0.02\;\text{m}$, and the remaining material parameters given in \cref{tab:parameters}. Solid curves show the material responses. In (a), the black dotted line denotes the tensile strength $f_\text{t}$, and the coloured dotted lines denote the corresponding steady-creep stress limits in the absence of fracture, resulting from \cref{eq:material_point_creep_limit,eq:steady_creep_stress_limits}.}
    \label{fig:material_point_rates}
\end{figure}

\subsubsection{Material-point response}
\label{sec:material_point_response}

To illustrate the competition between creep and loss of cohesion, we consider a single material point under uniaxial tensile stress, $\sigma_{xx}=\sigma_{xy}=0$, with a prescribed axial strain $\varepsilon_{yy}=\dot{\varepsilon}_0t$ imposed in the vertical direction, and no restrictions on the strain in horizontal direction. For this example, the intrinsic two-dimensional material is used with $G_\text{c}=1.0\;\text{J/m}^2$ and $\ell=0.02\;\text{m}$. The remaining parameters are given in \cref{tab:parameters}. The responses in \cref{fig:material_point_rates} are evaluated directly from the constitutive equations. Since the phase-field is spatially uniform, its gradient vanishes and the history formulation reduces to
\begin{equation}
    \label{eq:homogeneous_phase_field}
    \phi=\frac{2(1-\kappa)\mathcal{F}}{G_\text{c}/\ell+2(1-\kappa)\mathcal{F}}
    \qquad
    \mathcal{F}=\max_{\tau\in[0,t]}F_\text{d}(\m{\eta}(\tau))
\end{equation}

For this uniaxial stress state and using $n=3$, the creep law becomes $\dot{\varepsilon}_{\text{v},yy}=A_\text{Glen}\sigma_{yy}^3/8$. Before fracture, the imposed axial strain is shared between elastic strain and creep strain, so the stress evolves according to
\begin{equation}
    \label{eq:material_point_creep_limit}
    \dot{\sigma}_{yy}=E\left(\dot{\varepsilon}_0-\frac{A_\text{Glen}\sigma_{yy}^3}{8}\right)
    \qquad
    \sigma_{\text{v},\infty}=\left(\frac{8\dot{\varepsilon}_0}{A_\text{Glen}}\right)^{1/3}
\end{equation}
The coloured dotted lines in \cref{fig:material_point_stress} show these steady-creep stresses, providing the uniaxial counterparts of the dotted stress contours in \cref{fig:fracture_surface_2d}. Starting from zero stress, the response approaches its corresponding limit as the creep rate increases towards the imposed axial strain rate. These lines are limits for an intact material under the prescribed constant strain rates, with creep being active at stresses below them. The black dotted line gives the competing limit imposed by the intact tensile strength, $f_\text{t}=0.4\;\text{MPa}$, where no fractures develop when the stress state is below this limit, whereas upon reaching this limit, eigenstrains start to develop to represent the crack.

Comparing the two limits gives the transition rate
\begin{equation}
    \label{eq:material_point_critical_rate}
    \dot{\varepsilon}_\text{crit}=\frac{A_\text{Glen}f_\text{t}^3}{8}=8\cdot10^{-9}\;\text{s}^{-1}
\end{equation}
For $\dot{\varepsilon}_0<\dot{\varepsilon}_\text{crit}$, the steady-creep stress lies below the tensile strength. The five subcritical responses in \cref{fig:material_point_rates} therefore remain intact, with $\phi=0$ and $\eta_{yy}=0$. Their stresses gradually level off while creep accommodates the continued increase in total strain. At equal imposed strain, the lower loading rates allow for more time for creep to develop, hence obtaining similar creep strains at a lower stress.

For $\dot{\varepsilon}_0>\dot{\varepsilon}_\text{crit}$, the hypothetical steady-creep stress limit exceeds $f_\text{t}$. As a result, the four higher-rate responses reach the tensile strength before this balance can be established. Fracture eigenstrains then develop and increase the driving history in \cref{eq:homogeneous_phase_field}, causing $\phi$ to increase. The resulting reduction of the cohesive strength to $d(\phi)f_\text{t}$ produces the descending stress in \cref{fig:material_point_stress}, with the stress following the strength of the damaged material. Higher loading rates leave less time for creep and reach fracture onset at smaller total strains. After onset, the increasing fracture eigenstrain accommodates a growing share of the imposed strain, following $\varepsilon_{yy}=\sigma_{yy}/E+\varepsilon_{\text{v},yy}+\eta_{yy}$. The falling stress releases elastic strain and reduces the creep rate through its cubic stress dependence, so the creep-strain curves flatten while the fracture eigenstrain continues to grow. Together, this response shows how the competition between creep and the tensile strength controls fracture onset, and how phase-field degradation subsequently changes the dominant deformation mechanism from being driven by bulk creep to being driven by interface decohesion.

This single material point response illustrates the local mechanisms, but neglects non-local effects. For instance, the explicit dependence on $G_\text{c}/\ell$ in \cref{eq:homogeneous_phase_field} is a consequence of considering the single material point in isolation: spatial gradients are suppressed, and a localized fracture process zone cannot develop. In the full formulation, $\ell$ controls the width of the diffuse fracture zone, and both the local and gradient contributions enter the fracture-energy balance to control the amount of dissipated energy. The length-scale dependence of this isolated-point degradation curve therefore cannot be used to infer a corresponding dependence of the macroscopic fracture response. 

\subsection{Finite element implementation}
\label{sec:implementation}
The governing equations, \cref{eq:momentum_weak,eq:phase_weak}, are solved using the finite element method, using the open-source FEniCSx framework \citep{Baratta2023}. Displacements $\v{u}(\v{x})$ are discretized using quadratic triangular Bernstein elements, while linear triangular elements are used for the phase-field $\phi(\v{x})$. The displacement shape functions are collected in the matrix $\m{N}_\text{u}^{el}(\v{x})$, and the phase-field shape functions in the row vector $\v{N}_\phi^{el}(\v{x})$:
\begin{equation}
    \label{eq:fe_interpolation}
        \v{u}^{t+\Delta t}(\v{x})=\sum_{el}\m{N}_\text{u}^{el}(\v{x})\v{u}^{el,t+\Delta t} \qquad 
        \phi^{t+\Delta t}(\v{x})=\sum_{el}\v{N}_\phi^{el}(\v{x})\v{\upphi}^{el,t+\Delta t}
\end{equation}
Here, $\v{u}^{el}$ and $\v{\upphi}^{el}$ are the nodal values, and the superscript $el$ indicates element-local quantities. For brevity, this element label and the spatial argument $(\v{x})$ of the shape functions are omitted in the following. Superscripts $t$ and $t+\Delta t$ indicate the previous converged state and the state being solved for, respectively.

Mandel notation is used for the strain and stress vectors, $\v{\upvarepsilon}$ and $\v{\upsigma}$, with shear components scaled by $\sqrt{2}$ to preserve tensor contractions. The strain vector follows from the displacement coefficients through:
\begin{equation}
    \label{eq:fe_strain}
    \v{\upvarepsilon}^{t+\Delta t}=\m{L}\m{N}_\text{u}\v{u}^{t+\Delta t}
\end{equation}
where $\m{L}$ is the symmetric-gradient operator expressed in Mandel notation. Using these shape functions, the discretized weak form of the momentum balance becomes:
\begin{equation}
    \label{eq:discretized_u}
        \v{F}_\text{u}^{t+\Delta t}=\int_\Omega (\m{L}\m{N}_\text{u})^\top\v{\upsigma}^{t+\Delta t} +\rho\m{N}_\text{u}^\top\m{N}_\text{u} \ddot{\v{u}}^{t+\Delta t}+\frac{\rho c}{\Delta t} \m{N}_\text{u}^\top\m{N}_\text{u} \left(\v{u}^{t+\Delta t}-\v{u}^{t}\right) -\m{N}_\text{u}^\top\v{b}^{t+\Delta t}\;\text{d}\Omega -\int_\Gamma\m{N}_\text{u}^\top\v{\tau}^{t+\Delta t}\;\text{d}\Gamma=\v{0}
\end{equation}
Similarly, the discretized phase-field equation is given by:
\begin{equation}
    \label{eq:discretized_phi}
        \v{F}_\phi^{t+\Delta t}=\int_\Omega \frac{G_\text{c}}{\ell}\left(\v{N}_\phi^\top\v{N}_\phi\v{\upphi}^{t+\Delta t} +\ell^2(\nabla\v{N}_\phi)^\top\nabla\v{N}_\phi\v{\upphi}^{t+\Delta t}\right) -2(1-\kappa)\v{N}_\phi^\top \left(1-\v{N}_\phi\v{\upphi}^{t+\Delta t}\right)\mathcal{F}^{t+\Delta t}\;\text{d}\Omega=\v{0}
\end{equation}
The element contributions are assembled into the global displacement and phase-field residuals. The stresses $\v{\upsigma}^{t+\Delta t}$ are evaluated at the integration points using the strain, creep strain, and fracture eigenstrain at $t+\Delta t$, with more detail provided in \cref{sec:return_mapping}. 

The acceleration and stored velocity are obtained using Newmark time discretization \citep{Newmark1959}. Writing both updates directly in terms of the displacement coefficients $\v{u}^{t+\Delta t}$ and the known displacement, velocity, and acceleration at $t$, the velocity and acceleration are defined by:
\begin{align}
    \label{eq:newmark_velocity}
    \dot{\v{u}}^{t+\Delta t}&=\frac{\gamma}{\beta \Delta t}\left( \v{u}^{t+\Delta t}-\v{u}^t\right)-\left(\frac{\gamma}{\beta}-1\right) \dot{\v{u}}^t-\Delta t\left(\frac{\gamma}{2\beta}-1\right)\ddot{\v{u}}^t\\
    \label{eq:newmark_acceleration}
    \ddot{\v{u}}^{t+\Delta t}&= \frac{1}{\beta \Delta t^2}\left( \v{u}^{t+\Delta t}-\v{u}^t\right)-\frac{1}{\beta \Delta t}\dot{\v{u}}^t-\left(\frac{1}{2\beta}-1\right)\ddot{\v{u}}^t
\end{align}
where $\beta$ and $\gamma$ are the Newmark time-integration parameters. 

The resulting system of equations from \cref{eq:discretized_u,eq:discretized_phi} is solved using a multi-pass staggered scheme. Within each outer iteration, the displacement problem is solved at fixed phase-field using a Newton-Raphson scheme with a residual-based backtracking line search, restricting the step-length factor to $0.1\leq\alpha\leq1$. The local constitutive variables are updated at the integration points during the displacement solution step. The resulting crack-driving history is subsequently held fixed while the linear phase-field equation is solved. Due to the poor conditioning of the momentum balance as a result of the loss of stiffness during fracture, direct solvers are used for both equations, using MUMPS \citep{Amestoy2001} through PETSc \citep{Balay2023}.

Convergence is assessed using the normalized residual-based error $e/e_0\leq10^{-6}$, where $e$ is the residual norm and $e_0$ is its initial reference value for the corresponding field, kept fixed across the outer iterations. This criterion is used for the displacement Newton iterations, and both global residuals are re-evaluated after each complete outer iteration to assess coupled convergence. The staggered iterations are repeated until both fields satisfy the convergence criterion or a maximum of 10 outer iterations has been performed, in which case the latest iterate is retained.

After the staggered solution scheme has fully converged, adaptive mesh refinement is performed if required, as described in \cref{sec:adaptive_mesh_refinement}, followed by solving the system again on the refined mesh using the same convergence criterion and outer-iteration limit. Refinement and solution are repeated until no further refinement is required. All staggered iterations and repeated solves following mesh refinement refer to the same time increment $[t,t+\Delta t]$. The previous converged state at $t$ for the displacement and phase-field, as well as the quadrature point constitutive variables, remains fixed throughout and are advanced only after completing the increment.

\subsubsection{Return-mapping based scheme}
\label{sec:return_mapping}

The stress, creep strain, and fracture eigenstrain are evaluated through a local constitutive problem at each integration point, comparable to the return-mapping schemes used in plasticity \citep{SimoHughes1998}. Following the return-mapping formulation for cohesive fracture from \citep{Hageman2026b}, the fracture eigenstrain is represented by its multipliers $\lambda_i$ rather than introducing additional global degrees of freedom that would require solving monolithically with the displacement. The extension considered here is to solve the creep and fracture constitutive relations simultaneously, without placing priority on either of the dissipative mechanisms. Both mechanisms therefore respond to the same updated stress state within each time increment, including when creep relaxation and fracture nucleation occur during the same increment.

The local unknowns of this integration-point local scheme are collected in the state vector:
\begin{equation}
    \label{eq:rm_total_state}
    \v{s} = \begin{bmatrix}\v{\upsigma}^\top & \Delta \lambda_\text{v} & \v{\lambda}^\top \end{bmatrix}^\top
\end{equation}
where $\v{\upsigma}$ is the stress, $\Delta\lambda_\text{v}=\dot{\lambda}_\text{v}\Delta t$ is the creep multiplier increment, and $\v{\lambda}=[\lambda_1\;\lambda_2]^\top$ contains the current fracture multipliers, with $\lambda_i\geq0$. Mandel notation is used throughout this subsection for stress and strain quantities. The deviatoric projector and linear-elastic stiffness matrix are defined as:
\begin{equation}
    \label{eq:rm_elastic_projector}
    \m{P}_\text{dev}=\m{I}-\frac{1}{n_\text{d}}\v{i}\v{i}^{\top}
    \qquad
    \m{D}_\text{el}=\frac{\partial\v{\upsigma}}{\partial\v{\upvarepsilon}_\text{el}}
\end{equation}
with $\v{i}$ the volumetric Mandel vector, $\v{i}=[1\;\;1\;\;0]^\top$ in two dimensions and $\v{i}=[1\;\;1\;\;1\;\;0\;\;0\;\;0]^\top$ in three dimensions.

Using these matrices, we define the constitutive model for the elastic stresses, following from \cref{eq:stress_constitutive}, as:
\begin{equation} \label{eq:residual_stress}
    \v{r}_\sigma = \v{\upsigma} - \m{D}_\text{el}\left(\v{\upvarepsilon}-\v{\upvarepsilon}_\text{v}^t - \Delta\lambda_\text{v}\frac{\partial f_\text{v}}{\partial\v{\upsigma}} - \sum_{i=1}^{2}\lambda_i\v{G}_i\right) = \v{0}
\end{equation}
where all quantities are taken at the current time, $t+\Delta t$, unless explicitly noted otherwise. Similarly, the creep constitutive model from \cref{eq:creep_multiplier,eq:creep_flow_direction} can be expressed as a residual form:
\begin{equation} \label{eq:residual_creep}
    r_\text{v} = \Delta \lambda_\text{v} - A\Delta t \left(\frac{n_\text{d}}{2}\v{\upsigma}^{\top}\m{P}_\text{dev}\v{\upsigma}\right)^{n/2} = 0
\end{equation}
These relations use backward-Euler integration of the creep rate, $\v{\upvarepsilon}_\text{v}^{t+\Delta t}=\v{\upvarepsilon}_\text{v}^{t}+\Delta t\,\dot{\v{\upvarepsilon}}_\text{v}^{t+\Delta t}$.

The constitutive model for the fracture is driven by the combined elastic and fracture eigenstrains. For brevity, this strain state is denoted as:
\begin{equation}
    \label{eq:rm_fracture_strain}
    \v{\upvarepsilon}_\text{r} = \v{\upvarepsilon}-\v{\upvarepsilon}_\text{v}^t - \Delta \lambda_\text{v} \frac{\partial f_\text{v}}{\partial \v{\upsigma}}
\end{equation}
which determines the eigenstrain directions $\v{G}_i(\v{\upvarepsilon}_\text{r})$ and the tensile or compressive branch of the volumetric fracture criterion. Writing the combined strength potential as $F(\v{\upeta})=d(\phi)F_\text{d}(\v{\upeta})+F_\text{i}(\v{\upeta})$, the fracture constitutive model is expressed as:
\begin{equation} \label{eq:residual_fracture}
    r_i  = 0 = \begin{cases} \lambda_i \qquad &\text{if} \qquad \v{G}_i^\top\left(-\v{\upsigma} + \frac{\partial F}{\partial \v{\upeta}}\right) > 0 \\
\v{G}_i^\top\left(-\v{\upsigma} + \frac{\partial F}{\partial \v{\upeta}}\right) + \kappa K \lambda_i \qquad &\text{otherwise} \end{cases}\qquad i=1,2
\end{equation}
where $\kappa$ is a small dimensionless regularisation parameter that provides a residual stiffness after fracture. 

The residuals from \cref{eq:residual_stress,eq:residual_creep,eq:residual_fracture} are collected in $\v{r}=[\v{r}_\sigma^\top\;r_\text{v}\;r_1\;r_2]^\top$ and solved using a Newton-Raphson scheme at each integration point. Each local Newton correction $\Delta\v{s}$ satisfies $\m{K}\Delta\v{s}=-\v{r}$. With both fracture directions active, the residual Jacobian is:
\begin{equation}
    \label{eq:rm_updated_jacobian}
    \m{K}= \frac{\partial \v{r}}{\partial \v{s}} =
    \begin{bmatrix}
        \m{I}+\Delta\lambda_\text{v}\m{D}_\text{el}
        \left(\m{I}-\dfrac{\partial\v{\upeta}}{\partial\v{\upvarepsilon}_\text{r}}\right)
        \dfrac{\partial^2 f_\text{v}}{\partial\v{\upsigma}^2}
        &\m{D}_\text{el}
        \left(\m{I}-\dfrac{\partial\v{\upeta}}{\partial\v{\upvarepsilon}_\text{r}}\right)\dfrac{\partial f_\text{v}}{\partial \v{\upsigma}}
        &\m{D}_\text{el}\v{G}_1
        &\m{D}_\text{el}\v{G}_2\\[2ex]
        -An\Delta t\,f_\text{v}^{n-1}
        \left(\dfrac{\partial f_\text{v}}{\partial \v{\upsigma}}\right)^{\top}
        &1&0&0\\[2ex]
        -\v{G}_1^{\top}-\Delta\lambda_\text{v}
        \dfrac{\partial r_1}{\partial\v{\upvarepsilon}_\text{r}}
        \dfrac{\partial^2 f_\text{v}}{\partial\v{\upsigma}^2}
        &-\dfrac{\partial r_1}{\partial\v{\upvarepsilon}_\text{r}}
        \dfrac{\partial f_\text{v}}{\partial\v{\upsigma}}
        &\v{G}_1^{\top}\dfrac{\partial^2 F}{\partial\v{\upeta}^{2}}\v{G}_1+\kappa K
        &\v{G}_1^{\top}\dfrac{\partial^2 F}{\partial\v{\upeta}^{2}}\v{G}_2\\[2ex]
        -\v{G}_2^{\top}-\Delta\lambda_\text{v}
        \dfrac{\partial r_2}{\partial\v{\upvarepsilon}_\text{r}}
        \dfrac{\partial^2 f_\text{v}}{\partial\v{\upsigma}^2}
        &-\dfrac{\partial r_2}{\partial\v{\upvarepsilon}_\text{r}}
        \dfrac{\partial f_\text{v}}{\partial\v{\upsigma}}
        &\v{G}_2^{\top}\dfrac{\partial^2 F}{\partial\v{\upeta}^{2}}\v{G}_1
        &\v{G}_2^{\top}\dfrac{\partial^2 F}{\partial\v{\upeta}^{2}}\v{G}_2+\kappa K
    \end{bmatrix}
\end{equation}
The derivatives in $\m{K}$ are evaluated at fixed total strain, phase-field, and previous state, within the current constitutive branch. The derivatives appearing in the matrix are given by:
\begin{align*}
    &\frac{\partial\v{\upeta}}{\partial\v{\upvarepsilon}_\text{r}}
    =\lambda_1\frac{\partial\v{G}_1}{\partial\v{\upvarepsilon}_\text{r}}
    +\lambda_2\frac{\partial\v{G}_2}{\partial\v{\upvarepsilon}_\text{r}}\\
    &\frac{\partial r_i}{\partial\v{\upvarepsilon}_\text{r}}
    =\left(\frac{\partial F}{\partial\v{\upeta}}-\v{\upsigma}\right)^{\top}
    \frac{\partial\v{G}_i}{\partial\v{\upvarepsilon}_\text{r}}
    +\v{G}_i^{\top}\frac{\partial^2 F}{\partial\v{\upeta}^{2}}
    \frac{\partial\v{\upeta}}{\partial\v{\upvarepsilon}_\text{r}}
\end{align*}
Here, $\partial r_i/\partial\v{\upvarepsilon}_\text{r}$ is a row derivative taken at fixed stress and fracture multipliers. These terms retain the changes in the fracture directions caused by the creep update, including their effect on the strength potential. For an inactive direction, the corresponding fracture row of $\m{K}$ is replaced by zero entries except for a unit entry in the column of $\lambda_i$. The resulting model-specific Jacobian is given in \ref{app:model_jacobian}, substituting the specific fracture strength surface.

Within each local Newton iteration, the linear system is solved through successive Schur reductions. Partitioning $\m{K}$ into stress, creep, and fracture blocks, the stress corrections can first be eliminated. For $a,b\in\{\text{v},\lambda\}$, define:
\begin{equation}
    \label{eq:rm_schur_stress}
    \m{S}_{ab}=\m{K}_{ab}-\m{K}_{a\sigma}\m{K}_{\sigma\sigma}^{-1}\m{K}_{\sigma b}
    \qquad
    \widehat{\v{b}}_a=\v{b}_a-\m{K}_{a\sigma}\m{K}_{\sigma\sigma}^{-1}\v{b}_\sigma,
    \qquad \v{b}=-\v{r}
\end{equation}
The remaining creep block $S_{\text{vv}}$ is a scalar and can be eliminated next, leaving a two-by-two system for the fracture multipliers:
\begin{equation}
    \label{eq:rm_schur_fracture}
    \left(\m{S}_{\lambda\lambda}
    -\m{S}_{\lambda\text{v}}S_{\text{vv}}^{-1}\m{S}_{\text{v}\lambda}\right)
    \Delta\v{\lambda}
    =\widehat{\v{b}}_\lambda
    -\m{S}_{\lambda\text{v}}S_{\text{vv}}^{-1}\widehat b_\text{v}
\end{equation}
The creep and stress corrections then follow by back-substitution:
\begin{equation}
    \label{eq:rm_schur_backsubstitution}
    \begin{aligned}
        \Delta\Delta\lambda_\text{v}
        &=S_{\text{vv}}^{-1}\left(\widehat b_\text{v}
        -\m{S}_{\text{v}\lambda}\Delta\v{\lambda}\right)\\
        \Delta\v{\upsigma}
        &=\m{K}_{\sigma\sigma}^{-1}\left(\v{b}_\sigma
        -\m{K}_{\sigma\text{v}}\Delta\Delta\lambda_\text{v}
        -\m{K}_{\sigma\lambda}\Delta\v{\lambda}\right)
    \end{aligned}
\end{equation}
The inverse notation denotes the solution of the corresponding block systems. The resulting corrections provide the Newton update direction for $\v{s}$. This elimination retains the full coupling between the mechanisms while avoiding explicit inversion of the complete local matrix, extending the Schur-complement treatment used for cohesive fracture in \citep{Hageman2026b}. While not verified for viscoplasticity, this approach was seen to be needed due to the poor conditioning of the local Jacobian matrix, particularly when the material is fractured. 

To obtain a consistent tangent matrix for the global Newton iterations, we also define the matrix $\m{C}$:
\begin{equation}
    \label{eq:rm_updated_strain_derivative}
    \m{C} = \frac{\partial\v{r}}{\partial\v{\upvarepsilon}}
    =\begin{bmatrix}
        -\m{D}_\text{el}\left(\m{I}-\dfrac{\partial\v{\upeta}}{\partial\v{\upvarepsilon}_\text{r}}\right)\\[2ex]
        \v{0}^{\top}\\
        \dfrac{\partial r_1}{\partial\v{\upvarepsilon}_\text{r}}\\
        \dfrac{\partial r_2}{\partial\v{\upvarepsilon}_\text{r}}
    \end{bmatrix}
\end{equation}
The derivatives in $\m{C}$ are evaluated at fixed local state $\v{s}$, phase-field, and previous state. For an inactive fracture direction, the corresponding row of $\m{C}$ is zero. The consistent tangent for the global displacement Newton scheme is then obtained by differentiating the converged local equations \citep{SimoTaylor1985}. Since $\v{r}(\v{s},\v{\upvarepsilon})=\v{0}$, the derivatives satisfy:
\begin{equation}
    \label{eq:rm_consistent_tangent}
    \m{K}
    \begin{bmatrix}
        \displaystyle\frac{\partial\v{\upsigma}}{\partial\v{\upvarepsilon}}\\[1ex]
        \displaystyle\frac{\partial\Delta\lambda_\text{v}}{\partial\v{\upvarepsilon}}\\[1ex]
        \displaystyle\frac{\partial\v{\lambda}}{\partial\v{\upvarepsilon}}
    \end{bmatrix}
    =-\m{C}
    \qquad
    \m{D}_\text{tan}=\frac{\partial\v{\upsigma}}{\partial\v{\upvarepsilon}}
    =-\begin{bmatrix}\m{I}&\v{0}&\m{0}\end{bmatrix}\m{K}^{-1}\m{C}
\end{equation}
The same Schur reductions are used with the matrix of strain-derivative right-hand sides $-\m{C}$. This tangent includes the response of the creep increment, the fracture multipliers, and their directions to a change in total strain. If fracture is inactive, the equations reduce to the implicit power-law creep update, whereas for the reference cases where creep is suppressed, they reduce to the cohesive return-mapping formulation. Thus, the coupled constitutive response enters the global momentum balance solely through the stress and its consistent tangent.

The scheme is given here for the two fracture multipliers of the criterion of \cref{eq:r2_2l_potentials,eq:r2_2l_directions}. For the opening-only criterion of \cref{sec:opening_criterion}, the same scheme is used with three fracture multipliers and the direction tensors of \cref{eq:opening_directions}, which adds the third multiplier and its condition to the local system of equations.

After the quadrature constitutive problem has converged, the stress and tangent are returned to the global Newton solver, used within the weak form from \cref{eq:discretized_u}. The fracture driving energy is also updated using these new eigenstrains as:
\begin{equation}
    \label{eq:rm_history_update}
    \mathcal{F}^{t+\Delta t}
    =\max\left(\mathcal{F}^{t},F_\text{d}\bigl(\v{\upeta}^{t+\Delta t}\bigr)\right)
\end{equation}
The updated integration-point quantities are re-evaluated at every global Newton iteration using the same previous state at $t$.

\subsubsection{Adaptive mesh refinement}
\label{sec:adaptive_mesh_refinement}

Adaptive mesh refinement is used to resolve developing fractures while retaining coarser elements elsewhere \citep{Heister2015,FreddiMingazzi2023}. Error-estimator-based $hp$-adaptivity has been developed for phase-field fracture \citep{Bird2023}, while adaptive implementations of stress-based models have been demonstrated in the damage mechanics challenge \citep{Gupta2024}. Adaptive frameworks have been developed to use recursive refinement within the staggered solution process to resolve glacier crevasses without prescribing their paths \citep{Gupta2026}. Physical indicators based on local stability or damage have also been used to guide refinement \citep{Hirshikesh2021,Feng2026AMR}. Here, the fracture eigenstrain is used as the refinement indicator. It is zero in intact material, including during creep, but becomes non-zero during the displacement solve before the phase-field is updated. It can therefore indicate fracture initiation without requiring an appreciable phase-field value or a prescribed crack path. For each element, the indicator is defined by the maximum over its integration points:
\begin{equation}
    \label{eq:mesh_refinement_indicator}
    \mathcal{I}^{el,t+\Delta t}
    =\max_{q\in el}\left\lVert\v{\upeta}_q^{t+\Delta t}\right\rVert
    =\max_{q\in el}\sqrt{\sum_i\left(\lambda_{i,q}^{t+\Delta t}\right)^2}
\end{equation}
The final equality follows from the orthonormal eigenstrain directions, and their dependence on the unit vectors $\v{G}$. Elements are marked when $\mathcal{I}^{el,t+\Delta t}\geq\eta_\text{tol}$ and their characteristic size satisfies $h^\text{el}>\ell/4$, where $\eta_\text{tol}$ is a small positive tolerance. The marked elements and a surrounding buffer are refined towards a target size of $h^\text{el}\leq\ell/4$ to resolve the diffuse phase-field profile and provide a refined region ahead of the fracture. The local strength criterion is independent of element size and phase-field length scale \citep{Hageman2026b}, allowing eigenstrains to develop in elements with $h^\text{el}\gg\ell$. Refinement can therefore be triggered on a coarse mesh before the phase-field is fully resolved. The unrefined mesh is chosen to resolve the geometry, stress gradients, and creep response accurately, whereas only upon refinement it is able to capture the phase-field and crack propagation.

Refinement is checked after each staggered solve and performed using the Plaza-Carey algorithm in FEniCSx \citep{PlazaCarey2000}. The displacement, velocity, acceleration, phase-field, and their stored values at the old timestep are transferred by finite-element interpolation onto the refined mesh. For a scalar field component with coefficient vector $\v{a}$, the transfer satisfies:
\begin{equation}
    \label{eq:mesh_field_transfer}
    \v{N}_\text{new}^{el_\text{c}}(\v{x})\v{a}_\text{new}^{el_\text{c},\tau}
    =\v{N}_\text{old}^{el_\text{p}}(\v{x})\v{a}_\text{old}^{el_\text{p},\tau}
    \qquad \v{x}\in el_\text{c}\subset el_\text{p}
\end{equation}
where $el_\text{p}$ and $el_\text{c}$ denote a parent and child element, $\tau\in\{t,t+\Delta t\}$ denotes the timestep of the data, and old and new refer to the meshes themselves. This same method is used for vector fields, transferring each individual component independently.

Integration-point fields, including the creep strain, fracture eigenstrain, and fracture driving history, require a separate transfer \citep{Peric1996,Kumar2015}. For each scalar component $z$ and stored time level $\tau$, a polynomial $z_\text{new}^\tau(\v{x})$ is fitted within each parent element by minimizing the weighted error at the original integration points. The transferred values are then obtained by evaluating this polynomial at the child integration points:
\begin{equation}
    \label{eq:mesh_quadrature_transfer}
        \min_{z_\text{new}^\tau}
        \sum_{q\in el_\text{p}}|w_q|
        \left(z_\text{new}^\tau(\v{x}_{q,\text{old}})-z_{q,\text{old}}^\tau\right)^2 \qquad z_{q,\text{new}}^\tau
        =z_\text{new}^\tau(\v{x}_{q,\text{new}})
\end{equation}
The $w_q$ are the parent quadrature weights. The highest complete polynomial order whose coefficients can be determined uniquely from the parent-element data is used. This interpolation is only performed for refined elements, whereas values on unchanged elements are copied to the new mesh directly.

Once refined, the coupled problem is solved again over the same interval $[t,t+\Delta t]$, using the transferred state at $t$ as the fixed previous state. Refinement and solution are repeated until the target resolution is reached, without accumulating additional creep increments between repeated solves.  Refined regions are retained during subsequent unloading or crack closure, as no mesh coarsening algorithm is considered.

\subsubsection{Element-size correction of the fracture energy}
\label{sec:Gc_correction}

Once a crack has localised, the fracture eigenstrain is confined to a single layer of integration points, while the phase-field retains its diffuse profile of width $\ell$. The degraded strength transmitted by the opened integration points, $d(\phi)f_\text{t}$, is evaluated with the interpolated phase-field at those points, which lies below the peak of the profile by an amount proportional to the element size. As a result the discrete cohesive law dissipates more energy than $G_\text{c}$. For the closely related $\Omega^2$ model, Feng et al.\ \citep{Feng2026AMR} showed that this error is first order in the element size and that it leaves the shape of the traction-separation law unchanged, only scaling its energy dissipation:
\begin{equation}
    \label{eq:Gc_num}
    G_\text{c}^\text{num}=G_\text{c}\left(1+c\,\frac{h^\text{el}}{\ell}\right)+\mathcal{O}\!\left(\frac{(h^\text{el})^2}{\ell^2}\right)
\end{equation}
whereas the diffuse part of the phase-field profile only contributes second-order errors. The same analysis for the present eigenstrain-based model is given in \ref{app:Gc_correction}, where the coefficient is shown to depend on the position of the opened integration point within the element, $c=2\xi(1-\xi)$, giving $c=1/3$ for the Gauss points used here (and recovering $c=1/2$ of \citep{Feng2026AMR} for a crack through the element centre). This value is confirmed by a calibration on a notched bar in \ref{app:Gc_correction}. Since the cohesive law is shape-preserving, the leading error is removed by prescribing a corrected fracture energy in the crack-density terms of \cref{eq:phase_weak}:
\begin{equation}
    \label{eq:Gc_inp}
    G_\text{c}^\text{inp}=\frac{G_\text{c}}{1+c\,\bar{h}/\ell} \qquad c=\frac{1}{3}
\end{equation}
Following \citep{Feng2026AMR}, the correction is applied non-locally: $\bar{h}$ is the mean size of the elements carrying fracture eigenstrain, $\mathcal{I}^{el}\geq\eta_\text{tol}$, within a distance $2\ell$ of the element considered, so that a single corrected value is used across the diffuse band around each crack. Elements without an opened element nearby retain $G_\text{c}$, and $\bar{h}$ is re-evaluated at the start of each increment and after each refinement. On the uniform meshes of \cref{sec:element_size_sensitivity} this reduces to a constant factor, while under adaptive refinement the correction follows the average local element size of the crack core. The strength surface, the eigenstrain update, and therefore the fracture initiation are unaffected, as only the energy dissipated during loss of cohesion changes. Unless stated otherwise, the correction is used in all simulations.

\section{Benchmark case}
\label{sec:results_2D_plate}

\begin{table}[tbp]
    \centering
    \begin{minipage}[t]{0.58\linewidth}
        \vspace{30pt}
        \centering
        \setlength{\tabcolsep}{4pt}
    \begin{tabular}{|l|c|c|}
        \hline
        \multicolumn{3}{|c|}{General and elastic Parameters}\\
        \hline
        Density & $\rho$ & $910\;\text{kg/m}^3$ \\
        Young's modulus & $E$ & $9.5\;\text{GPa}$ \\
        Poisson's ratio & $\nu$ & $0.35$ \\
        Damping coefficient & $c$ & $1.10\cdot10^6\;\text{s}^{-1}$ \\
        \hline
        \multicolumn{3}{|c|}{Creep Parameters}\\
        \hline
        Glen's law exponent & $n$ & $3$ \\
        Glen's law rate factor & $A_\text{Glen}$ & $1.0\cdot10^{-24}\;\text{s}^{-1}\text{Pa}^{-3}$ \\
        Creep coefficient, $n_\text{d}=2$ & $A$ & $5.0\cdot10^{-25}\;\text{s}^{-1}\text{Pa}^{-3}$ \\
        Creep coefficient, $n_\text{d}=3$ & $A$ & $2.2\cdot10^{-25}\;\text{s}^{-1}\text{Pa}^{-3}$ \\
        \hline
        \multicolumn{3}{|c|}{Cohesive Fracture Parameters}\\
        \hline
        Tensile strength & $f_\text{t}$ & $0.4\;\text{MPa}$ \\
        Compressive strength & $f_\text{c}$ & $0.6\;\text{MPa}$ \\
        Residual strength & $\kappa$ & $10^{-3}$ \\
        Fracture energy$^*$ & $G_\text{c}$ & $1-10\;\text{J/m}^2$ \\
        Phase-field length scale$^*$ & $\ell$ & $0.005-0.1\;\text{m}$ \\
        \hline
    \end{tabular}
    \caption{Material parameters used to describe the viscous and fracture response of ice. Parameters marked with an asterisk are varied in the sensitivity studies.}
    \label{tab:parameters}
    \end{minipage}\hfill
    \begin{minipage}[t]{0.4\linewidth}
        \vspace{0pt}
        \centering
        \captionsetup{type=figure}
        \includegraphics{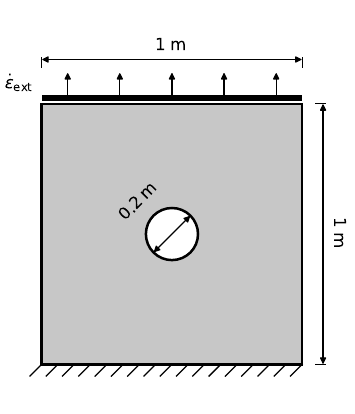}
        \caption{Geometry and boundary conditions used for the parameter sensitivity studies, using an intrinsic two-dimensional model.}
        \label{fig:plateWithHole}
    \end{minipage}
\end{table}

To study the competition between bulk creep and cohesive fracture, and the sensitivity of the results to the element size and phase-field length scale, we consider a square plate with a central hole, shown in \cref{fig:plateWithHole}. The plate has a side length of $L_x=L_y=1\;\text{m}$ and a hole of radius $R=0.1\;\text{m}$. The bottom edge is fixed in both directions, while the top edge is free to move horizontally and is displaced vertically at a constant rate, $U_\text{ext}=\dot{\varepsilon}_\text{ext}L_y t$. An intrinsic two-dimensional material is used, so that the complete stress state can be visualized against the strength surface. The finite stress concentrations at the sides of the hole separate fracture initiation from the subsequent propagation. The load-displacement curves report the average vertical stress $\overline{\sigma}_{yy}$ against the imposed displacement.

The material parameters used throughout are given in \cref{tab:parameters}, and approximate those of near-melting ice \citep{SchulsonDuval2009}. Following the convention used for ice \citep{CuffeyPaterson2010}, the table gives Glen's rate factor $A_\text{Glen}$, which is related to the coefficient $A$ of \cref{eq:creep_constitutive} by:
\begin{equation}
    \label{eq:glen_rate_conversion}
    A=\frac{2}{n_\text{d}^{(n+1)/2}}A_\text{Glen}
\end{equation}
This conversion uses the equivalent-stress normalization from \cref{eq:creep_flow_direction}. The resulting values of $A$ are given in \cref{tab:parameters} for the two-dimensional material description used in this section, and the three-dimensional description used for the application cases in \cref{sec:applications}.

The fracture energy is varied from $G_\text{c}=1$ to $10\;\text{J/m}^2$, ranging from a mostly brittle to an extended cohesive response. The associated cohesive length follows from the elastic modulus, fracture energy, and tensile strength \citep{Hageman2026b}:
\begin{equation}
    \ell_\text{phys} \approx \frac{E G_\text{c}}{f_\text{t}^2}
\end{equation}
giving $\ell_\text{phys}\approx0.06-0.6\;\text{m}$ for the parameter range considered. This is an elastic estimate, as the extent of the process zone also depends on creep and loading history. Three external strain rates are considered, $\dot{\varepsilon}_\text{ext}=1.0\cdot10^{-9}$, $1.5\cdot10^{-9}$, and $2.5\cdot10^{-9}\;\text{s}^{-1}$, covering creep-dominated deformation and the transition to fracture.

Unless stated otherwise, the simulations use adaptive refinement with a target element size of $h^\text{el}\leq\ell/4$, starting from a background element size of $0.03\;\text{m}$ with a pre-refined region of element size $0.002\;\text{m}$ around the hole. The eigenstrain indicator of \cref{sec:adaptive_mesh_refinement} is used with $\eta_\text{tol}=10^{-12}$ and a buffer of width $6\cdot(\ell/4)$ around the marked elements. One refinement stage is permitted per increment, after which the increment is resolved on the refined mesh, so the target size can be reached over several increments and the eigenstrains can properly localize in the smallest elements. Newmark parameters $\beta=0.5625$ and $\gamma=1$ are used. The damping coefficient of \cref{tab:parameters} is chosen large enough to suppress inertial effects, enforcing quasi-static crack propagation without branching, despite the short timescales associated with brittle fracture. The time increment is limited by a maximum change of the integrated crack density of $0.0025\;\text{m}$ per increment and a maximum imposed strain increment of $\dot{\varepsilon}_\text{ext}\Delta t=3\cdot10^{-7}$, which restricts the fracture growth per increment while allowing large increments during creep. The coupled response is first illustrated for a representative case, followed by studies of the element size at fixed $\ell$, of the fracture-energy correction, and of the phase-field length scale.

\subsection{Competition between creep and fracture}

\begin{figure}
    \centering
    \begin{subfigure}{8cm}
        \includegraphics{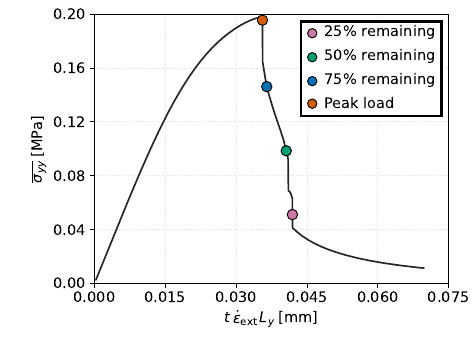}
        \caption{}
    \end{subfigure}
    \begin{subfigure}{8cm}
        \includegraphics{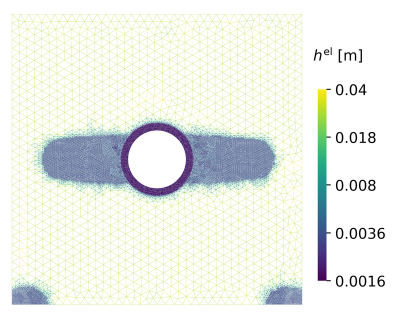}
        \caption{}
    \end{subfigure}
    \caption{Load-displacement response for $G_\text{c}=5\;\text{J/m}^2$, $\ell=0.02\;\text{m}$, and $\dot{\varepsilon}_\text{ext}=1.5\cdot10^{-9}\;\text{s}^{-1}$, indicating the states shown in the subsequent figures (a), and the adaptive mesh at $75\%$ remaining load (b).}
    \label{fig:LoadDispDetailed}
\end{figure}

\begin{figure}
    \centering
    \vspace{-4cm}
    \begin{subfigure}{8cm}
        \centering
        \includegraphics{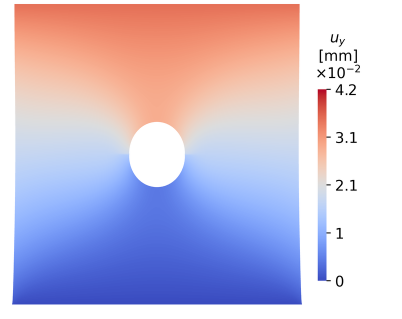}
        \caption{}
    \end{subfigure}
    \begin{subfigure}{8cm}
        \centering
        \includegraphics{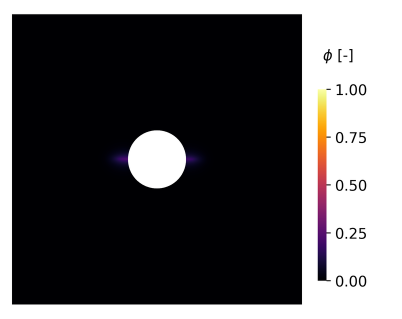}
        \caption{}
    \end{subfigure}
    \begin{subfigure}{8cm}
        \centering
        \includegraphics{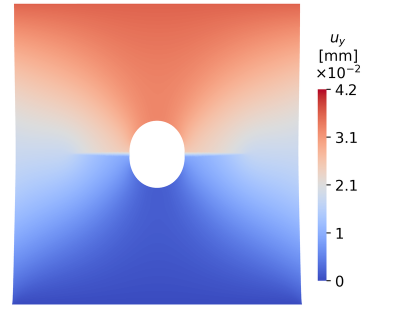}
        \caption{}
    \end{subfigure}
    \begin{subfigure}{8cm}
        \centering
        \includegraphics{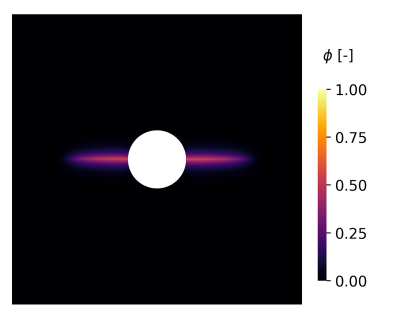}
        \caption{}
    \end{subfigure}
    \begin{subfigure}{8cm}
        \centering
        \includegraphics{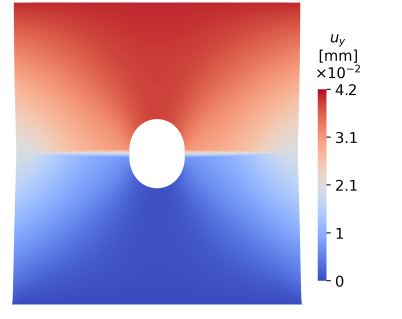}
        \caption{}
    \end{subfigure}
    \begin{subfigure}{8cm}
        \centering
        \includegraphics{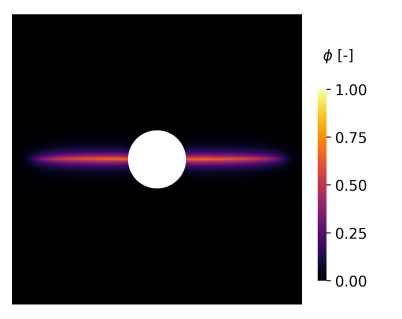}
        \caption{}
    \end{subfigure}
        \begin{subfigure}{8cm}
        \centering
        \includegraphics{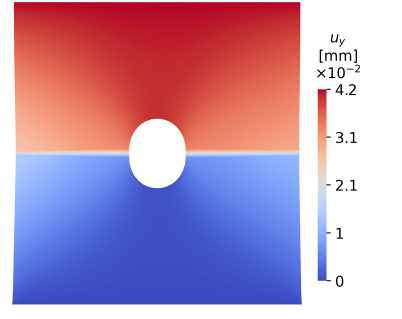}
        \caption{}
    \end{subfigure}
    \begin{subfigure}{8cm}
        \centering
        \includegraphics{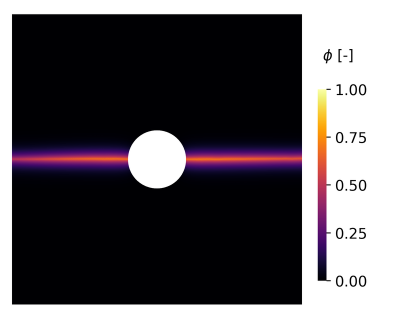}
        \caption{}
    \end{subfigure}
    \caption{Vertical displacement (left column) and phase-field (right column) near the peak load and at $75\%$, $50\%$, and $25\%$ remaining load on the descending branch, from top to bottom. Displacements in the deformed configurations are magnified by $\times1000$. The selected states are indicated in \cref{fig:LoadDispDetailed}.}
    \label{fig:uy_phasefield}
\end{figure}

\begin{figure}
    \centering
    \vspace{-4cm}
    \begin{subfigure}{8cm}
        \centering
        \includegraphics{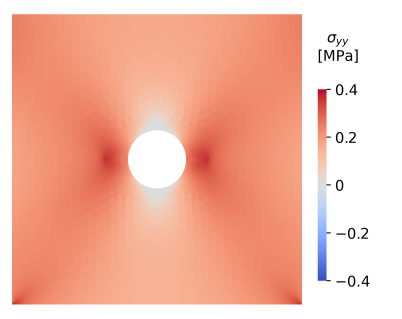}
        \caption{}
    \end{subfigure}
    \begin{subfigure}{8cm}
        \centering
        \includegraphics{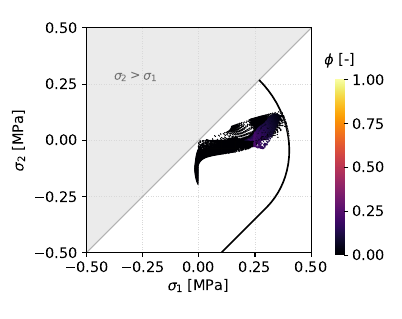}
        \caption{}
    \end{subfigure}
    \begin{subfigure}{8cm}
        \centering
        \includegraphics{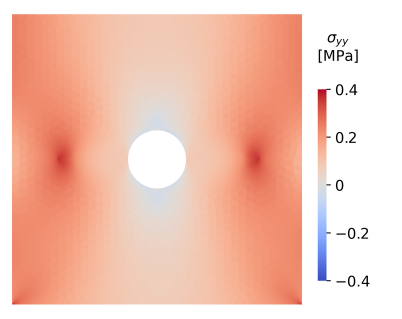}
        \caption{}
    \end{subfigure}
    \begin{subfigure}{8cm}
        \centering
        \includegraphics{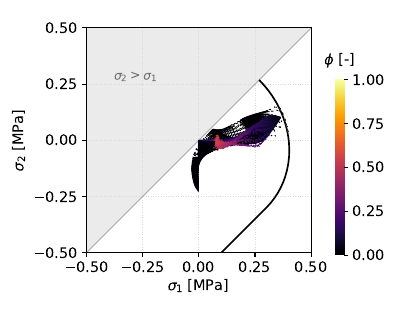}
        \caption{}
    \end{subfigure}
    \begin{subfigure}{8cm}
        \centering
        \includegraphics{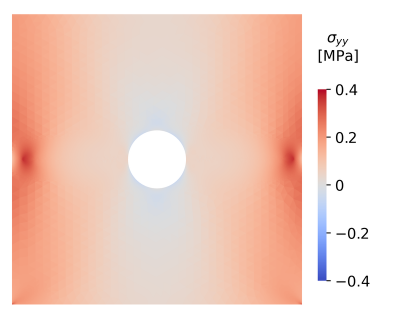}
        \caption{}
    \end{subfigure}
    \begin{subfigure}{8cm}
        \centering
        \includegraphics{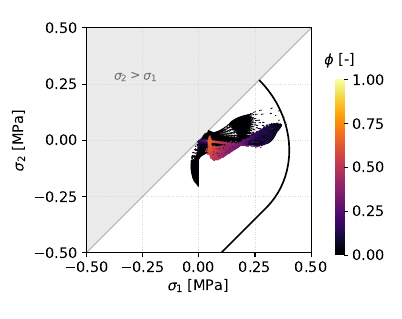}
        \caption{}
    \end{subfigure}
        \begin{subfigure}{8cm}
        \centering
        \includegraphics{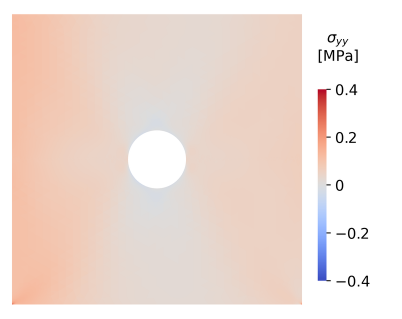}
        \caption{}
    \end{subfigure}
    \begin{subfigure}{8cm}
        \centering
        \includegraphics{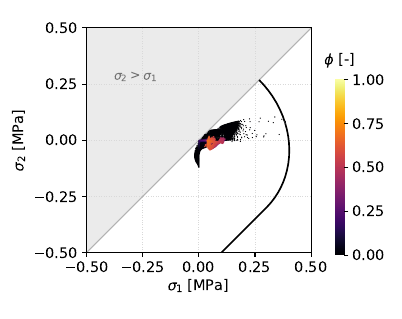}
        \caption{}
    \end{subfigure}
    \caption{Vertical stress $\sigma_{yy}$ (left column) and integration-point principal stresses, coloured by the phase-field and compared against the undegraded strength surface (right column). The rows correspond to the states from \cref{fig:uy_phasefield}. Principal stresses are ordered such that $\sigma_1\geq\sigma_2$.}
    \label{fig:stresses}
\end{figure}

We first consider the $G_\text{c}=5\;\text{J/m}^2$, $\ell=0.02\;\text{m}$, and $\dot{\varepsilon}_\text{ext}=1.5\cdot10^{-9}\;\text{s}^{-1}$ case. The load-displacement response and a snapshot of the adaptive mesh are shown in \cref{fig:LoadDispDetailed}, and the displacement, phase-field, and stress fields at the four states indicated in the load-displacement curve in \cref{fig:uy_phasefield,fig:stresses}.

The rate of applied stress required for the constant strain rate slows as creep develops, reaches a peak, and then decreases first sharply and then along a gradual softening branch. The fields in \cref{fig:uy_phasefield} show how the softening develops: At the onset of fractures, the displacement varies smoothly around the hole and the phase-field remains close to zero. Fractures then develop from the left and right sides of the hole, where the stresses are highest, and propagate horizontally towards the boundaries. During this propagation, the displacement discontinuity concentrates across the fracture while the upper and lower parts of the plate unload. Cohesive tractions continue to be transmitted, so the load decreases progressively even after the fracture path connects the hole to the boundaries. As the phase-field describes the loss of cohesion along this path, rather than a traction-free crack, the phase-field variable slowly increases during this decohesion, rather than being one at the crack interface from the onset of fracture. The mesh in \cref{fig:LoadDispDetailed}b is also refined near the fixed bottom corners: fracture eigenstrains start to develop at the stress concentrations there, but cease to grow as soon as the cracks developing from the hole unload the plate.

The stress fields in \cref{fig:stresses} show the initial concentrations at the sides of the hole moving towards the advancing fracture tips, while the material behind the tips carries progressively less vertical stress. In principal-stress space, intact material approaches the strength surface, whereas points with increasing phase-field move inwards to lower stresses, consistent with the strength reduction through $d(\phi)$. At no point do any of the integration points exceed the stress surface. The same stress state controls creep: since the creep rate of \cref{eq:creep_constitutive} depends on the current stress alone, the stress concentrations in \cref{fig:stresses} indicate where creep is active, and the migration of these stress concentrations with the propagating fracture tips is accompanied by the same migration of the creep activity, resulting in the highest viscous dissipation rate being just ahead of the crack tip. Similarly, the lowest rate of creep are at the crack, where the stress state supported by the strength surface approaches zero.

The effect of the loading rate and fracture energy on this competition is summarized in \cref{tab:rate_energy} for $\ell=0.02\;\text{m}$, with the corresponding load-displacement curves included in \cref{fig:elementSizeSensitivity,fig:elementSizeSensitivity_Gc10,fig:lengthScaleSensitivity}. Up to the first non-zero fracture eigenstrains (the ``activation'' of crack propagation), the response is governed by elasticity and creep alone, such that the displacement at activation does not depend on the fracture energy. From there on, creep in the bulk and the loss of cohesion at the crack compete to dissipate the work done by the imposed displacement. At the lowest rate, creep relaxes the stresses around the hole faster than the loading raises them: the eigenstrains remain negligible, and the applied stress settles on a creep plateau which is the same for all fracture energies. At the higher rates a cohesive zone develops, with the fracture energy determining the failure mode. For the smallest $G_\text{c}$, the cohesive tractions decay over a negligible displacement, resulting in brittle failure shortly after activation, whereas for larger $G_\text{c}$ the crack already extends while the load still increases, delaying the peak and lengthening the descending branch. Notably, the peak load does not depend monotonically on the loading rate. For brittle failure, slower loading allows creep to relax the stress concentration and delay the localization of the eigenstrains into a crack, such that the peak load decreases with increasing rate. In contrast, for an extended cohesive zone the reduced creep relaxation at the higher rate keeps the material ahead of the crack stiffer, and the peak load increases with the rate.

\begin{table}[tbp]
    \centering
    \caption{Displacement at activation of the fracture eigenstrain $u_\text{act}$ and at peak load $u_\text{peak}$, and peak external stress $\overline{\sigma}_\text{peak}$, using $\ell=0.02\;\text{m}$ and $\ell/h^\text{el}=4$. At the lowest rate no load drop occurs, with the stress at the end of the loading interval given instead ($^*$).}
    \label{tab:rate_energy}
    \begin{tabular}{lccccccccc}
        \hline
        & \multicolumn{3}{c}{$G_\text{c}=1\;\text{J/m}^2$} & \multicolumn{3}{c}{$G_\text{c}=5\;\text{J/m}^2$} & \multicolumn{3}{c}{$G_\text{c}=10\;\text{J/m}^2$} \\
        $\dot{\varepsilon}_\text{ext}\;[\text{s}^{-1}]$ & $u_\text{act}$ & $u_\text{peak}$ & $\overline{\sigma}_\text{peak}$ & $u_\text{act}$ & $u_\text{peak}$ & $\overline{\sigma}_\text{peak}$ & $u_\text{act}$ & $u_\text{peak}$ & $\overline{\sigma}_\text{peak}$ \\
        & [mm] & [mm] & [MPa] & [mm] & [mm] & [MPa] & [mm] & [mm] & [MPa] \\
        \hline
        $1.0\cdot10^{-9}$ & -- & --$^*$ & 0.185$^*$ & -- & --$^*$ & 0.185$^*$ & -- & --$^*$ & 0.185$^*$ \\
        $1.5\cdot10^{-9}$ & 0.030 & 0.030 & 0.190 & 0.030 & 0.035 & 0.198 & 0.030 & 0.046 & 0.205 \\
        $2.5\cdot10^{-9}$ & 0.020 & 0.021 & 0.165 & 0.020 & 0.026 & 0.189 & 0.020 & 0.035 & 0.210 \\
        \hline
    \end{tabular}
\end{table}

\FloatBarrier

\subsection{Element-size sensitivity}
\label{sec:element_size_sensitivity}

\begin{figure}
    \centering
    \begin{subfigure}{8cm}
        \includegraphics{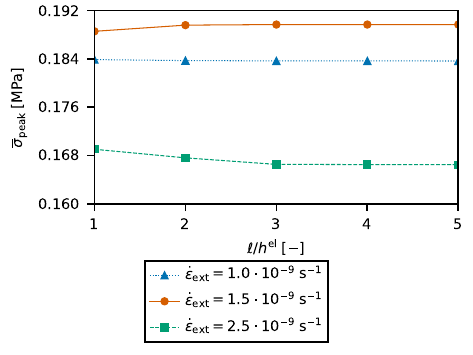}
        \caption{}
    \end{subfigure}
    \begin{subfigure}{8cm}
        \includegraphics{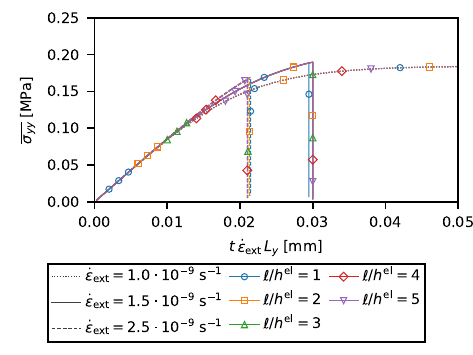}
        \caption{}
    \end{subfigure}
    \caption{Sensitivity of the peak load (a) and the full load-displacement response (b) to the element size $h^\text{el}$ for a fixed phase-field length scale of $\ell=0.02\;\text{m}$ and fracture energy $G_\text{c}=1\;\text{J/m}^2$ ($\ell_\text{phys}\approx 0.06\;\text{m}$). }
    \label{fig:elementSizeSensitivity}
\end{figure}

\begin{figure}
    \centering
    \begin{subfigure}{8cm}
        \includegraphics{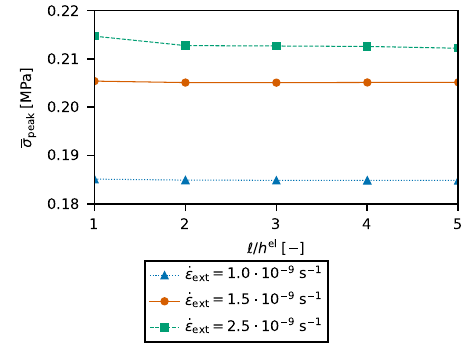}
        \caption{}
    \end{subfigure}
    \begin{subfigure}{8cm}
        \includegraphics{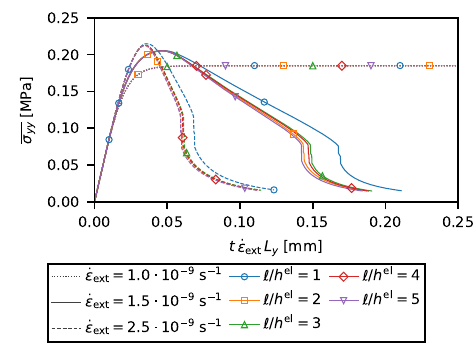}
        \caption{}
    \end{subfigure}
    \caption{Sensitivity of the peak load (a) and the full load-displacement response (b) to the element size $h^\text{el}$ for a fixed phase-field length scale of $\ell=0.02\;\text{m}$ and fracture energy $G_\text{c}=10\;\text{J/m}^2$ ($\ell_\text{phys}\approx 0.6\;\text{m}$). }
    \label{fig:elementSizeSensitivity_Gc10}
\end{figure}

The effect of the element size is studied at a fixed length scale of $\ell=0.02\;\text{m}$ on prescribed uniform meshes with $\ell/h^\text{el}=1-5$. This study is done without adaptive refinement, so that the spatial resolution is separated from changes of the mesh during fracture. Results are shown for $G_\text{c}=1\;\text{J/m}^2$ in \cref{fig:elementSizeSensitivity} and for $G_\text{c}=10\;\text{J/m}^2$ in \cref{fig:elementSizeSensitivity_Gc10}. 

For the lower fracture energy, the results for all element sizes are near-indistinguishable at all three loading rates, \cref{fig:elementSizeSensitivity}, with the peak external stress differing by at most $1.7\%$ over the complete range of element sizes. The lowest loading rate results in the same creep plateau on all meshes, while at the two higher rates the load drops near-instantly once fracture develops, such that no effect of the element size on the unloading behaviour is observed.

The higher fracture energy gives an extended cohesive response, \cref{fig:elementSizeSensitivity_Gc10}, allowing the post-peak behaviour to be assessed separately from the peak load. The peak loads are again insensitive to the element size, differing by at most $1.2\%$ between the coarsest and finest mesh. The softening shows a clear effect of the element size for the coarsest mesh, $\ell/h^\text{el}=1$, which cannot resolve the phase-field profile and as a result retains its load-carrying capacity for longer. For $\ell/h^\text{el}\geq2$, the descending branches show no systematic ordering with the element size. The displacement at which the load has decreased to half its peak value, which is used as a measure of the softening interval as the area under the load-displacement curve also contains creep dissipation, differs by at most $4\%$ between these meshes, comparable to the mesh-dependent scatter observed in the $G_\text{c}$ correction scheme of \ref{app:Gc_correction}.

These observations follow from the manner in which the two stages of fracture enter the formulation. Nucleation is decided pointwise, by comparing the integration-point stress to the strength surface, without the phase-field or its length scale being involved. As long as the mesh resolves the stress concentration at the hole, the peak load is therefore reproduced irrespective of the element size. In contrast, the loss of cohesion requires the phase-field profile to be resolved and, with the eigenstrain confined to a single layer of integration points, introduces the first-order error in the dissipated energy that is removed by the correction of \cref{eq:Gc_inp}. The same resolution is needed for creep, which is driven by the stress concentration ahead of the crack. As a result, refinement can be triggered by the eigenstrain on a coarse mesh and confined to the region where cohesion is being lost, as exploited by the indicator of \cref{sec:adaptive_mesh_refinement}. While the target ratio of $\ell/h^\text{el}=4$ is slightly conservative, with the peak load and the softening interval changing by less than $0.2\%$ and $3\%$ upon further refinement, this ratio will be used in the following sections to ensure no effect of the mesh resolution itself on the remaining studies.

\FloatBarrier

\subsection{Effect of the fracture-energy correction}
\label{sec:Gc_correction_effect}

\begin{figure}
    \centering
    \begin{subfigure}{8cm}
        \includegraphics{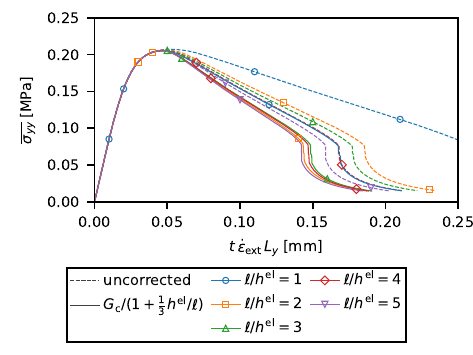}
        \caption{}
    \end{subfigure}
    \begin{subfigure}{8cm}
        \includegraphics{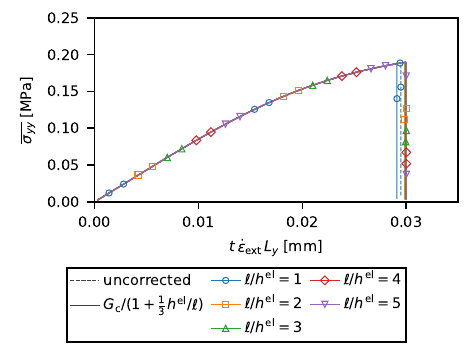}
        \caption{}
    \end{subfigure}
    \caption{Effect of the element-size correction of the fracture energy, \cref{eq:Gc_inp}, on the load-displacement response for $G_\text{c}=10\;\text{J/m}^2$ (a) and $G_\text{c}=1\;\text{J/m}^2$ (b), using $\ell=0.02\;\text{m}$ and $\dot{\varepsilon}_\text{ext}=1.5\cdot10^{-9}\;\text{s}^{-1}$.}
    \label{fig:GcCorrectionEffect}
\end{figure}

The mesh-insensitive descending branches of \cref{fig:elementSizeSensitivity_Gc10} rely on the correction of \cref{eq:Gc_inp}. Without it, the discrete cohesive law dissipates $G_\text{c}(1+h^\text{el}/(3\ell))$, \cref{eq:Gc_num}, an excess of $17\%$ to $7\%$ for $\ell/h^\text{el}=2$-$5$, such that each mesh effectively simulates a different material with increased dissipation. This is shown in \cref{fig:GcCorrectionEffect}a: the uncorrected descending branches are ordered by element size and shift monotonically towards the converged response upon refinement, as expected for a first-order discretization error, whereas the corrected branches show no such ordering and only differ by the scatter caused by where the crack passes relative to the nodes. Notably, the $\ell/h^\text{el}=1$ mesh, which is far outside the asymptotic range of \cref{eq:Gc_num}, reproduces the uncorrected $\ell/h^\text{el}=4$ result once corrected, confirming that the error is governed by the dissipated energy and not by the shape of the under-resolved band. The peak loads change by less than $1\%$ for both the corrected and uncorrected cases, confirming that the correction solely acts on the cohesive law and leaves the strength-based nucleation unaffected.

Without the correction, a mesh-refinement study thus only converges linearly and from one side, with the converged response having a shorter softening interval than any of the uncorrected ones. Under adaptive refinement this is more severe, as the element size varies along the crack path and over time, such that the error would act as a spatially and temporally varying fracture energy. This is prevented by the non-local evaluation of $\bar{h}$ in \cref{eq:Gc_inp} \citep{Feng2026AMR}. The relevance of the correction depends on the failure mode: for $G_\text{c}=10\;\text{J/m}^2$, the cohesive length is comparable to the plate and the crack advances quasi-statically, such that an error in the dissipated energy directly alters the softening, which is further amplified by the additional time that creep has to relax the stresses ahead of the crack. For $G_\text{c}=1\;\text{J/m}^2$, propagation is driven by an elastic energy release that greatly exceeds the fracture energy, and the corrected and uncorrected curves coincide to within $0.3\%$, \cref{fig:GcCorrectionEffect}b. The correction is therefore of most relevance in the regime where fracture is slow enough to interact with creep, while being inconsequential for brittle failure.

\FloatBarrier

\subsection{Length-scale sensitivity}
\label{sec:length_scale_sensitivity}

\begin{figure}
    \centering
    \begin{subfigure}{8cm}
        \includegraphics{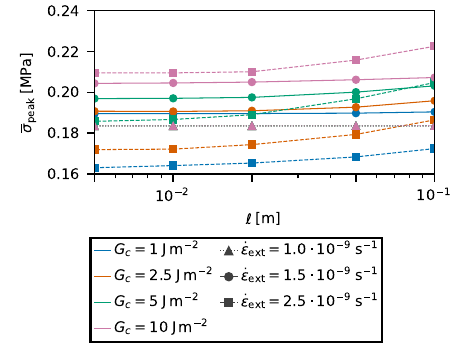}
        \caption{}
    \end{subfigure}
    \begin{subfigure}{8cm}
        \includegraphics{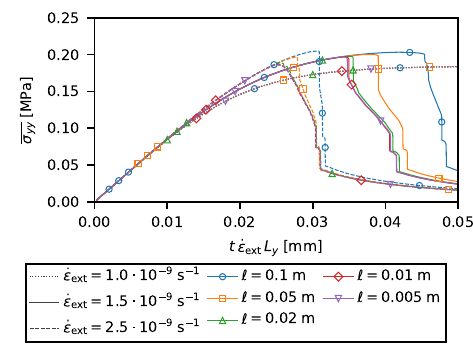}
        \caption{}
    \end{subfigure}
    \caption{Sensitivity of the peak stress to the phase-field length scale $\ell$ (a), and load-displacement response for $G_\text{c}=5\;\text{J/m}^2$ (b).}
    \label{fig:lengthScaleSensitivity}
\end{figure}

Finally, the effect of the phase-field length scale is considered for $\ell=0.005$-$0.1\;\text{m}$ and $G_\text{c}=1$, $2.5$, $5$, and $10\;\text{J/m}^2$ at the three loading rates, \cref{fig:lengthScaleSensitivity}. As the adaptive refinement targets $h^\text{el}\leq\ell/4$, changing $\ell$ changes both the width of the diffuse decohesion region and the element size along the propagating fracture. This study therefore also verifies the adaptive refinement against the prescribed meshes of \cref{sec:element_size_sensitivity}: for $G_\text{c}=10\;\text{J/m}^2$ and $\ell=0.02\;\text{m}$, the adaptively refined mesh reproduces the peak load of the uniform mesh with the same $\ell/h^\text{el}$ in \cref{fig:elementSizeSensitivity_Gc10} to within $0.1\%$, while its descending branch shows a slightly earlier loss of load-carrying capacity, in line with the refined elements along the crack path being equal to or smaller than the target element size.

As no fracture develops at the lowest loading rate, identical results are obtained for all length scales and fracture energies. For the other loading rates, the peak load decreases with decreasing $\ell$, with this sensitivity increasing with the loading rate: the peak changes by a few percent over the full range of $\ell$ at the intermediate rate, and by up to ten percent at the highest rate. In all cases the peak loads converge upon using smaller $\ell$, changing by less than one percent between the smallest length scales. The post-peak response converges more slowly, \cref{fig:lengthScaleSensitivity}b, with the larger length scales delaying the load drop and extending the descending branch, whereas the curves for the three smallest length scales lie close together.

While the prescribed strength surface is independent of $\ell$, this does not imply an identical structural response for all length scales. The largest value, $\ell=0.1\;\text{m}$, equals the hole radius and exceeds the cohesive length for $G_\text{c}=1\;\text{J/m}^2$, such that the diffuse region interacts with the geometry and with the developing cohesive zone. Creep amplifies the resulting differences in the stress distribution, as the creep rate scales with the third power of the equivalent stress, explaining the larger sensitivity at the highest rate, where the stress concentration is least relaxed when fracture initiates. In practice, $\ell$ therefore needs to be small relative to the geometric features to ensure that the stress state is correctly captured for nucleation, especially close to the creep stress limit. Furthermore, as creep interacts with the stresses at the crack tip, the length scale also needs to resolve the cohesive zone, and thus be comparable to the physical length scale.

\FloatBarrier

\section{Large-scale applications}
\label{sec:applications}
To demonstrate the model beyond the one-metre plate of \cref{sec:results_2D_plate}, we consider two cases on domains that are hundreds of phase-field length scales in size: the fragmentation of a sea-ice floe, which involves dynamic and branching fracture, and the growth of crevasses in an ice cliff, where fracture is driven by gravity and the two strength criteria of \cref{sec:fracture_constitutive,sec:opening_criterion} are compared. Both cases use the material parameters of \cref{tab:parameters} unless stated otherwise, together with the adaptive mesh refinement of \cref{sec:adaptive_mesh_refinement} and the element-size correction of \cref{sec:Gc_correction}.

\subsection{Sea-ice fragmentation}
\label{sec:sea_ice}

\begin{figure}[tbp]
    \centering
    \begin{subfigure}{8cm}
        \includegraphics{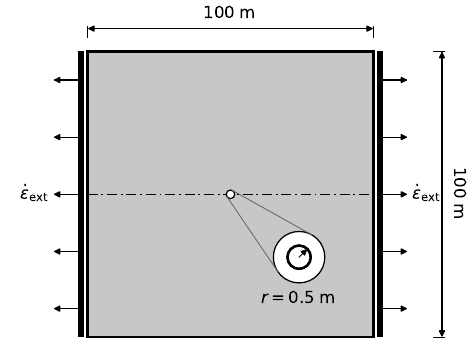}
        \caption{}
    \end{subfigure}
    \begin{subfigure}{8cm}
        \centering
        \includegraphics{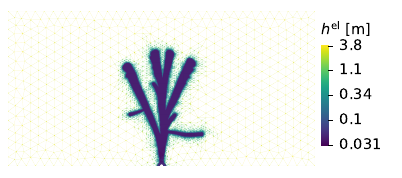}
        \caption{}
    \end{subfigure}
    \caption{Geometry and boundary conditions of the sea-ice floe (a), and adaptive mesh of the simulated upper half during fragmentation (b).}
    \label{fig:seaice_setup}
\end{figure}

We first consider a square sea-ice floe with a side length of $L=100\;\text{m}$ and a thickness of $h=1\;\text{m}$, which is pulled apart at a constant rate, \cref{fig:seaice_setup}a. The left and right edges are displaced in opposite horizontal directions, imposing a far-field strain rate $\dot{\varepsilon}_\text{ext}$, while the top and bottom edges are free. A circular defect of radius $R=0.5\;\text{m}$ at the centre of the floe provides a finite stress concentration, from which fractures initiate and propagate towards the free edges. As the geometry and loading are symmetric about the horizontal centreline, only the upper half of the floe is simulated. The three-dimensional constitutive model is used under plane-stress conditions, in contrast to the intrinsic two-dimensional material of \cref{sec:results_2D_plate}. Results are reported through the nominal stress $F/(Lh)$, with $F$ the reaction force, and the far-field strain $\varepsilon_\text{ext}=\dot{\varepsilon}_\text{ext}t$. The crack length is estimated by integrating the regularized crack surface, counted for the full floe, with the nucleation of the crack taken as the last increment before this length exceeds $0.1\;\text{m}$.

The material parameters of \cref{tab:parameters} are used with $G_\text{c}=5\;\text{J/m}^2$, giving $\ell_\text{phys}\approx0.3\;\text{m}$, and a phase-field length scale of $\ell=0.25\;\text{m}$. This length scale is comparable to the cohesive length and equal to half the defect radius, following \cref{sec:length_scale_sensitivity}, while the floe spans $400\,\ell$. Three far-field strain rates are considered, $\dot{\varepsilon}_\text{ext}=1.0\cdot10^{-9}$, $2.5\cdot10^{-9}$, and $1.0\cdot10^{-8}\;\text{s}^{-1}$, referred to as the slow, intermediate, and fast rate, together with an elastic reference without creep at the fast rate. The damping coefficient is reduced to $c=11\;\text{s}^{-1}$, so that stress waves decay over timescales longer than the fracture process, and the propagation of the cracks is governed by inertia.

The background mesh has an element size of $3\;\text{m}$, with the region within $1\;\text{m}$ of the defect pre-refined to $h^\text{el}=\ell/4$. Adaptive refinement is used with a target element size of $h^\text{el}\leq\ell/4$. As for the benchmarks in the prior section, the time increment is limited by an imposed strain increment of $3\cdot10^{-7}$, and in addition by a change of the integrated crack density of $2\ell$ per increment. The elastic and intermediate cases are stopped once the crack propagates dynamically, the slow case is continued up to a far-field strain of $10^{-4}$, and the fast case is continued after nucleation to demonstrate the ability to capture crack branching.

\begin{figure}[tbp]
    \centering
    \includegraphics{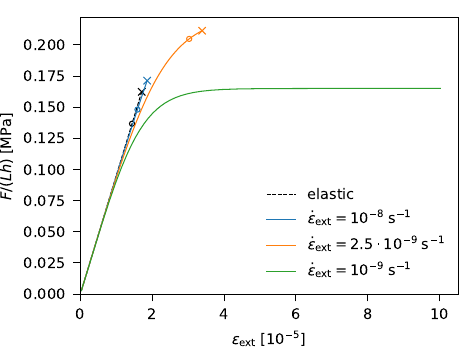}
    \caption{Nominal stress against far-field strain up to the nucleation of the crack (crosses), with circles indicating where the strength surface is first reached.}
    \label{fig:seaice_rates}
\end{figure}

\begin{figure}[!tbp]
    \centering
    \begin{subfigure}{8cm}
        \centering
        \includegraphics{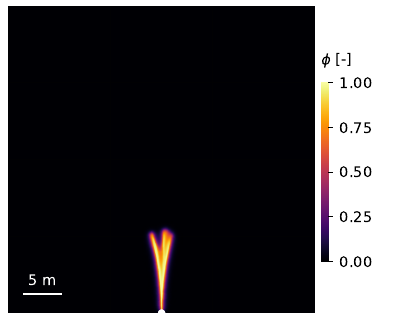}
        \caption{}
    \end{subfigure}
    \begin{subfigure}{8cm}
        \centering
        \includegraphics{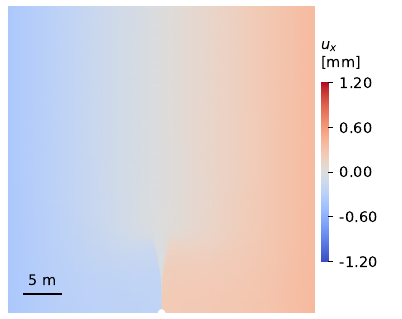}
        \caption{}
    \end{subfigure}
    \begin{subfigure}{8cm}
        \centering
        \includegraphics{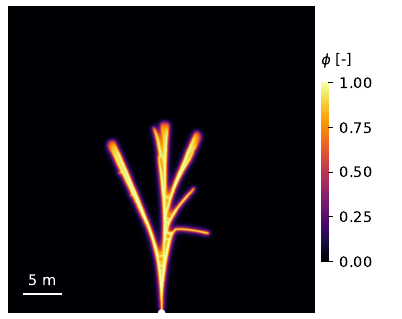}
        \caption{}
    \end{subfigure}
    \begin{subfigure}{8cm}
        \centering
        \includegraphics{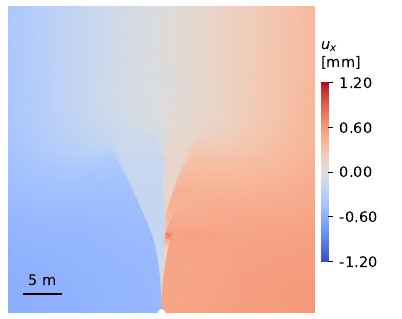}
        \caption{}
    \end{subfigure}
    \begin{subfigure}{8cm}
        \centering
        \includegraphics{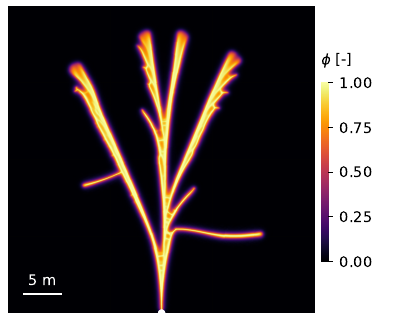}
        \caption{}
    \end{subfigure}
    \begin{subfigure}{8cm}
        \centering
        \includegraphics{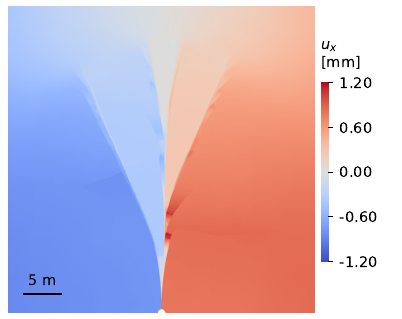}
        \caption{}
    \end{subfigure}
    \caption{Phase-field (left) and horizontal displacement $u_x$ (right) for the fast loading rate at $7.0$, $15.3$, and $22.3\;\text{ms}$ after nucleation (top to bottom), shown for a $40\times40\;\text{m}$ region above the defect.}
    \label{fig:seaice_patterns}
\end{figure}

At the fast rate, the crack nucleates at the defect after half an hour of loading, at a stress of less than half the tensile strength. From there on the crack propagates unstably, with the complete fracture process shown in \cref{fig:seaice_patterns} taking place within milliseconds. The crack initially propagates straight from the defect, and splits into three branches after a few metres. These branches fan out and develop secondary branches, some of which arrest shortly after forming. The horizontal displacements confirm the main branches to be open cracks, with the wedges of ice between them moving independently. This branching is a result of the available energy: within a few metres of the defect, the energy release rate of a single crack exceeds $G_\text{c}$ by an order of magnitude, with the excess being converted into additional crack surface \citep{Borden2012}. As a result, the crack length grows at an increasing rate, exceeding six times the length of a single straight crack at the end of the simulation. As expected for an unstable process, the obtained pattern is not symmetric about the vertical axis, even though the geometry and loading are. Furthermore, repeated simulations of the same geometry obtained different crack patterns. 

Notably, the fronts of the leading branches advance at $0.95$ times the Rayleigh wave speed, well above the speeds at which cracks in brittle solids are observed to branch \citep{RaviChandar1984,Fineberg1999}. This is a consequence of the strong over-driving in combination with the strength-based nucleation: the stresses ahead of each front reach the strength surface over a region that increases with the distance travelled, such that cohesion is lost over a widening band rather than at a single crack tip, visible as the wedges of partial damage behind the fronts in \cref{fig:seaice_patterns}. These wedges widen until reaching approximately twice the width of the phase-field band, upon which new branches develop. 

The effect of creep is contained in the loading stage, shown in \cref{fig:seaice_rates}. Without creep, the strength surface is first reached at a nominal stress close to $f_\text{t}/3$, in line with the stress concentration at a circular hole. As the creep rate scales with the third power of the stress, creep relaxes the stress concentration at the defect faster than the far-field stress. The nominal stress at which the strength is reached, and thus the elastic energy stored in the floe at nucleation, therefore increases as the loading rate decreases, being $50\%$ above the elastic value at the intermediate rate. At the slow rate, the nominal stress saturates at the stress for which creep alone accommodates the imposed strain rate, and the floe deforms without the strength surface being reached. The transition from a fracture-dominated to a creep-dominated response of \cref{sec:results_2D_plate} is thus also obtained at the scale of a floe.

This case also shows the range of scales covered by the adaptive refinement and time stepping. The initial mesh of the half floe consists of $2.8\cdot10^3$ elements, which suffices for the loading stage and for the complete slow case. During fragmentation, the mesh grows to $4.1\cdot10^5$ elements, with the refinement confined to bands around the cracks, as shown for the end of the simulation in \cref{fig:seaice_setup}b, whereas a uniform mesh of the same resolution would require several million elements. Similarly, the time increment decreases from $30\;\text{s}$ during loading to a few microseconds during fragmentation. As a result, the computational cost is almost entirely determined by the fracture process: the $22\;\text{ms}$ after nucleation require $2\cdot10^3$ increments and $67\;\text{h}$ on $52$ cores, whereas the slow case completes in less than ten minutes on $8$ cores.

\subsection{Crevasse growth in an ice cliff}
\label{sec:cliff}

\begin{figure}[tbp]
    \centering
    \includegraphics{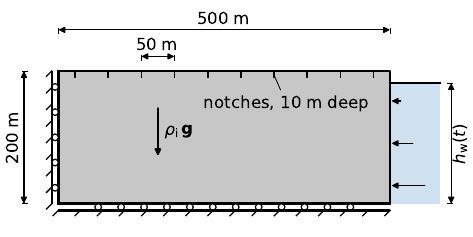}
    \caption{Geometry and boundary conditions of the ice cliff.}
    \label{fig:cliff_setup}
\end{figure}

The second case considers surface crevasses in a grounded glacier that terminates in the ocean, \cref{fig:cliff_setup}, where fracture is driven by gravity. The cliff is represented by a domain with a height of $H=200\;\text{m}$ and a length of $500\;\text{m}$, using plane-strain conditions. The horizontal displacement is constrained on the left edge and the vertical displacement on the bottom, while the top is free of tractions. Gravity acts on the complete domain from the start of the simulation, as a body force of magnitude $\rho_\text{i}g$ with $g=9.81\;\text{m/s}^2$, which is not degraded by the phase-field: ice that has lost its cohesion retains its weight. The right edge is supported by the hydrostatic pressure of sea water, with a density of $\rho_\text{w}=1025\;\text{kg/m}^3$, up to a level $h_\text{w}$ above the bed. This water level is used as the loading parameter: lowering it removes the support of the face and brings the upper part of the cliff into horizontal tension.

The water level is held at $h_\text{w}=182\;\text{m}$ during an initialization  period of $100\;\text{h}$, for which the water pressure nearly balances the pressure that the weight of the ice causes over the height of the cliff. It is then lowered to zero at a constant rate over one hour, starting at $t=0$, after which the simulation is continued for a further $600\;\text{s}$. Two cliffs are compared, which only differ in their rheology: a viscoelastic cliff, using the creep parameters of \cref{tab:parameters}, and an elastic reference without creep. Without creep, the ice is laterally confined under its own weight, with the horizontal stress increasing with depth at the fraction $\nu/(1-\nu)\approx0.54$ of the rate at which the vertical stress increases. With creep, the deviatoric stresses relax during the initialization and the stress approaches a lithostatic state, such that the surface of the viscoelastic cliff is under a larger tensile stress for the same water level. During the lowering, creep in addition relaxes part of the stresses that develop as the water level falls.

Ten notches with a depth of $10\;\text{m}$ and a width of $0.5\;\text{m}$ are placed in the surface at a spacing of $50\;\text{m}$, providing the positions at which crevasses can start. The material parameters of \cref{tab:parameters} are used with $G_\text{c}=10\;\text{J/m}^2$, giving $\ell_\text{phys}\approx0.6\;\text{m}$, and a phase-field length scale of $\ell=1\;\text{m}$. Two parameters deviate from the table. The compressive strength is increased to $f_\text{c}=1.5\;\text{MPa}$, as with $f_\text{c}=0.6\;\text{MPa}$ the two-multiplier criterion would be reached in compression by the weight of the intact ice in the lower part of the elastic cliff. The damping coefficient is set to $c=1.1\cdot10^3\;\text{s}^{-1}$, for which the cliff as a whole responds to a change in loading within approximately ten seconds. The background mesh consists of $1.2\cdot10^3$ elements with a size of $25\;\text{m}$, and is refined adaptively to a target element size of $h^\text{el}\leq\ell/4$, and a buffer of width $4\cdot(\ell/4)$. The time increment is one hour during the initialization and at most $30\;\text{s}$ during the lowering, and is reduced where needed to limit the change of the integrated crack density to $5\ell$ per increment.

In contrast to the preceding cases, material that has lost its cohesion is still loaded in compression: the ice beside an open crevasse is free of horizontal stress, but carries the weight of the ice above it. This case is therefore simulated with both criteria of \cref{sec:fracture_constitutive,sec:opening_criterion}. For the two-multiplier criterion, the strength surface is closed in compression, with the deviatoric stress that can be carried reducing with the loss of cohesion, \cref{fig:opening_surface_2d}. As the regularized crack leaves the ice beside a crevasse partly damaged over a width set by $\ell$, and the deviatoric part of the overburden increases linearly with depth, the degraded strength surface is reached in compression at sufficient depth, without any tensile stress being present.

\begin{figure}[tbp]
    \centering
    \begin{subfigure}{8cm}
        \centering
        \includegraphics{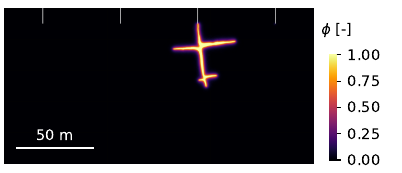}
        \caption{}
    \end{subfigure}
    \begin{subfigure}{8cm}
        \centering
        \includegraphics{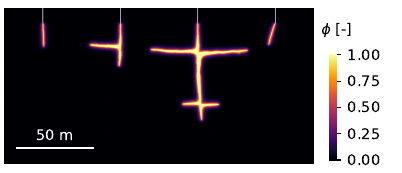}
        \caption{}
    \end{subfigure}
    \begin{subfigure}{8cm}
        \centering
        \includegraphics{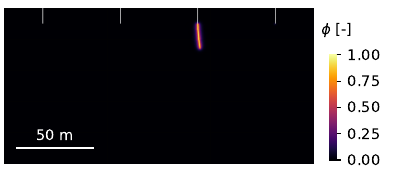}
        \caption{}
    \end{subfigure}
    \begin{subfigure}{8cm}
        \centering
        \includegraphics{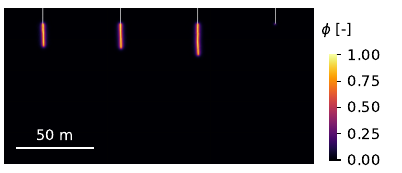}
        \caption{}
    \end{subfigure}
    \caption{Phase-field near the cliff face for the two-multiplier criterion (a, b) and the opening-only criterion (c, d), for the elastic cliff at $h_\text{w}=155\;\text{m}$ (a, c) and the viscoelastic cliff at $h_\text{w}=164\;\text{m}$ (b, d).}
    \label{fig:cliff_criteria}
\end{figure}

This is confirmed by the phase-field obtained for both cliffs, shown in \cref{fig:cliff_criteria}a and b, with all simulations in this figure limiting the crack growth to $\ell/4$ per increment. The first crevasse initially grows straight down as the water level is lowered. Once it reaches a depth of just under $30\;\text{m}$, horizontal arms develop from both of its walls, extending $15$ to $30\;\text{m}$ to either side within approximately one second. Across these arms, the ice above moves downwards relative to the ice below: they are not opening cracks, but bands in which the partly damaged ice beside the crevasse has failed under the overburden. These two simulations were not continued beyond the states shown.

\begin{figure}[tbp]
    \centering
    \begin{subfigure}{8cm}
        \centering
        \includegraphics{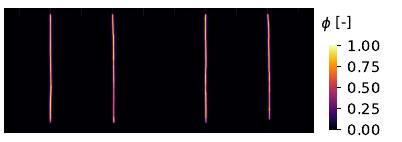}
        \caption{}
    \end{subfigure}
    \begin{subfigure}{8cm}
        \centering
        \includegraphics{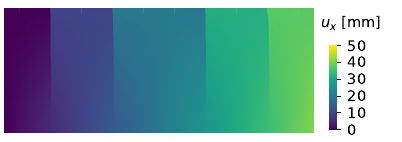}
        \caption{}
    \end{subfigure}
    \begin{subfigure}{8cm}
        \centering
        \includegraphics{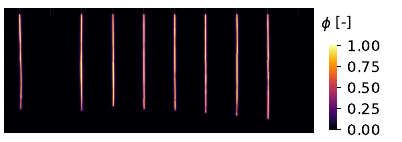}
        \caption{}
    \end{subfigure}
    \begin{subfigure}{8cm}
        \centering
        \includegraphics{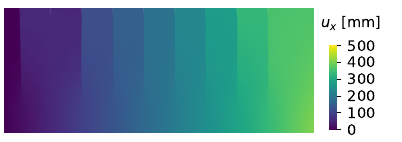}
        \caption{}
    \end{subfigure}
    \caption{Phase-field (a, c) and horizontal displacement $u_x$ (b, d) at the end of the simulation for the elastic (a, b) and viscoelastic (c, d) cliff, using the opening-only criterion.}
    \label{fig:cliff_final}
\end{figure}

With the opening-only criterion, the crevasses pass through the depth at which the arms formed without any change in their growth, \cref{fig:cliff_criteria}c and d. Up to that depth, the two criteria result in the same crevasses at the same water level, as their strength surfaces coincide for the tensile stress states under which the crevasses nucleate and grow. Below it, the crevasses remain straight over the complete lowering of the water level, \cref{fig:cliff_final}, with the horizontal displacement being discontinuous across each of them. Where the cohesion has been lost, the horizontal and shear stresses are of the order of kilopascals, whereas the vertical stress equals the weight of the ice above it. The damaged ice thus retains its weight and transfers it to the ice below, without transmitting any tension across the crevasse. This is obtained without degrading the body force or decomposing the elastic energy into degraded and non-degraded parts, both of which are needed in phase-field models of crevasses in which the phase-field degrades the elastic energy \citep{Sun2021,Clayton2022,Clayton2025,Hageman2026}. The eigenstrain-based formulation is thus able to include gravity directly, provided that it is combined with a strength surface which accounts for the compressive stresses that gravity causes in damaged material. The results that follow all use the opening-only criterion.

\begin{figure}[tbp]
    \centering
    \includegraphics{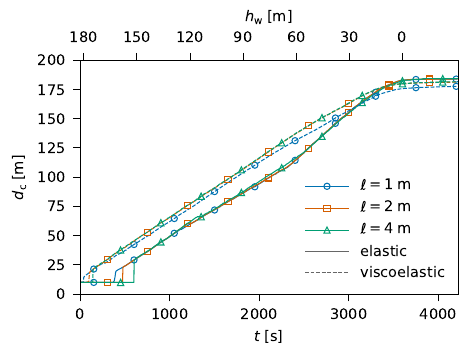}
    \caption{Depth $d_\text{c}$ of the deepest crevasse in the elastic and the viscoelastic cliff for three phase-field length scales, using the opening-only criterion.}
    \label{fig:cliff_depth}
\end{figure}

When and where the crevasses form follows from the stress state at the end of the initialization period. In the viscoelastic cliff, the surface is under tension along the full length of the cliff, such that the first crevasse starts to grow within the first minute of the lowering and eight of the ten notches develop into crevasses, which subsequently grow together. In the elastic cliff, the surface away from the face is initially under compression, and a single crevasse first forms close to the face, starting with an unstable phase in which it nearly doubles the depth of the notch within two seconds. As each crevasse relieves the tensile stress in the ice beside it, only three further crevasses develop inland, at a spacing of $100$ to $150\;\text{m}$. Once started, the crevasses of both cliffs grow in a stable manner, at a rate that is set by the lowering of the water level. For the elastic cases, the crevasses arrest once the water level has reached zero, \cref{fig:cliff_depth}. In contrast, the viscoelastic cases retain a small propagation rate due to stress relaxation once the water level reaches zero, causing the crevasse to grow by about $2\;\text{mm}/\text{s}$ (against $\approx 50\;\text{mm}/\text{s}$ during water-level change). 

To verify that these results do not depend on the phase-field length scale, both cliffs have also been simulated with $\ell=2\;\text{m}$ and $\ell=4\;\text{m}$, up to seven times $\ell_\text{phys}$, \cref{fig:cliff_depth}. As the strength surface is independent of $\ell$, the fracture eigenstrain first develops at the notches at the same water level for all three length scales. However, the growth of a crevasse from its notch starts later for a larger length scale, in the elastic cliff at a water level of $152\;\text{m}$ for $\ell=4\;\text{m}$ compared to $163\;\text{m}$ for $\ell=1\;\text{m}$. We attribute this to the notches, which are $0.5\;\text{m}$ wide and thus not resolved by the phase-field, in line with \cref{sec:length_scale_sensitivity}, where $\ell$ was required to be small relative to the geometric features at which fracture nucleates. Once a crevasse grows, its depth no longer depends on $\ell$: the depths in the elastic cliff remain within approximately $3\;\text{m}$ of each other, up to the arrest of the deepest crevasse at $184\;\text{m}$ for all three length scales. The later start does alter which of the notches develop into crevasses. In the viscoelastic cliff, the crevasses that start first relieve the tensile stress at the neighbouring notches before these have started to grow, such that only every second notch develops into a crevasse for $\ell=2$ and $4\;\text{m}$, giving five crevasses instead of eight. Being spaced further apart, these grow up to $8\;\text{m}$ deeper than the crevasses for $\ell=1\;\text{m}$, while agreeing with each other to within $1\;\text{m}$ up to the end of the simulation, at which they have reached a depth of $181\;\text{m}$. The number of crevasses thus depends on accurately resolving the crack nucleation at notches, whereas their growth is independent of the length scale.

As for the floe, the computational cost is determined by the fracture process. The mesh of the elastic cliff grows from $1.2\cdot10^3$ to $2.9\cdot10^5$ elements, all of which are added along the $0.8\;\text{km}$ of crevasses that have formed at the end of the simulation. The lowering requires $3.7\cdot10^2$ increments and $20\;\text{h}$ on $16$ cores, with a typical time increment of $10\;\text{s}$ which reduces to $10\;\text{ms}$ when a crevasse starts. With eight crevasses, the viscoelastic cliff ends with $1.7\;\text{km}$ of crevasses and $5.3\cdot10^5$ elements, and requires $7.3\cdot10^2$ increments and $116\;\text{h}$ on $16$ cores.

\FloatBarrier
\section{Conclusions}
In this paper, we have presented a cohesive phase-field formulation in which power-law creep strains and fracture eigenstrains are internal variables of a single material model. Cracks nucleate once the stress reaches a prescribed strength surface, the phase-field governs the subsequent loss of cohesion, and creep responds to the same stress state, without any further coupling between the two mechanisms being introduced: no degradation function acts on the creep law, and creep does not enter the driving force of the phase-field. Both sets of internal variables are updated together by a local return-mapping scheme with a consistent tangent, such that no priority is given to either mechanism and no global degrees of freedom are added. Furthermore, application of recently developed methods for other phase-field formulations, adaptive refinement based on decohesion and energy corrections, were shown to also be advantageous for eigenstrain-based phase-field. 

The benchmark cases show that the interaction between creep and fracture follows from the shared stress state alone: under slow loading, creep relaxes the stresses before the strength surface is reached and no fracture develops, whereas under faster loading the fracture energy determines whether cohesion is lost abruptly, or gradually while creep continues to relax the stresses ahead of the crack. As nucleation is decided pointwise by the strength surface, the load at which cracks nucleate is independent of the element size, allowing the mesh refinement to be triggered by the fracture eigenstrain on a coarse mesh. In contrast, the energy dissipated during the loss of cohesion exceeds the fracture energy by an amount that is of first order in the element size. Correcting the fracture energy for the local element size removes this error, which is most important in the regime where fracture is slow enough to interact with creep. Furthermore, the phase-field length scale needs to be small compared to the geometry and comparable to the cohesive length, as creep makes the response sensitive to the stress distribution within the cohesive zone.

The two applications demonstrate that these properties carry over to large-scale applications, considering domains which are hundreds of length scales in size. For the sea-ice floe, the loading rate determines whether the floe deforms by creep alone or fractures, with the adaptive refinement and time stepping covering the transition from hours of loading to dynamic, branching fracture within milliseconds. For the ice cliff, gravity is included as a body force that is not degraded by the phase-field, which requires a suitable strength surface: when this surface closes in compression, the partly damaged ice beside a crevasse fails under the weight of the ice above it, whereas restricting the fracture eigenstrain to opening results in straight crevasses. These crevasses grow in a stable manner, with the phase-field length scale influencing which of the notches develop into crevasses, but not the subsequent growth of these crevasses.

The presented results are limited to two-dimensional cases in which notches or defects prescribe where cracks can start, power-law creep with constant parameters, and, for the opening-only criterion, tensile failure. However, as the eigenstrains and creep strains are resolved at the integration points, extending the model to three dimensions, to temperature-dependent creep parameters, or to other creep laws only involves the local constitutive model, while the global solution scheme, the adaptive refinement, and the fracture-energy correction remain unchanged. Similarly, other strength surfaces can be included through the potentials of the fracture eigenstrain, allowing the failure criterion to be chosen to suit the loading conditions of the problem considered. The presented formulation thus provides a basis for simulating problems in which slow creep and rapid fracture act within the same domain, such as the calving of glaciers and ice shelves, the break-up of sea ice, and creep fracture of metals, polymers, and rocks at elevated temperatures, with the fracture process resolved only where and when it occurs.

\section*{Data availability}
\noindent The finite element code used within this study is available at \url{https://github.com/T-Hageman/FEMCubes}, and all input files required to reproduce the results presented in this paper are available at \url{https://github.com/T-Hageman/Paper_Cohesive_ViscoElastic}.

\section*{Acknowledgements}
\noindent No external funding was received for this research. The authors acknowledge computational resources provided by the University of Oxford Advanced Research Computing Service \citep{Richards2015}.

\section*{Declaration of competing interest}
\noindent The author declares that they have no known competing financial interests or personal relationships that could have appeared to influence the work reported in this paper.

\section*{Declaration of generative AI in the writing process}
\noindent During the preparation of this manuscript, the author used codex (GPT-6.1 Sol, OpenAI) and claude code (Opus 5.5, Anthropic) to assist with improving the language and readability of the text, and provide critical feedback on draft versions of the manuscript. AI was more heavily used for Appendix B and C, to assist with the derivations presented therein. The author reviewed and edited all AI-generated suggestions and takes full responsibility for the content of the published work.

\FloatBarrier

\appendix
\makeatletter
\@addtoreset{figure}{section}
\@addtoreset{table}{section}
\makeatother
\section{Jacobian for the power-law creep and two-direction fracture model}
\label{app:model_jacobian}

This appendix provides the matrices $\m{K}$ and $\m{C}$ from \cref{eq:rm_updated_jacobian,eq:rm_updated_strain_derivative} for the specific combination of the power-law creep model and the fracture potential and directions from \cref{eq:r2_2l_potentials,eq:r2_2l_directions}. The block rows and columns of $\m{K}$ follow the ordering: $\v{\upsigma}$, $\Delta\lambda_\text{v}$, $\lambda_1$, and $\lambda_2$, with the same block-row order used for $\m{C}$. All current quantities are evaluated at $t+\Delta t$, with the time superscript omitted. The phase-field and previous state are held fixed, and the derivatives are taken within the current tensile or compressive branch. The matrix $\m{K}$ is evaluated at fixed total strain, whereas $\m{C}$ is evaluated at fixed local state $\v{s}$.

The equivalent creep stress and its derivatives in Mandel notation are:
\begin{equation}
    \label{eq:app_creep_derivatives}
        f_\text{v}
        =\sqrt{\frac{n_\text{d}}{2}\v{\upsigma}^{\top}\m{P}_\text{dev}\v{\upsigma}}
        \qquad
        \frac{\partial f_\text{v}}{\partial\v{\upsigma}}
        =\frac{n_\text{d}}{2f_\text{v}}\m{P}_\text{dev}\v{\upsigma} \qquad
        \frac{\partial^2 f_\text{v}}{\partial\v{\upsigma}^{2}}
        =\frac{1}{f_\text{v}}\left(
        \frac{n_\text{d}}{2}\m{P}_\text{dev}
        -\frac{\partial f_\text{v}}{\partial\v{\upsigma}}
        \left(\frac{\partial f_\text{v}}{\partial\v{\upsigma}}\right)^{\top}\right)
\end{equation}
Here, $n_\text{d}=2$ for the intrinsic two-dimensional model, and $n_\text{d}=3$ in three dimensions or under plane-stress and plane-strain conditions. The corresponding derivatives of the fracture directions and eigenstrain are:
\begin{equation}
    \label{eq:app_direction_derivatives}
        \frac{\partial\v{G}_1}{\partial\v{\upvarepsilon}_\text{r}}
        =\m{0}
        \qquad
        \frac{\partial\v{G}_2}{\partial\v{\upvarepsilon}_\text{r}}
        =\frac{\m{P}_\text{dev}-\v{G}_2\v{G}_2^{\top}}
        {\lVert\m{P}_\text{dev}\v{\upvarepsilon}_\text{r}\rVert}\qquad 
        \frac{\partial\v{\upeta}}{\partial\v{\upvarepsilon}_\text{r}}=\lambda_2\frac{\partial\v{G}_2}{\partial\v{\upvarepsilon}_\text{r}}
\end{equation}
The volumetric direction is constant within each branch (but swaps between compression and tension), whereas the deviatoric direction changes with the strain remaining after creep to match the direction of the deviatoric strain.

Substituting $\v{\upeta}=\lambda_1\v{G}_1+\lambda_2\v{G}_2$ into the strength potential gives:
\begin{equation}
    \label{eq:app_strength_potential}
    F=
    \begin{cases}
        d(\phi)Q,&\v{i}^{\top}\v{\upvarepsilon}_\text{r}\geq0\\
        10^6f_\text{t}^*\sqrt{n_\text{d}}\lambda_1
        +d(\phi)f_\text{s}^*\lambda_2
        &\v{i}^{\top}\v{\upvarepsilon}_\text{r}<0
    \end{cases}
    \qquad
    Q=\sqrt{n_\text{d}{f_\text{t}^*}^{2}\lambda_1^2
    +{f_\text{s}^*}^{2}\lambda_2^2}
\end{equation}
\begin{samepage}
The factor $\sqrt{n_\text{d}}$ follows from $\v{i}^{\top}\v{G}_1=\pm\sqrt{n_\text{d}}$, while $\v{G}_2$ has unit deviatoric norm. The projected fracture resistances therefore satisfy:
\begin{equation}
    \label{eq:app_strength_gradient}
    \frac{\partial F}{\partial\v{\lambda}}
    =\begin{bmatrix}
        \v{G}_1^{\top}\dfrac{\partial F}{\partial\v{\upeta}}\\
        \v{G}_2^{\top}\dfrac{\partial F}{\partial\v{\upeta}}
    \end{bmatrix}
    =\begin{cases}
        \dfrac{d(\phi)}{Q}
        \begin{bmatrix}
            n_\text{d}{f_\text{t}^*}^{2}\lambda_1\\
            {f_\text{s}^*}^{2}\lambda_2
        \end{bmatrix}
        &\v{i}^{\top}\v{\upvarepsilon}_\text{r}\geq0\\
        \begin{bmatrix}
            10^6f_\text{t}^*\sqrt{n_\text{d}}\\
            d(\phi)f_\text{s}^*
        \end{bmatrix}
        &\v{i}^{\top}\v{\upvarepsilon}_\text{r}<0
    \end{cases}
\end{equation}
At fixed $\v{\upvarepsilon}_\text{r}$, the projected Hessian has entries $\partial^2 F/\partial\lambda_i\partial\lambda_j=\v{G}_i^{\top}(\partial^2 F/\partial\v{\upeta}^{2})\v{G}_j$ and is given explicitly by:
\begin{equation}
    \label{eq:app_strength_hessian}
    \frac{\partial^2 F}{\partial\v{\lambda}^{2}}
    =\begin{cases}
        \dfrac{d(\phi)n_\text{d}{f_\text{t}^*}^{2}{f_\text{s}^*}^{2}}{Q^3}
        \begin{bmatrix}
            \lambda_2^2&-\lambda_1\lambda_2\\
            -\lambda_1\lambda_2&\lambda_1^2
        \end{bmatrix}
        &\v{i}^{\top}\v{\upvarepsilon}_\text{r}\geq0,\\[3ex]
        \m{0}&\v{i}^{\top}\v{\upvarepsilon}_\text{r}<0
    \end{cases}
\end{equation}
\end{samepage}
The tensile branch couples the two fracture multipliers through this Hessian. On the compressive branch, the resistances are constant and the projected Hessian vanishes. At zero eigenstrain, fracture initiation is checked using the strength criterion; the tensile derivatives above apply for $Q>0$.

Within each branch, the projected resistances in \cref{eq:app_strength_gradient} depend only on the multipliers, so their derivatives with respect to $\v{\upvarepsilon}_\text{r}$ vanish at fixed multipliers. The remaining direction dependence of the active fracture residuals is therefore:
\begin{equation}
    \label{eq:app_residual_derivatives}
    \frac{\partial r_1}{\partial\v{\upvarepsilon}_\text{r}}
    =\v{0}^{\top}
    \qquad
    \frac{\partial r_2}{\partial\v{\upvarepsilon}_\text{r}}
    =-\v{\upsigma}^{\top}
    \frac{\partial\v{G}_2}{\partial\v{\upvarepsilon}_\text{r}}
\end{equation}
Substitution into \cref{eq:rm_updated_jacobian} gives the specialized Jacobian with both fracture directions active:
\begingroup
\small
\setlength{\arraycolsep}{3pt}
\begin{equation}
    \label{eq:app_local_jacobian}
    \m{K}=
    \begin{bmatrix}
        \m{I}+\Delta\lambda_\text{v}\m{D}_\text{el}
        \left(\m{I}-\dfrac{\partial\v{\upeta}}{\partial\v{\upvarepsilon}_\text{r}}\right)
        \dfrac{\partial^2 f_\text{v}}{\partial\v{\upsigma}^{2}}
        &\m{D}_\text{el}
        \left(\m{I}-\dfrac{\partial\v{\upeta}}{\partial\v{\upvarepsilon}_\text{r}}\right)
        \dfrac{\partial f_\text{v}}{\partial\v{\upsigma}}
        &\m{D}_\text{el}\v{G}_1
        &\m{D}_\text{el}\v{G}_2\\[2ex]
        -An\Delta t\,f_\text{v}^{n-1}
        \left(\dfrac{\partial f_\text{v}}{\partial\v{\upsigma}}\right)^{\top}
        &1&0&0\\[2ex]
        -\v{G}_1^{\top}&0
        &\dfrac{\partial^2 F}{\partial\lambda_1^2}+\kappa K
        &\dfrac{\partial^2 F}{\partial\lambda_1\partial\lambda_2}\\[2ex]
        -\v{G}_2^{\top}+\Delta\lambda_\text{v}\v{\upsigma}^{\top}
        \dfrac{\partial\v{G}_2}{\partial\v{\upvarepsilon}_\text{r}}
        \dfrac{\partial^2 f_\text{v}}{\partial\v{\upsigma}^{2}}
        &\v{\upsigma}^{\top}
        \dfrac{\partial\v{G}_2}{\partial\v{\upvarepsilon}_\text{r}}
        \dfrac{\partial f_\text{v}}{\partial\v{\upsigma}}
        &\dfrac{\partial^2 F}{\partial\lambda_2\partial\lambda_1}
        &\dfrac{\partial^2 F}{\partial\lambda_2^2}+\kappa K
    \end{bmatrix}
\end{equation}
\endgroup
The corresponding strain-derivative matrix follows from \cref{eq:rm_updated_strain_derivative,eq:app_direction_derivatives,eq:app_residual_derivatives}:
\begin{equation}
    \label{eq:app_strain_derivative}
    \m{C}=\frac{\partial\v{r}}{\partial\v{\upvarepsilon}}
    =\begin{bmatrix}
        -\m{D}_\text{el}\left(\m{I}-\lambda_2
        \dfrac{\partial\v{G}_2}{\partial\v{\upvarepsilon}_\text{r}}\right)\\[2ex]
        \v{0}^{\top}\\[1ex]
        \v{0}^{\top}\\[1ex]
        -\v{\upsigma}^{\top}
        \dfrac{\partial\v{G}_2}{\partial\v{\upvarepsilon}_\text{r}}
    \end{bmatrix}
\end{equation}
All derivatives in \cref{eq:app_local_jacobian,eq:app_strain_derivative} are supplied by \cref{eq:app_creep_derivatives,eq:app_direction_derivatives,eq:app_strength_hessian}. The terms involving $\partial\v{G}_2/\partial\v{\upvarepsilon}_\text{r}$ retain the change in deviatoric fracture direction during the creep update. For an inactive fracture direction, its residual is $r_i=\lambda_i$ and the corresponding row of $\m{K}$ is zero except for a unit entry in the column of $\lambda_i$, while its row of $\m{C}$ is zero. Expressions involving vanishing creep or deviatoric-strain invariants are evaluated using their limiting or regularized forms.

\section{Element-size error of the discrete cohesive law}
\label{app:Gc_correction}

This appendix derives the coefficient $c$ of the element-size error in \cref{eq:Gc_num} for the discretisation used in this work. It follows the one-dimensional analysis that Feng et al.\ \citep{Feng2026AMR} performed for the $\Omega^2$ model \citep{Feng2026}, in which the fracture is also represented by a sharply localised damage variable within a diffuse phase-field. The difference between the two coefficients arises from where the opened integration point sits relative to the phase-field nodes. The analysis is then generalised to phase-field elements of arbitrary polynomial order, which shows that the error remains first order in the element size for all orders, whereas its coefficient depends strongly on the interpolation order.

\subsection{Continuum reference}

We consider a bar in uniaxial tension as a one-dimensional domain, containing a single crack at $x=x_\text{c}$ with crack opening $\delta=\int\eta\,\text{d}x$. Once the crack has localised, the eigenstrain acts as a Dirac measure at the crack, and the phase-field evolution of \cref{eq:phase_weak} reduces (for $\kappa\to0$) to
\begin{equation}
    \label{eq:app_pf_dirac}
    \phi-\ell^2\frac{\text{d}^2\phi}{\text{d}x^2} = 2(1-\phi)\,\frac{f_\text{t}\,\delta\,\ell}{G_\text{c}}\,\delta_\text{D}(x-x_\text{c})
\end{equation}
where the driving force $F_\text{d}(\m{\eta})=f_\text{t}\eta$ follows from the strength potential in uniaxial tension. Away from the crack the phase-field decays exponentially, $\phi=\phi_\text{c}\,\text{e}^{-|x-x_\text{c}|/\ell}$, and integrating \cref{eq:app_pf_dirac} across the crack relates the crack opening to the peak value $\phi_\text{c}$:
\begin{equation}
    \label{eq:app_jump}
    \frac{f_\text{t}\,\delta}{G_\text{c}} = \frac{\phi_\text{c}}{1-\phi_\text{c}}
\end{equation}
At the crack the stress equals the degraded strength, $\sigma=f_\text{t}(1-\phi_\text{c})^2$. Eliminating $\phi_\text{c}$ gives the traction-separation law of the localised crack:
\begin{equation}
    \label{eq:app_tsl}
    \sigma(\delta) = f_\text{t}\left(\frac{G_\text{c}}{G_\text{c}+f_\text{t}\delta}\right)^2 \qquad \int_0^\infty \sigma\,\text{d}\delta = G_\text{c}
\end{equation}
This is the cohesive law obtained for eigenstrain-based phase-field models in the limit of a sharp crack \citep{Vicentini2026}, which dissipates exactly $G_\text{c}$ per unit crack area.

\subsection{Discrete problem with linear elements}

To show how this exact fracture energy changes by using the  finite element methods, we use linear elements with a constant length. The phase-field is interpolated linearly between nodes $x_i=ih$, using an element spanning $x=[0,h]$, and the integrals over each element are evaluated with two Gauss points. Its integration points lie at $\xi h$ and $(1-\xi)h$ within each element, with $\xi=\frac{1}{2}(1-1/\sqrt{3})$. The stress is uniform along the bar, while the strength $d(\phi)f_\text{t}$ at each integration point is evaluated with the phase-field interpolated at that point. Only the integration point closest to the phase-field maximum can therefore reach the strength surface, and all other points remain elastic. The eigenstrain occupies a single Gauss point, which carries the full crack opening $\delta=\tfrac{1}{2}h\,\eta_\text{q}$. This differs from the $\Omega^2$ model \citep{Feng2026AMR}, where the damage variable is constant within an element and the crack is taken through the centre of an element, with both of its nodes at the peak phase-field value. Here, the phase-field maximum is taken at node $0$ and the opened Gauss point lies at $x_\text{q}=\xi h$ in the adjacent element $[0,h]$, where the interpolated phase-field is
\begin{equation}
    \phi_\text{q} = (1-\xi)\phi_0+\xi\phi_1
\end{equation}

Outside the opened element the nodal equations for the phase-field are driving-force free, and their decaying solution is a geometric sequence on either side of the crack:
\begin{equation}
    \label{eq:app_decay}
    \phi_{-i}=\phi_0 r^i \qquad \phi_{i}=\phi_1 r^{i-1}\;(i\geq 1) \qquad r=\text{e}^{-kh} \qquad k=\frac{1}{\ell}\left(1+\frac{s^2}{24}\right)
\end{equation}
with $s=h/\ell$ and where terms of fourth order in $s$ are omitted from the decay rate $k$. The discrete profile thus matches the exponential decay of the continuum solution up to second-order terms, so the diffuse part of the phase-field only contributes second-order errors. The first-order error comes from the two nodes of the opened element. Their nodal equations, obtained with the source evaluated at the opened Gauss point, become after division by $h$:
\begin{align}
    \label{eq:app_node0}
    a\,\phi_0 - b\,\phi_1 &= 2(1-\phi_\text{q})(1-\xi)\,\hat{\delta}\\
    \label{eq:app_node1}
    -b\,\phi_0 + a\,\phi_1 &= 2(1-\phi_\text{q})\,\xi\,\hat{\delta}
\end{align}
with $a=(2-r)/s^2+(4+r)/6$, $b=1/s^2-1/6$, and the normalised crack opening $\hat{\delta}=f_\text{t}\delta\ell/(G_\text{c}h)$. The factors $(1-\xi)$ and $\xi$ are the values of the two nodal shape functions at the opened Gauss point. The sum and difference of these two equations give:
\begin{align}
    \label{eq:app_sum}
    (a-b)\,(\phi_0+\phi_1) &= 2(1-\phi_\text{q})\,\hat{\delta}\\
    \label{eq:app_diff}
    (a+b)\,(\phi_0-\phi_1) &= 2(1-\phi_\text{q})(1-2\xi)\,\hat{\delta}
\end{align}
Expanding the coefficients for small $s$ gives $a-b=1/s+1/2+s/24$, which combines the secant approximation of the gradient jump at the crack with the capacity contributions of the adjacent elements, and $a+b=2/s^2$ to leading order. From \cref{eq:app_diff}, the difference between the two nodal values is $\phi_0-\phi_1=(1-2\xi)\,s\,\bar{\phi}$, with $\bar{\phi}=\tfrac{1}{2}(\phi_0+\phi_1)$ the mean of the two nodal values: the nodes sample the exponential profile centred at the opened point, located a distance $\xi h$ from node $0$ and $(1-\xi)h$ from node $1$. The phase-field interpolated at the opened quadrature point follows as
\begin{equation}
    \phi_\text{q} = \bar{\phi}+\left(\tfrac{1}{2}-\xi\right)(\phi_0-\phi_1) = \bar{\phi}\left[1+\tfrac{1}{2}(1-2\xi)^2 s\right]
\end{equation}
Substituting this into \cref{eq:app_sum} and keeping terms up to first order in $s$ gives the discrete counterpart of \cref{eq:app_jump}:
\begin{equation}
    \label{eq:app_discrete_jump}
    \frac{f_\text{t}\,\delta}{G_\text{c}} = s\,\hat{\delta} = \frac{\bar{\phi}}{1-\phi_\text{q}}\left(1+\frac{s}{2}\right)
    = \frac{\phi_\text{q}}{1-\phi_\text{q}}\Bigl[1+ 2\xi(1-\xi)\,s\Bigr]
\end{equation}
The opened point transmits exactly the degraded strength, $\sigma=f_\text{t}(1-\phi_\text{q})^2$, so comparing \cref{eq:app_discrete_jump} with \cref{eq:app_jump,eq:app_tsl} shows that the discrete traction-separation law has the same form as the continuum law, but with the fracture energy replaced by
\begin{equation}
    \label{eq:app_Gc_num}
    G_\text{c}^\text{num} = G_\text{c}\left(1+c\,\frac{h}{\ell}\right) \qquad c = 2\xi(1-\xi)=\frac{1}{3}
\end{equation}
The shape of the cohesive law is thus preserved and only its dissipated energy is scaled, with the coefficient depending solely on the position $\xi$ of the opened integration point within the element. To first order, $c$ is the amount by which the linear interpolation underestimates the peak of the exponential profile when this peak lies a fraction $\xi$ into an element, $\phi^h(x_\text{c})=\phi_\text{c}[1-2\xi(1-\xi)s]$, after accounting for the capacity and gradient contributions. For a crack through the centre of an element, $\xi=1/2$ and $c=1/2$, recovering the coefficient obtained for the $\Omega^2$ model \citep{Feng2026AMR}, whereas for the present formulation the crack sits at a Gauss point and $c=1/3$. Solving the discrete equations \cref{eq:app_node0,eq:app_node1,eq:app_decay} numerically, without any truncation, gives $(G_\text{c}^\text{num}/G_\text{c}-1)\,\ell/h = 0.329$, $0.330$, and $0.332$ for $h/\ell=1/2$, $1/4$, and $1/8$ at the Gauss point, independent of the value of $\phi_\text{q}$, and $0.521$, $0.510$, and $0.505$ for $\xi=1/2$, matching the $1/2+s/24$ of the element-centred crack.

\begin{figure}
    \centering
    \begin{subfigure}{8cm}
        \includegraphics{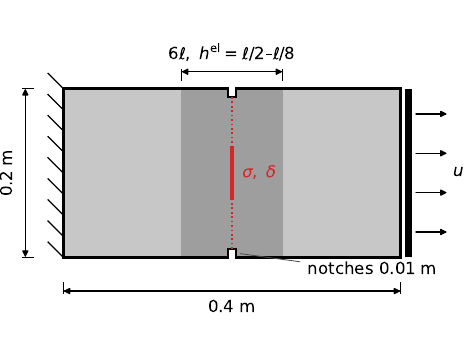}
        \caption{}
    \end{subfigure}
    \begin{subfigure}{8cm}
        \includegraphics{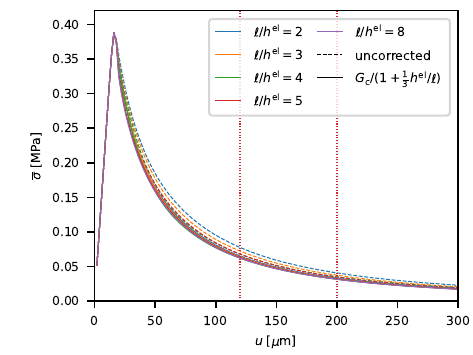}
        \caption{}
    \end{subfigure}
    \begin{subfigure}{8cm}
        \includegraphics{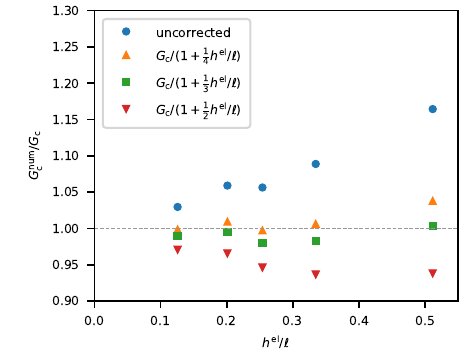}
        \caption{}
    \end{subfigure}
    \caption{Calibration of the element-size correction on a double-edge-notched bar: (a) geometry, with coarse elements away from the fracture and the fine mesh centre region shaded. Traction $\sigma$ and opening $\delta$  are extracted  along the crack line. (b) Stress against end displacement for $\ell/h^\text{el}=2$-$8$ with the uncorrected (dashed) and corrected (solid) fracture energy, with the dotted vertical lines marking the displacements at which the local cohesive law is read for figure (c). (c) resulting effective fracture energy of the discrete cohesive law as a function of the mean element size in the fracture zone, without correction and with the correction of \cref{eq:Gc_inp}.}
    \label{fig:GcCorrectionCalibration}
\end{figure}

\subsubsection{Calibration on a notched bar and correction}

On the triangular meshes used in this work, the crack crosses the elements at varying positions relative to the phase-field nodes, and the degree-three quadrature places the opened integration points at barycentric distances of $0.109$ and $0.232$ of the element height from the nearest edge. The one-dimensional analysis therefore only brackets the coefficient, $0\leq c\leq 1/2$, with its value along a crack being an average over the crossing positions. To verify \cref{eq:app_Gc_num} for the actual discretisation, the double-edge-notched bar of \cref{fig:GcCorrectionCalibration}a ($0.4\times0.2\;\text{m}$, notches of $0.01\;\text{m}$, $G_\text{c}=30\;\text{J/m}^2$ so that no snap-back occurs, no creep or inertia, and otherwise the material parameters of \cref{tab:parameters}) is loaded in tension until the bar has fractured, using uniform unstructured meshes with $\ell/h^\text{el}=2$, $3$, $4$, $5$, and $8$ in the centre region. The resulting load-displacement curves are shown in \cref{fig:GcCorrectionCalibration}b. The effective fracture energy of the discrete cohesive law is obtained at mid-height of the cross-section from the crack-line traction $\sigma$ and the crack opening $\delta$, with the opening taken as the jump of the displacement extrapolated linearly from $1.5$-$3.5\,\ell$ on either side of the crack. Inverting \cref{eq:app_tsl} gives $G_\text{c}^\text{num}=f_\text{t}\delta/(\sqrt{f_\text{t}/\sigma}-1)$, which is evaluated at the two end displacements marked in \cref{fig:GcCorrectionCalibration}b and averaged. \Cref{fig:GcCorrectionCalibration}c shows the resulting $G_\text{c}^\text{num}/G_\text{c}$ against the mean element size in the fracture zone. Without correction, the effective fracture energy increases linearly with $h^\text{el}/\ell$. With the corrected fracture energy of \cref{eq:Gc_inp} and $c=1/3$, the effective fracture energy is mostly mesh independent within some scatter, and the spread of the load-displacement curves of the bar for $\ell/h^\text{el}=2$-$8$ along the softening branch reduces from $11$-$23\%$ to $4$-$7\%$, without a systematic dependence on the element size. Two other coefficients are included in \cref{fig:GcCorrectionCalibration}c for comparison: with $c=1/4$ a residual overestimate remains that grows to $4\%$ on the coarsest mesh, whereas $c=1/2$ underestimates the fracture energy by $3$-$6\%$ with the error increasing with $h^\text{el}/\ell$. We note that, despite this correction, a small error persists due to the location of the crack within the element varying element-to-element. Furthermore, as $G_\text{c}$ only governs decohesion, not the strength surface, this correction does not impact the peak strength as shown in \cref{fig:GcCorrectionCalibration}(b).

\subsection{Discrete problem with elements of order $p$}
\label{app:Gc_order}

While only linear phase-field elements are used in this work, the extension to higher interpolation orders follows directly from the preceding analysis, and is included here to support future work on eigenstrain-based phase-field models. The analysis for linear elements solves the nodal equations of the opened element explicitly, which becomes impractical for higher interpolation orders. The same result can however be obtained without reference to the individual shape functions, which generalises it to phase-field elements of arbitrary polynomial order $p$. As before, the eigenstrain is confined to a single integration point $x_\text{q}$, which carries the full crack opening $\delta$ and is located a fraction $\xi$ into an element of length $h$, and the integrals of the phase-field terms are evaluated exactly. The discrete counterpart of \cref{eq:app_pf_dirac} is then to find $\phi^h$ in the finite element space $\mathcal{V}^h$ such that, for all test functions $w\in\mathcal{V}^h$:
\begin{equation}
    \label{eq:app_weak_p}
    a(\phi^h,w) = 2(1-\phi_\text{q})\,\frac{f_\text{t}\,\delta\,\ell}{G_\text{c}}\,w(x_\text{q}) \qquad a(\phi,w)=\int \phi\,w+\ell^2\,\frac{\text{d}\phi}{\text{d}x}\frac{\text{d}w}{\text{d}x}\;\text{d}x
\end{equation}
with $\phi_\text{q}=\phi^h(x_\text{q})$ the phase-field interpolated at the opened point. The left-hand side is linear in $\phi^h$ and the source acts at a single point, so the solution is a multiple of the discrete Green's function $g^h\in\mathcal{V}^h$ of the phase-field operator:
\begin{equation}
    \label{eq:app_green}
    a(g^h,w) = \ell\,w(x_\text{q}) \quad \forall\, w\in\mathcal{V}^h \qquad \phi^h = 2(1-\phi_\text{q})\,\frac{f_\text{t}\,\delta}{G_\text{c}}\,g^h
\end{equation}
Evaluating this solution at the opened point, with $g_\text{q}=g^h(x_\text{q})$, gives the relation between the crack opening and the phase-field at the opened point, and comparing it with \cref{eq:app_jump} gives the fracture energy of the discrete cohesive law:
\begin{equation}
    \label{eq:app_Gc_green}
    \frac{f_\text{t}\,\delta}{G_\text{c}} = \frac{\phi_\text{q}}{1-\phi_\text{q}}\,\frac{1}{2g_\text{q}} \qquad G_\text{c}^\text{num} = \frac{G_\text{c}}{2g_\text{q}}
\end{equation}
This result involves no truncation in $h/\ell$. For any interpolation order, element size, and position of the opened point, the discrete traction-separation law retains the shape of \cref{eq:app_tsl}, and the discretisation only enters through the value of the discrete Green's function at the opened point. In the continuum limit, the Green's function is $g=\tfrac{1}{2}\text{e}^{-|x-x_\text{q}|/\ell}$, for which $g(x_\text{q})=1/2$ and the fracture energy $G_\text{c}$ is recovered.

The continuum Green's function satisfies $a(g,w)=\ell\,w(x_\text{q})$ for all admissible $w$, including those in $\mathcal{V}^h$, so that $a(g-g^h,w)=0$ for all $w\in\mathcal{V}^h$. The discrete Green's function is therefore the best approximation of $g$ within $\mathcal{V}^h$, as measured in the energy norm $\|w\|_a^2=a(w,w)$. Using this orthogonality together with $a(g,g)=\ell/2$ and $a(g^h,g^h)=\ell\,g_\text{q}$ gives:
\begin{equation}
    \label{eq:app_energy_error}
    \frac{1}{2}-g_\text{q} = \frac{1}{\ell}\,\|g-g^h\|_a^2 = \frac{1}{\ell}\min_{w\in\mathcal{V}^h}\|g-w\|_a^2 \qquad \frac{G_\text{c}^\text{num}}{G_\text{c}} = \left(1-\frac{2}{\ell}\,\|g-g^h\|_a^2\right)^{-1}
\end{equation}
The discrete cohesive law thus always dissipates more energy than $G_\text{c}$, irrespective of the interpolation order, and the excess is set by how well the finite element space is able to represent the continuum phase-field profile around the opened point.

Away from the opened point, $g$ is smooth and its contribution to the error decreases with $h^{2p}$. Within the opened element, however, the profile has a kink at $x_\text{q}$, where its gradient jumps from $1/(2\ell)$ to $-1/(2\ell)$, and this kink cannot be represented by a polynomial of any order. To leading order in $h/\ell$, the error is therefore that of approximating this jump in the gradient term of the opened element, $x\in[0,h]$:
\begin{equation}
    \label{eq:app_projection}
    \|g-g^h\|_a^2 \approx \ell^2\int_0^h\left(\frac{\text{d}g}{\text{d}x}-\Pi_{p-1}\frac{\text{d}g}{\text{d}x}\right)^2\text{d}x = \frac{h}{8}\int_{-1}^{1}\Bigl(\text{sgn}(z-z_\text{q})-\Pi_{p-1}\,\text{sgn}(z-z_\text{q})\Bigr)^2\,\text{d}z
\end{equation}
with the element coordinate $z=2x/h-1$, the opened point at $z_\text{q}=2\xi-1$, and $\Pi_{p-1}$ the $L_2$-projection onto the polynomials of degree $p-1$, the space containing the gradients of the shape functions. Since this projection preserves the mean gradient over the element, the best approximation interpolates $g$ at both ends of the opened element, and the adjacent elements are not affected. Expanding the step function in the Legendre polynomials $P_k$, the projection removes its first $p$ terms:
\begin{equation}
    \label{eq:app_legendre}
    \frac{1}{2}\int_{-1}^{1}\Bigl(\text{sgn}(z-z_\text{q})-\Pi_{p-1}\,\text{sgn}(z-z_\text{q})\Bigr)^2\,\text{d}z = 1-\sum_{k=0}^{p-1}(2k+1)\,a_k^2 \qquad a_k=\frac{1}{2}\int_{-1}^{1}\text{sgn}(z-z_\text{q})\,P_k(z)\;\text{d}z
\end{equation}
with $a_0=-z_\text{q}$ and $a_k=(1-z_\text{q}^2)\,P_k'(z_\text{q})/(k(k+1))$ for $k\geq1$. Substituting \cref{eq:app_projection,eq:app_legendre} into \cref{eq:app_energy_error} gives the element-size error for elements of order $p$, to first order in $h/\ell$:
\begin{equation}
    \label{eq:app_Gc_num_p}
    G_\text{c}^\text{num} = G_\text{c}\left(1+c_p\,\frac{h}{\ell}\right) \qquad c_p = 2\xi(1-\xi)\left[1-4\xi(1-\xi)\sum_{k=1}^{p-1}\frac{2k+1}{k^2(k+1)^2}\Bigl(P_k'(2\xi-1)\Bigr)^2\right]
\end{equation}
For linear elements the sum is empty, and the coefficient of \cref{eq:app_Gc_num} is recovered. For quadratic and cubic elements, \cref{eq:app_Gc_num_p} results in:
\begin{equation}
    \label{eq:app_c_p23}
    c_2 = 2\xi(1-\xi)\Bigl[1-3\xi(1-\xi)\Bigr] \qquad c_3 = 2\xi(1-\xi)\Bigl[1-8\xi(1-\xi)+20\xi^2(1-\xi)^2\Bigr]
\end{equation}
On unstructured meshes the crack crosses the elements at varying positions, for which the mean of the coefficient over all positions within the element is the relevant measure:
\begin{equation}
    \label{eq:app_c_mean}
    \bar{c}_p = \int_0^1 c_p\;\text{d}\xi = \frac{p}{4p^2-1}
\end{equation}
giving $\bar{c}_p=1/3$, $2/15$, $3/35$, and $4/63$ for $p=1$-$4$.

\begin{figure}
    \centering
    \begin{subfigure}{8cm}
        \includegraphics{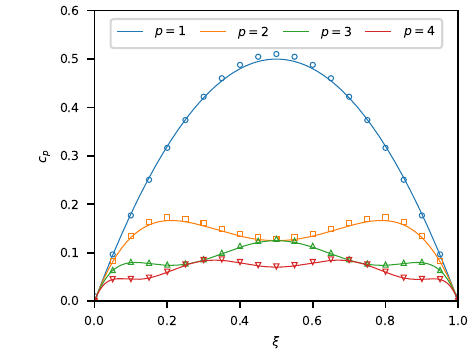}
        \caption{}
    \end{subfigure}
    \begin{subfigure}{8cm}
        \includegraphics{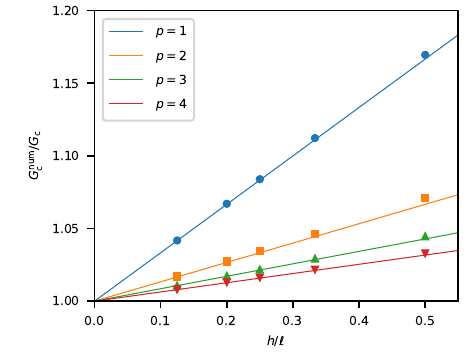}
        \caption{}
    \end{subfigure}
    \caption{Element-size error of the discrete cohesive law for phase-field elements of order $p$: (a) coefficient $c_p$ as a function of the position $\xi$ of the opened point within the element, with the lines showing \cref{eq:app_Gc_num_p} and the symbols the numerical solution of the discrete problem, \cref{eq:app_Gc_green}, for $h/\ell=1/4$. (b) Effective fracture energy averaged over all positions of the opened point as a function of the element size, with the symbols showing the numerical solution and the lines the first-order result $1+\bar{c}_p\,h/\ell$ using \cref{eq:app_c_mean}.}
    \label{fig:GcCorrectionOrder}
\end{figure}

The coefficients for $p=1$-$4$ are shown in \cref{fig:GcCorrectionOrder}a, together with the coefficients obtained by solving the discrete problem of \cref{eq:app_weak_p} numerically, without the leading-order approximation of \cref{eq:app_projection}. The numerical coefficients follow \cref{eq:app_Gc_num_p} for all orders and positions, with a deviation that is itself proportional to $h/\ell$ and thus only contributes to the second-order error. \Cref{fig:GcCorrectionOrder}b shows the resulting effective fracture energy, averaged over the position of the opened point, as a function of the element size. Two observations follow. First, the error remains first order in $h/\ell$ for all interpolation orders. Increasing the order improves the approximation of the smooth parts of the phase-field profile, but not of the kink at the crack, which lies inside an element as the integration points never coincide with the element boundaries. Higher-order elements therefore do not remove the need for the correction of \cref{eq:Gc_inp}, nor do they improve the rate at which an uncorrected simulation converges upon mesh refinement. Second, the coefficient depends strongly on the interpolation order, decreasing as $\bar{c}_p\approx1/(4p)$. A correction using the coefficient of linear elements would thus be more than twice as large as the error of quadratic elements, and the coefficient needs to be matched to the order of the phase-field interpolation. Expressed in terms of the node spacing $h/p$, the error coefficient $p\,\bar{c}_p$ only decreases from $1/3$ for linear elements towards $1/4$ for high orders, such that increasing the interpolation order reduces the error by about as much as refining a mesh of linear elements to the same node spacing.

The dependence on the position of the opened point also weakens with increasing order. For linear elements the coefficient varies between zero and $1/2$ across the element, whereas for quadratic elements it remains below $1/6$. For linear elements, the mean coefficient furthermore coincides with the coefficient at the Gauss point, $\bar{c}_1=1/3$, consistent with the calibration of \cref{fig:GcCorrectionCalibration}c, in which the crack crosses the triangular elements at varying positions. As for linear elements, the coefficients for $p>1$ are derived for a one-dimensional bar, and should be verified for the actual discretisation with a calibration similar to that of \cref{fig:GcCorrectionCalibration} before being used on unstructured two-dimensional meshes.

\section{Variational consistency of the eigenstrain directions}
\label{app:directions}

In \cref{sec:fracture_constitutive}, the fracture eigenstrain was expressed through a set of multipliers and direction tensors, \cref{eq:eigenstrain_directions}, with the direction tensors of \cref{eq:r2_2l_directions,eq:opening_directions} depending on the strain remaining after viscous creep. As in \cref{eq:rm_fracture_strain}, this strain is denoted as $\m{\varepsilon}_\text{r}=\m{\varepsilon}-\m{\varepsilon}_\text{v}$ for brevity. As the direction tensors change with both the strain and the creep strain, and two separate definitions of these eigenstrain directions are used, it is not directly evident that the resulting constitutive model still follows from the variational principle of \cref{eq:variational_principle}, and is thermodynamically consistent.

To demonstrate this, we consider the local energy density $\psi=\psi_\text{el}(\m{\varepsilon}_\text{r}-\m{\eta})+d(\phi)F_\text{d}(\m{\eta})+F_\text{i}(\m{\eta})$, and take its variation with respect to the strain while keeping the multipliers fixed:
\begin{equation}
    \label{eq:app_dir_question}
    \left.\delta\psi\right|_{\lambda_i} = \m{\sigma}:\delta\m{\varepsilon}_\text{r} + \m{X}:\sum_i\lambda_i\left(\frac{\partial\m{G}_i}{\partial\m{\varepsilon}_\text{r}}:\delta\m{\varepsilon}_\text{r}\right) \qquad \m{X}=\frac{\partial\psi}{\partial\m{\eta}}=-\m{\sigma}+d(\phi)\frac{\partial F_\text{d}}{\partial\m{\eta}}+\frac{\partial F_\text{i}}{\partial\m{\eta}}
\end{equation}
where the matrix $\m{X}$ contains the terms of the fracture constitutive model from \cref{eq:fracture_weak}. The first term is the stress used in the momentum balance, \cref{eq:stress_constitutive}, and, for variations of the creep strain, the stress driving the creep in \cref{eq:creep_weak}. The second term results from the direction tensors changing with the strain, and is included in neither of these relations. If this term were non-zero, the stress would no longer be the derivative of the energy, and the model would not be variationally consistent. In this appendix, we show that this term vanishes for both criteria used in this work. This is a result of the constitutive choice being made through the set of fracture eigenstrains that are admissible, which differs between the two criteria and determines the strength surface. The direction tensors are not chosen independently of this set: they are the directions of the eigenstrain tensor that fulfils \cref{eq:fracture_weak} for all admissible variations within it. Furthermore, we derive the dissipation of the model, and discuss the requirements under which these results hold.

\subsection{Eigenstrain problem}
We first consider the fracture eigenstrain $\m{\eta}$ itself as the local variable, without prescribing its directions. This eigenstrain is restricted to a set of admissible eigenstrains $\mathcal{K}$. For the two-multiplier criterion of \cref{eq:r2_2l_potentials}, this set contains all symmetric tensors, with interpenetration being penalized through $F_\text{i}$ rather than excluded. For the opening-only criterion of \cref{sec:opening_criterion}, the set only contains the positive semidefinite symmetric tensors. This set is the constitutive choice that distinguishes the two criteria. Both sets are convex and invariant under rotations, such that a rotated admissible eigenstrain is again admissible. The admissible variations are then given by the difference between any admissible eigenstrain $\m{\eta}^*$ and the current eigenstrain, $\delta\m{\eta}=\m{\eta}^*-\m{\eta}$, for which the local form of \cref{eq:fracture_weak} becomes:
\begin{equation}
    \label{eq:app_dir_inequality}
    0\leq\m{X}:\left(\m{\eta}^*-\m{\eta}\right) \qquad \forall\,\m{\eta}^*\in\mathcal{K}
\end{equation}
We note that the strength potentials are not differentiable at $\m{\eta}=\m{0}$ and at $\text{tr}(\m{\eta})=0$. At these states, the derivatives in $\m{X}$ are interpreted as derivatives in the direction of the variation, similar to the limit used to obtain the strength surface in \cref{eq:r2_strength_onset_energy}.

The elastic energy of \cref{eq:elastic_energy} is strictly convex in the fracture eigenstrain, and the potentials $F_\text{d}$ and $F_\text{i}$ are convex. As a result, an eigenstrain fulfils \cref{eq:app_dir_inequality} if, and only if, it minimizes the local energy density over the set of admissible eigenstrains, with this minimizer being unique:
\begin{equation}
    \label{eq:app_dir_minimisation}
    \m{\eta}=\underset{\m{\eta}^*\in\mathcal{K}}{\arg\min}\;\Big(\psi_\text{el}(\m{\varepsilon}_\text{r}-\m{\eta}^*)+d(\phi)F_\text{d}(\m{\eta}^*)+F_\text{i}(\m{\eta}^*)\Big)
\end{equation}
where the strain, creep strain, and phase-field are kept fixed at their current values. The fracture constitutive model can therefore equivalently be stated as a local minimization problem, in which the fracture eigenstrain reduces the elastic energy as far as allowed by the energetic cost of the strength potentials.

\subsection{Directions of the minimizer}
Next, we show that the directions of this minimizer follow from the strain remaining after viscous creep. The elastic energy and both strength potentials are isotropic, depending on their arguments solely through invariants. Considering an orthogonal tensor $\m{Q}$ that leaves the strain unchanged, $\m{Q}\m{\varepsilon}_\text{r}\m{Q}^\top=\m{\varepsilon}_\text{r}$, the rotated eigenstrain $\m{Q}\m{\eta}\m{Q}^\top$ is admissible and, due to the isotropy, results in the same energy in \cref{eq:app_dir_minimisation} as $\m{\eta}$. As the minimizer is unique, both need to be equal, $\m{Q}\m{\eta}\m{Q}^\top=\m{\eta}$. Using the reflections $\m{Q}=\m{I}-2\v{n}_i\otimes\v{n}_i$, with $\v{n}_i$ the unit eigenvectors of $\m{\varepsilon}_\text{r}$, this requires each eigenvector of $\m{\varepsilon}_\text{r}$ to also be an eigenvector of $\m{\eta}$. The fracture eigenstrain thus shares its principal directions with the strain remaining after viscous creep:
\begin{equation}
    \label{eq:app_dir_coaxial}
    \m{\eta}=\sum_{i=1}^{3}\eta_i\,\v{n}_i\otimes\v{n}_i
\end{equation}
with $\eta_i$ the principal fracture eigenstrains. When two or more principal strains coincide, any set of orthogonal eigenvectors can be used, as the corresponding principal eigenstrains are then equal as well.

For the opening-only criterion, the admissible eigenstrains have non-negative principal values, $\eta_i\geq0$. \Cref{eq:app_dir_coaxial} is then directly the multiplier form of \cref{eq:eigenstrain_directions}, with the direction tensors of \cref{eq:opening_directions} and $\lambda_i=\eta_i$.

For the two-multiplier criterion, this representation can be reduced further, as the potentials of \cref{eq:r2_2l_potentials} only depend on the eigenstrain through $\text{tr}(\m{\eta})$ and $\lVert\text{dev}(\m{\eta})\rVert$. For given values of these two invariants, expanding the elastic energy of \cref{eq:elastic_energy} shows that the only contribution depending on the orientation of the eigenstrain is $-2\mu\,\text{dev}(\m{\varepsilon}_\text{r}):\text{dev}(\m{\eta})$. This term is minimized when the deviatoric part of the eigenstrain is aligned with the deviatoric part of the strain, giving the direction $\m{G}_2$. Similarly, a volumetric eigenstrain with a sign opposite to that of $\text{tr}(\m{\varepsilon}_\text{r})$ would increase the volumetric elastic energy without reducing either of the potentials, such that the trace of the eigenstrain takes the sign of $\text{tr}(\m{\varepsilon}_\text{r})$, giving the two branches of $\m{G}_1$. The minimizer is therefore given by $\m{\eta}=\lambda_1\m{G}_1+\lambda_2\m{G}_2$ with $\lambda_i\geq0$, using the direction tensors of \cref{eq:r2_2l_directions}.

For both criteria, the constitutive relations of \cref{eq:fracture_multiplier_derivative} are the optimality conditions of \cref{eq:app_dir_minimisation} when only eigenstrains of the form of \cref{eq:eigenstrain_directions} are considered. As this restricted problem is also convex, and contains the minimizer of the full problem, both result in the same fracture eigenstrain. Introducing the multipliers thus does not weaken the fracture constitutive model: \cref{eq:app_dir_inequality} is fulfilled for variations in all components of the eigenstrain tensor, and the stress remains within the strength surface for all stress components, rather than solely in the directions $\m{G}_i$.

The two definitions of the direction tensors thus do not represent two choices of directions within the same model, but two choices of the set of admissible eigenstrains, with each set of direction tensors, taken over all strain states and non-negative multipliers, spanning its set. It is this set that determines the strength surface. Evaluating \cref{eq:app_dir_inequality} for $\m{\eta}=\m{0}$, in the same manner as \cref{eq:r2_strength_onset_energy}, the admissible stress states are given by:
\begin{equation}
    \label{eq:app_dir_strength_surface}
    \m{\sigma}:\m{\eta}^*\leq d(\phi)F_\text{d}(\m{\eta}^*)+F_\text{i}(\m{\eta}^*) \qquad \forall\,\m{\eta}^*\in\mathcal{K}
\end{equation}
Each admissible eigenstrain thus imposes a limit on the stress, with the strength surface being the envelope of these limits. Removing eigenstrains from $\mathcal{K}$ removes their limits, and thereby enlarges the strength surface. The opening-only criterion uses the same potential $F_\text{d}$ as the two-multiplier criterion, but excludes all eigenstrains with a negative principal value. This removes all limits on stress states without a tensile principal stress, resulting in the strength surface being open in compression, \cref{fig:opening_surface}. The set of admissible eigenstrains, and thereby the direction tensors, can therefore be chosen to obtain the required strength surface without affecting the variational consistency, as the derivation above holds for any set that is convex and invariant under rotations. In contrast, the orientation of the eigenstrain within this set cannot be chosen, as it follows from $\m{\varepsilon}_\text{r}$ through \cref{eq:app_dir_coaxial}. The direction tensors therefore need to be the principal directions of $\m{\varepsilon}_\text{r}$, or combinations of these that the potentials allow for, as is the case for $\m{G}_1$ and $\m{G}_2$.

\subsection{Stress, creep driving force, and dissipation}
With the directions established, we return to the second term of \cref{eq:app_dir_question}. For the two-multiplier criterion, the derivatives of the direction tensors are given in \cref{eq:app_direction_derivatives}: $\m{G}_1$ is constant within each branch, while changes in $\m{G}_2$ are deviatoric and orthogonal to $\m{G}_2$ itself. In contrast, the stress and the derivatives of the strength potentials, and therefore $\m{X}$, only contain components in the directions $\m{I}$ and $\m{G}_2$. When switching between the tensile and compressive branches of $\m{G}_1$, $\text{tr}(\m{\varepsilon}_\text{r})=0$ and the volumetric multiplier is zero, such that the fracture eigenstrain remains continuous. For the opening-only criterion with distinct principal strains $\varepsilon_{\text{r},i}$, a variation in the strain changes the direction tensors by:
\begin{equation}
    \label{eq:app_dir_dyad_variation}
    \delta\left(\v{n}_i\otimes\v{n}_i\right)=\sum_{j\neq i}\frac{\v{n}_i^\top\delta\m{\varepsilon}_\text{r}\v{n}_j}{\varepsilon_{\text{r},i}-\varepsilon_{\text{r},j}}\left(\v{n}_i\otimes\v{n}_j+\v{n}_j\otimes\v{n}_i\right)
\end{equation}
which, when expressed in the principal directions, only contains off-diagonal components, whereas $\m{X}$ shares its principal directions with $\m{\varepsilon}_\text{r}$ and thus only contains diagonal components. As a result, for both criteria:
\begin{equation}
    \label{eq:app_dir_orthogonality}
    \m{X}:\frac{\partial\m{G}_i}{\partial\m{\varepsilon}_\text{r}}=\m{0}
\end{equation}
and the second term of \cref{eq:app_dir_question} vanishes. Physically, a change in the directions only alters the fracture eigenstrain in components for which the energy is stationary. The stress from \cref{eq:stress_constitutive} is therefore the complete derivative of the energy with respect to the strain, including the changes in the fracture eigenstrain caused by this strain. Similarly, $-\m{\sigma}$ is the complete derivative with respect to the creep strain, as used in the creep constitutive model of \cref{eq:creep_weak}. For a fixed creep strain and phase-field, the material response is thus hyperelastic, with the minimum of \cref{eq:app_dir_minimisation} acting as its energy potential.

The dissipation of the model is obtained in a similar manner. Taking the time derivative of the local energy density, using the multiplier form of \cref{eq:eigenstrain_directions}, gives:
\begin{equation}
    \label{eq:app_dir_energy_rate}
    \dot{\psi}=\m{\sigma}:\left(\m{\dot{\varepsilon}}-\m{\dot{\varepsilon}}_\text{v}\right)+\m{X}:\sum_i\left(\dot{\lambda}_i\m{G}_i+\lambda_i\dot{\m{G}}_i\right)+\frac{\partial d}{\partial\phi}F_\text{d}(\m{\eta})\,\dot{\phi}
\end{equation}
where the second term contains the contributions of the fracture eigenstrain, and is equal to zero. The changes in the directions do not contribute due to \cref{eq:app_dir_orthogonality}, while the changes in the multipliers do not contribute as \cref{eq:fracture_multiplier_derivative} holds as an equality, $\m{X}:\m{G}_i=0$, for all directions in which the multiplier is non-zero or starts to increase. Subtracting the remaining terms from the stress power $\m{\sigma}:\m{\dot{\varepsilon}}$, and using the incompressibility of the creep strain and the degradation function of \cref{eq:strength_degradation}, results in the dissipation:
\begin{equation}
    \label{eq:dissipation}
    \m{\sigma}:\m{\dot{\varepsilon}}-\dot{\psi}=\m{\sigma}_\text{dev}:\m{\dot{\varepsilon}}_\text{v}+2(1-\kappa)(1-\phi)F_\text{d}(\m{\eta})\,\dot{\phi}\;\geq0
\end{equation}
The first term is the creep dissipation, which is non-negative following \cref{eq:creep_constitutive}. The second term is the energy released from the strength potential by the loss of cohesion, which acts as the driving term for the phase-field in \cref{eq:phase_weak}, and is non-negative as the phase-field can only increase. The fracture eigenstrain itself does not dissipate energy: at a fixed phase-field, it is fully recovered upon unloading, with all dissipation caused by either the creep or the loss of cohesion.

\subsection{Time-discrete problem and requirements}
Finally, we consider the time-discrete setting of the return-mapping scheme in \cref{sec:return_mapping}. Here, the directions are evaluated from the strain remaining after creep at the end of the time increment, \cref{eq:rm_fracture_strain}, which includes the creep increment that is solved for simultaneously with the fracture multipliers. The residuals of \cref{eq:residual_stress,eq:residual_creep,eq:residual_fracture} are then the optimality conditions of the incremental minimization problem:
\begin{equation}
    \label{eq:app_dir_incremental}
    \min_{\Delta\m{\varepsilon}_\text{v},\;\m{\eta}\in\mathcal{K}}\;\;\psi_\text{el}\left(\m{\varepsilon}-\m{\varepsilon}_\text{v}^t-\Delta\m{\varepsilon}_\text{v}-\m{\eta}\right)+d(\phi)F_\text{d}(\m{\eta})+F_\text{i}(\m{\eta})+\Delta t\,\frac{n}{n+1}A^{-1/n}\left(\frac{2}{3}\frac{\Delta\m{\varepsilon}_\text{v}:\Delta\m{\varepsilon}_\text{v}}{\Delta t^2}\right)^{\frac{n+1}{2n}}
\end{equation}
subject to $\text{tr}(\Delta\m{\varepsilon}_\text{v})=0$, where the last term is the creep potential of \cref{eq:creep_potential}, evaluated for the backward-Euler creep rate $\Delta\m{\varepsilon}_\text{v}/\Delta t$. This incremental problem is isotropic, and strictly convex in the combination of the creep increment and the fracture eigenstrain. Following the same arguments as above, both the creep increment and the fracture eigenstrain share their principal directions with the trial strain $\m{\varepsilon}-\m{\varepsilon}_\text{v}^t$, and therefore with $\m{\varepsilon}_\text{r}$. As a result, the stress is the derivative of the minimum of \cref{eq:app_dir_incremental} with respect to the strain, and the consistent tangent of the combined creep and fracture update is symmetric. The residual stiffness included in \cref{eq:residual_fracture} corresponds to an additional energy contribution $\frac{1}{2}\kappa K\,\m{\eta}:\m{\eta}$, which is isotropic and convex, and thus does not alter these arguments.

It is stressed that these results rely on three requirements: the elastic response and the strength potentials need to be isotropic, the set of admissible eigenstrains needs to be invariant under rotations, and the directions need to be evaluated from the current strain remaining after creep. If any of these requirements is not fulfilled, for instance for an anisotropic strength, or when the directions are evaluated from the total strain or from the previous time increment, the minimizer of \cref{eq:app_dir_minimisation} can no longer be expressed through \cref{eq:eigenstrain_directions} using these directions, and \cref{eq:app_dir_orthogonality} no longer holds. In such cases, either the second term of \cref{eq:app_dir_question} needs to be included in the stress and in the creep constitutive model, or \cref{eq:app_dir_minimisation} needs to be solved for the full eigenstrain tensor.

\bibliography{References}

\end{document}